\documentclass[acmtog,conference,nonacm]{acmart}

\usepackage{booktabs} %
\usepackage{color}
\usepackage[english]{babel}
\usepackage{multirow}
\usepackage{multicol}
\usepackage{makecell}
\usepackage{graphicx}
\usepackage{url}
\usepackage{subfigure}
\usepackage{dsfont}
\usepackage[toc,page]{appendix}
\usepackage{arydshln}
\usepackage{amssymb}
\usepackage{hyperref}
\usepackage{amsmath,tikz}
\usetikzlibrary{arrows,chains,matrix,positioning,scopes}
\usepackage{lipsum}
\usepackage[ruled]{algorithm2e} %
\usepackage{xspace}

\copyrightyear{2019}
\acmPrice{15.00}
\setcopyright{acmcopyright}
\acmConference[SIGGRAPH ASIA'19]{ACM Conference}{November 2019}{Brisbane, Australia}
\acmSubmissionID{108}
\setcitestyle{square} %
\acmBooktitle{}
\acmPrice{}
\acmDOI{}
\acmISBN{}

\newcommand{\name}{\textsc{SDM-NET}\xspace}

\renewcommand{\footnoterule}{%
	\kern -3pt
	\hrule width 0.477\textwidth height 0.2pt
	\kern 2pt
}

\SetAlFnt{\small}
\SetAlCapFnt{\small}
\SetAlCapNameFnt{\small}
\SetAlCapHSkip{0pt}

\newcommand{\rz}[1]{{\color{black}#1}}
\newcommand{\rv}[1]{{\color{black}#1}}
\newcommand{\YLN}[1]{{\color{black}#1}}
\newcommand{\GL}[1]{{\color{black}#1}}
\newcommand{\YL}[1]{{\color{black}#1}}

\newcommand{\yyja}[1]{{\color{black}#1}}

\newcommand{\hongbo}[1]{{\color{black}#1}}

\newcommand{\tabincell}[2]{\begin{tabular}{@{}#1@{}}#2\end{tabular}}

\begin{document}

\title{\name: Deep Generative Network for Structured Deformable Mesh}
\author{Lin Gao}
\affiliation{%
	\institution{Institute of Computing Technology, Chinese Academy of Sciences (CAS)}}
\email{gaolin@ict.ac.cn}

\author{Jie Yang}
\affiliation{%
	\institution{Institute of Computing Technology, CAS and University of Chinese Academy of Sciences}}
\email{yangjie01@ict.ac.cn}

\author{Tong Wu}
\affiliation{%
	\institution{Institute of Computing Technology, CAS and University of Chinese Academy of Sciences}}
\email{wutong@ict.ac.cn}

\author{Yu-Jie Yuan}
\affiliation{%
	\institution{Institute of Computing Technology, CAS and University of Chinese Academy of Sciences}}
\email{yuanyujie@ict.ac.cn}

\author{Hongbo Fu}
\affiliation{%
	\institution{\hongbo{School of Creative Media, }City University of Hong Kong}}

\author{Yu-Kun Lai}
\affiliation{%
	\institution{\hongbo{School of Computer Science and Informatics}, Cardiff University}}

\author{Hao Zhang}
\affiliation{%
	\institution{\hongbo{School of Computing Science}, Simon Fraser University}}
\authorsaddresses{Webpage: \url{http://geometrylearning.com/sdm-net/} \\This is the author's version of the work. It is posted here for your personal use. Not for redistribution. }
\renewcommand{\shortauthors}{Gao, Yang, Wu, Yuan, Fu, Lai and Zhang}

\begin{teaserfigure}
\centering
{
        \includegraphics[width=0.16\linewidth]{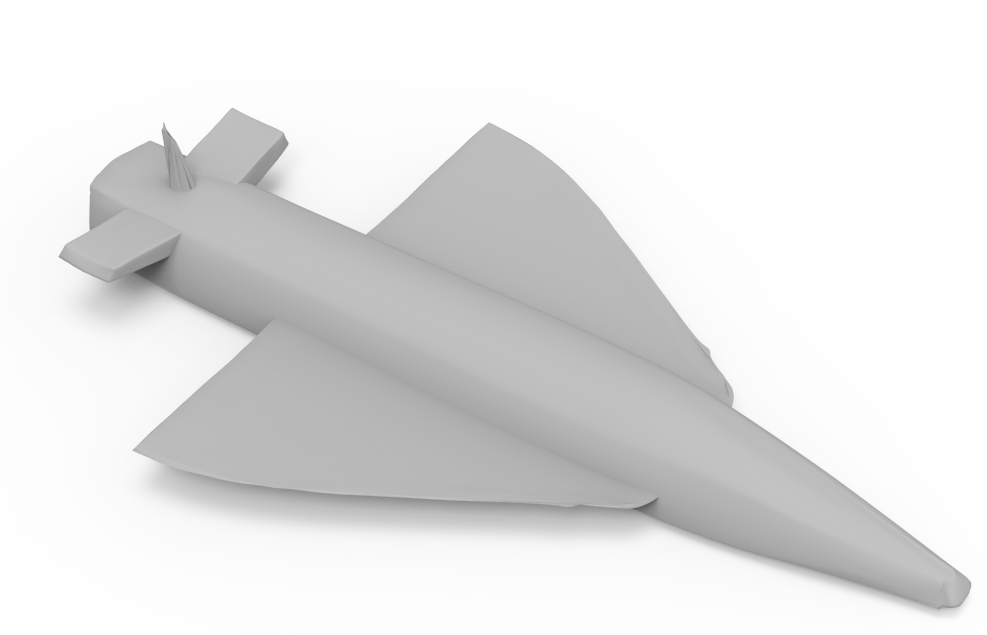}
		\includegraphics[width=0.16\linewidth]{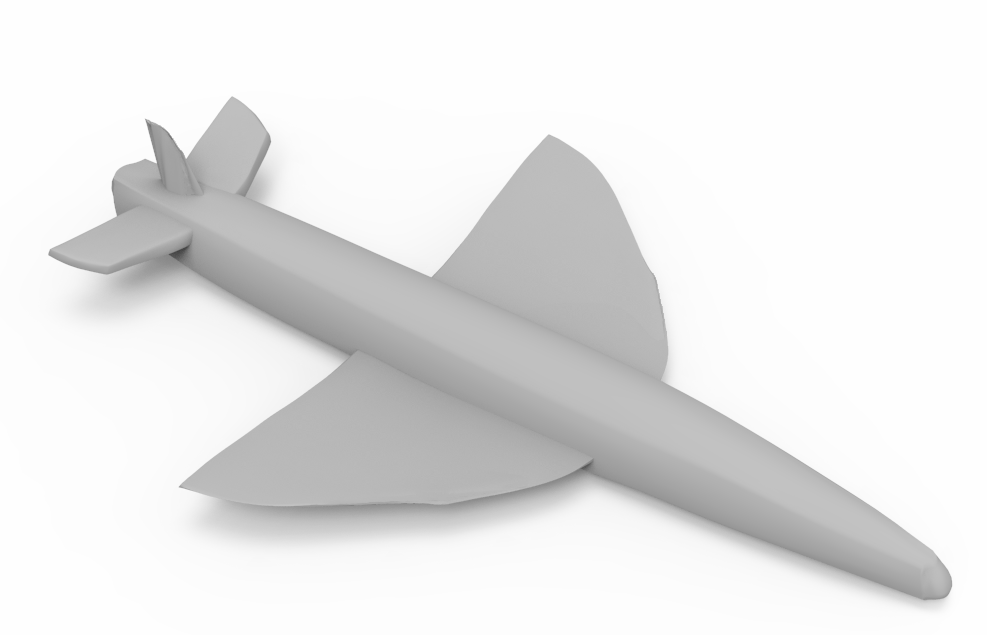}
		\includegraphics[width=0.16\linewidth]{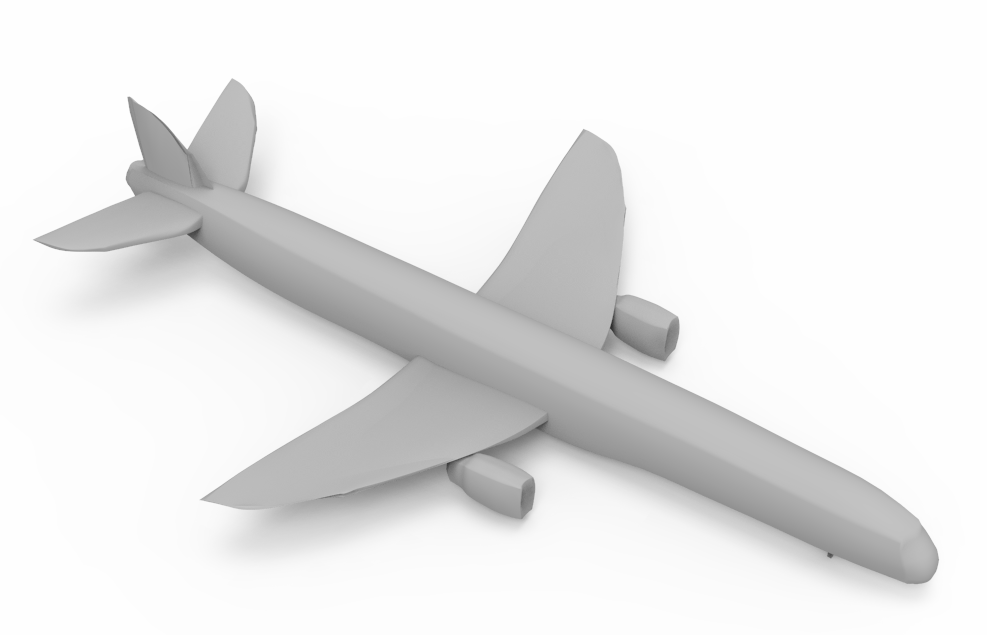}
		\includegraphics[width=0.16\linewidth]{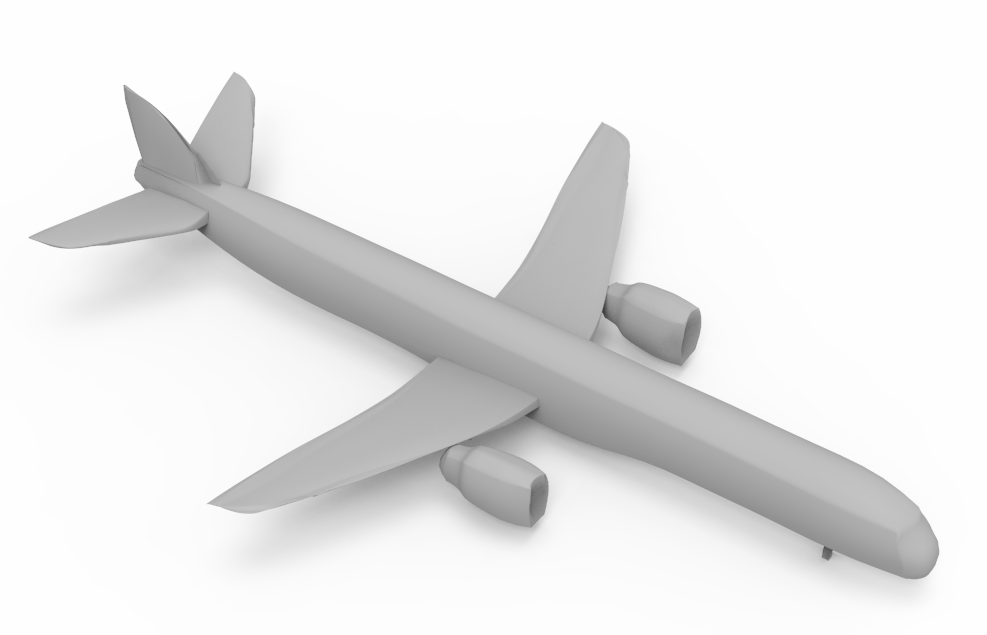}
		\includegraphics[width=0.16\linewidth]{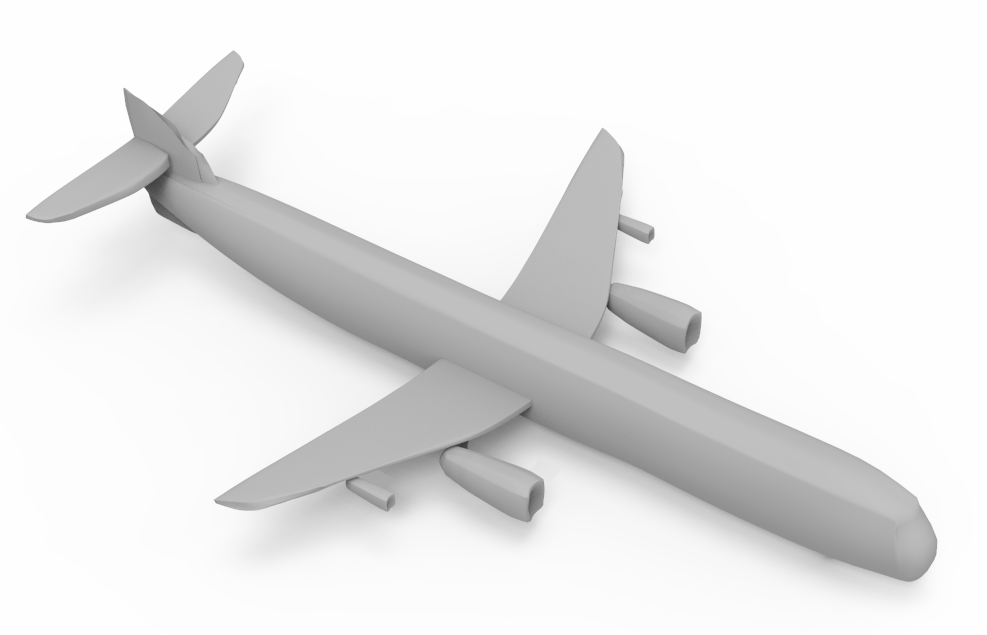}
		\includegraphics[width=0.16\linewidth]{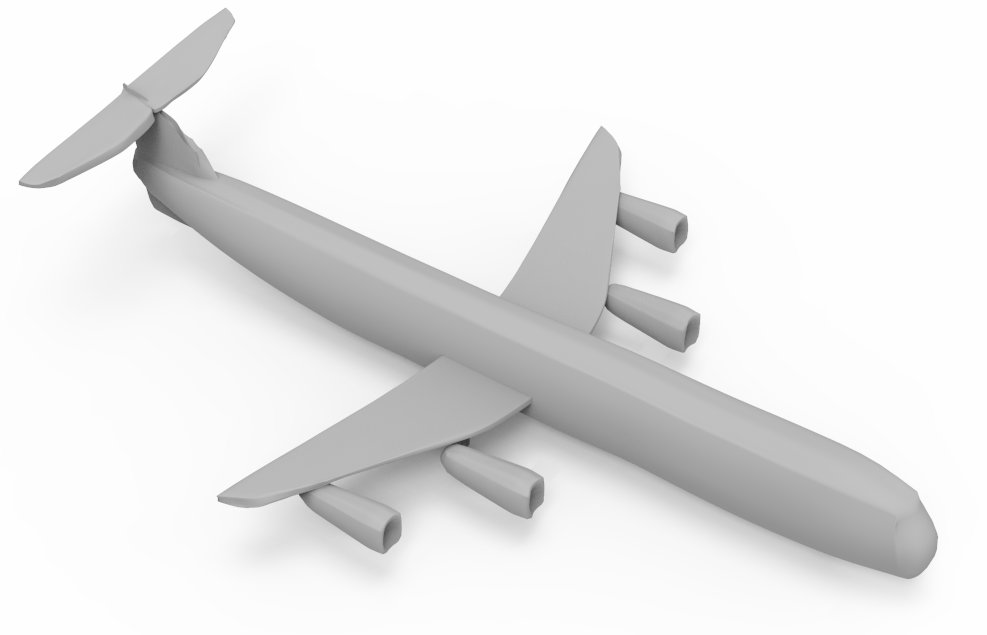}\\
		\includegraphics[width=0.16\linewidth]{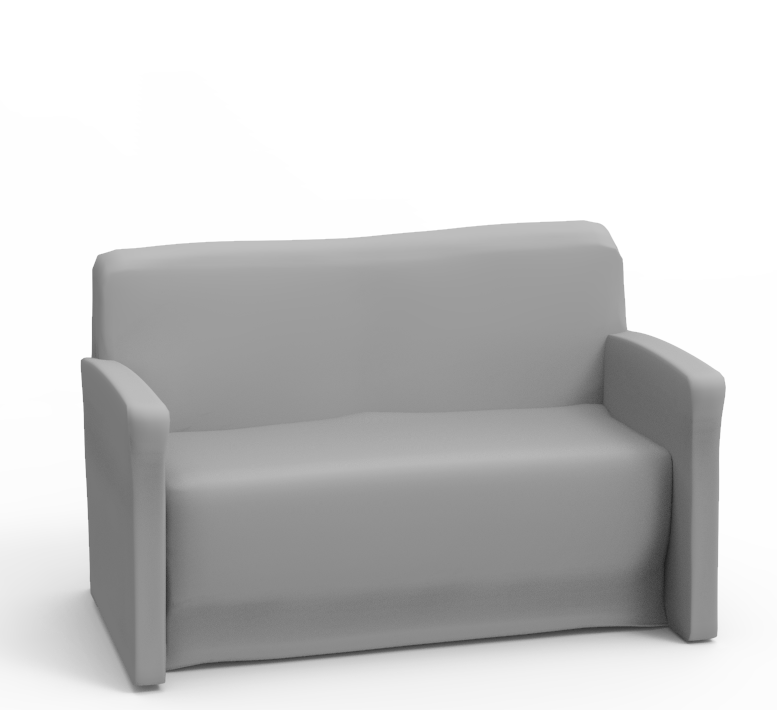}
		\includegraphics[width=0.16\linewidth]{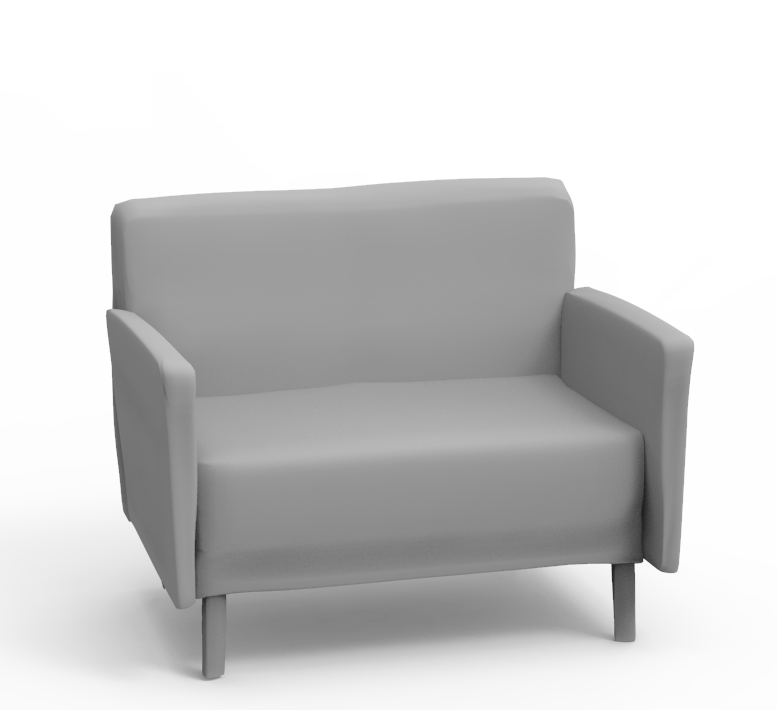}
		\includegraphics[width=0.16\linewidth]{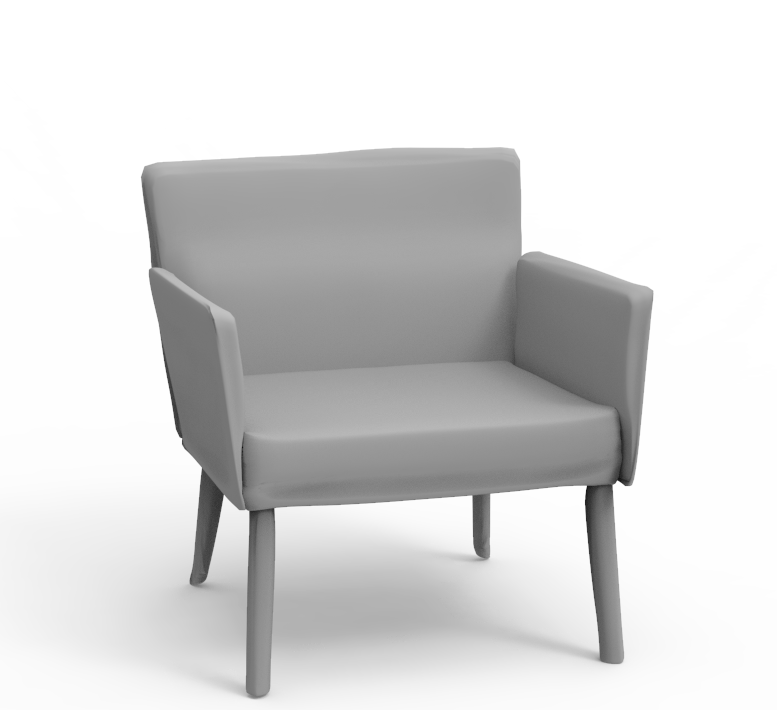}
		\includegraphics[width=0.16\linewidth]{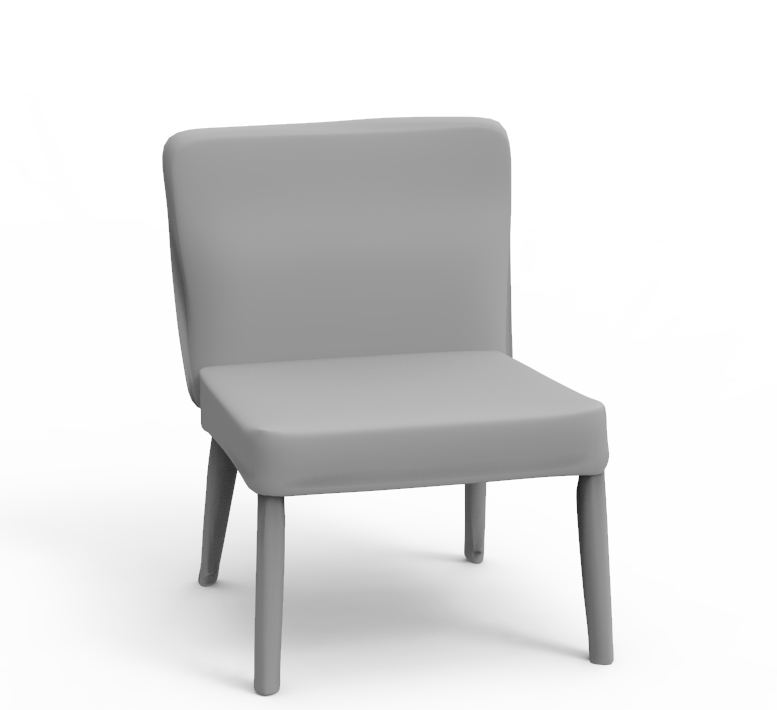}
		\includegraphics[width=0.16\linewidth]{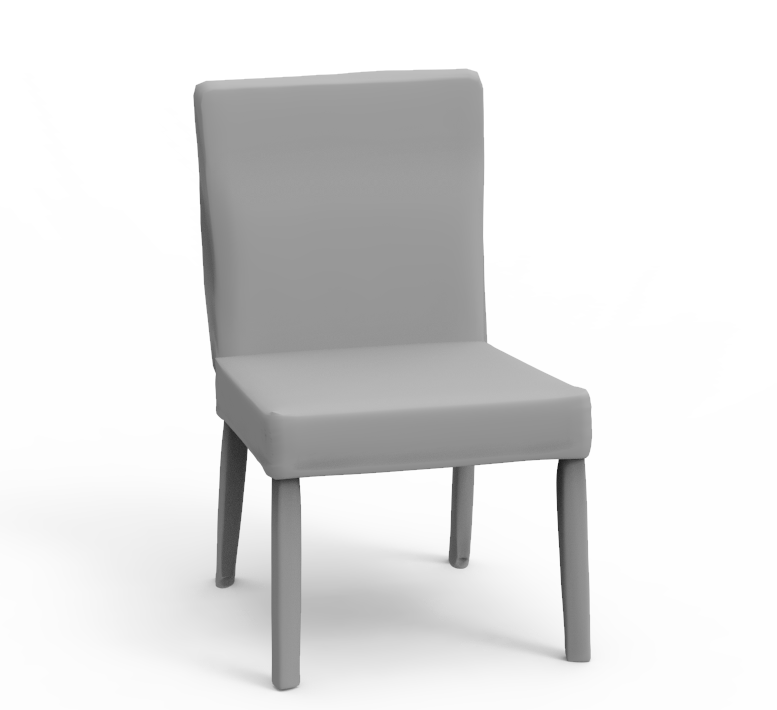}
		\includegraphics[width=0.16\linewidth]{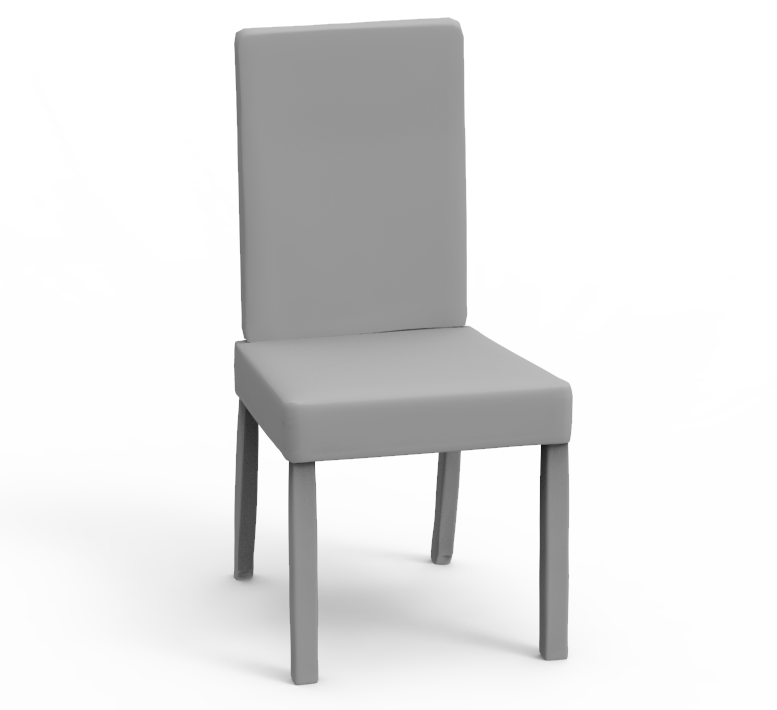}
}\caption{\YL{
Our deep generative neural network, SDM-NET, produces structured meshes composed of deformable parts. Part structures and geometries are jointly encoded into a latent space by an autoencoder, enabling quality 3D shape generation. We show shape interpolation results exhibiting flexible structure and fine geometric details. This is achieved by linearly interpolating airplane and chair latent codes and then reconstruction from the in-between codes. 
}}\label{fig:teaser}
\end{teaserfigure}

\begin{abstract}
We introduce SDM-NET, a deep generative neural network which produces {\em structured deformable meshes\/}. Specifically, the network is trained to generate a spatial arrangement of closed, deformable mesh parts, which respects the global part structure of a shape collection, e.g., chairs, airplanes, etc. Our key observation is that while the overall structure of a 3D shape can be complex, the shape can usually be decomposed into a set of parts, %
each homeomorphic to a box, and the finer-scale geometry of the part can be recovered by {\em deforming\/} the box.
The architecture of SDM-NET is that of a {\em two-level variational autoencoder\/} (VAE). At the part level, a PartVAE learns a deformable model of part geometries. At the structural level, we train a {Structured Parts VAE (SP-VAE)}, which {\em jointly\/} learns the part structure of a shape collection and the part geometries, ensuring \hongbo{the} \YLN{coherence} between global shape structure and surface details. Through extensive experiments and comparisons with the state-of-the-art deep generative models of shapes, we demonstrate the superiority of SDM-NET in generating meshes with visual quality, flexible topology, and meaningful structures, \hongbo{benefiting} shape interpolation and other \YLN{subsequent} modeling tasks.
\end{abstract}

\acmJournal{TOG}
\acmYear{2019}\acmVolume{38}\acmNumber{6}\acmArticle{243}\acmMonth{11} \acmDOI{10.1145/3355089.3356488}

\ccsdesc[500]{Computing methodologies~Shape modeling}

\keywords{Shape representation, variational autoencoder, structure, deformation, geometric details, generation, interpolation}
\maketitle

\section{Introduction}

Triangle meshes have been the dominant 3D shape representation in computer graphics, for modeling, rendering, manipulation, and animation. However, as deep learning becomes pervasive in visual computing, most deep convolutional neural networks (CNNs) developed for shape modeling and analysis have resorted to other representations including voxel grids~\cite{Girdhar16b,Qi2016,3dgan2016,Wu_2015_CVPR}, shape images~\cite{Su2015ICCV,Sinha2016DeepL3}, %
and point clouds~\cite{Qi2017cvpr,yin_sig18}.
\rz{One of the main reasons is that the non-uniformity and irregularity of triangle tessellations do not naturally support conventional convolution and pooling operations~\cite{Hanocka2019}. Yet, advantages of meshes over other shape representations should not be overlooked.}

Compared to voxels, meshes are more compact and better suited \YLN{to} representing finer surface details. Compared to points, meshes are more controllable and exhibit better visual quality. \rz{There 
have been recent attempts at developing mesh-specific convolutional operators designed for triangle tessellations~\cite{Poulenard2018,Hanocka2019}.} Current deep generative models for meshes are limited to either genus-zero meshes~\cite{Hamu2018,Maron2017TOG} or meshes sharing the same connectivity~\cite{Gao2018,meshvae2017}. Patch-based models which cover a shape with planar~\cite{Wang2018ocnn} or curved~\cite{AtlasNet2018} patches, are more adaptive, but surface quality is often {tampered} by visible seams and the patches are otherwise unstructured and incoherent.

In this paper, we introduce a novel deep generative neural network for meshes which overcomes the above limitations.
Our key observation is that while the overall structure of a 3D shape can be complex, the shape can usually be decomposed into a set of {\em parts\/}, 
each homeomorphic to a box, and the finer-scale geometry of the part can be recovered by {\em deforming\/} the box.
Hence, the architecture of our network is that of a {\em two-level\/} variational autoencoder (VAE)~\cite{kingma2013auto} which produces {\em structured deformable meshes\/} (SDM). At the part level, a PartVAE learns a deformable model of shape parts, by means of autoencoding fixed-connectivity, genus-zero meshes. At the structural level, we train a {Structured Parts VAE (SP-VAE)}, which {\em jointly\/} learns the part structure of a shape collection and the part geometries, ensuring \hongbo{the} \YLN{coherence} between global shape structure and surface details.

We call our network SDM-NET, as it is trained to generate structured deformable meshes, that is, a spatial arrangement of closed, deformable mesh parts, which respects the global part structure {(e.g., symmetry and support relations among shape parts)} of a shape collection, e.g., chairs, airplanes, etc. However, our network can generate shapes with a varying number of parts, up to a maximum count. Besides the advantages afforded by meshes mentioned above, a structured representation allows shapes generated by SDM-NET to be immediately reusable, e.g., for assembly-based modeling~\cite{Mitra2013}. In addition, the deformability of the mesh parts further facilitates editing and interpolation of the generated shapes.

SDM-NET is trained with a shape collection equipped with a consistent part structure, %
\YL{e.g., semantic segmentation}.
However, the shapes in the collection can be with arbitrary topologies and mesh connectivities. Such data sets are now widely available, e.g., ShapeNet~\cite{shapenet} and PartNet~\cite{mo2018partnet}, to name a few. While direct outputs from SDM-NET are not watertight meshes, each part is.

In summary, the main contributions of our work are:
\begin{itemize}
\item The first deep generative neural network which produces structured deformable meshes.
\item A novel network architecture corresponding to a two-level variational autoencoder which jointly learns shape structure and geometry. This is in contrast to the recent work, GRASS~\cite{li_sig17}, which learns shape structure and part geometry using separate networks.
\item A support-based part connection optimization to ensure the generation of plausible and physically valid shapes.
\end{itemize}

\YL{%
Figure~\ref{fig:teaser} demonstrates the  capability of our SDM-NET to reconstruct shapes with flexible structure and fine geometric details. By interpolating in the latent space, new plausible shapes with substantial structure change are generated.}

Through extensive experiments and comparisons with the state-of-the-art deep generative models of shapes, %
we demonstrate the superiority of SDM-NET in generating quality meshes and shape interpolations. 
We also show {the structured deformation meshes produced by SDM-NET enable other applications \YL{such as mesh editing},
which are not directly supported by the output from other contemporary deep neural networks.}

\section{Related Work}
\label{sec:related}

With the resurgence of deep neural networks, in particular CNNs, and an increasing availability of 3D shape collections~\cite{shapenet}, a steady stream of geometric deep learning methods have been developed for discriminative and generative processing of 3D shapes. In this section, we mainly discuss papers mostly related to our work, namely deep generative models of 3D shapes, and group them based on the underlying shape representations.

\noindent
\paragraph{Voxel grids}
The direct extension of pixels in 2D images to 3D is the voxel representation, which has a regular structure convenient for CNNs~\cite{Girdhar16b,3dgan2016,Qi2016}. Variational autoencoders (VAEs)~\cite{kingma2013auto} and Generative Adversarial Networks (GANs) \cite{goodfellow2014generative} can be built with this representation to produce new shapes. {Wu et al.~\shortcite{pageSAGnet19} utilize an autoencoder of two branches to encode geometry features and structure features separately, and fuse them into a single latent code to intertwine the two types of features for shape modeling.} However, these voxel based representations have huge memory and calculation costs, when the volumetric resolution is high.
To address this, sparse voxel-based methods use octrees to adaptively represent the geometry.  However, although such adaptive representations can significantly reduce the memory cost, their expressiveness of  geometric details is still limited by the resolution of leaf nodes of  octrees~\cite{Wang2017,Octree2017iccv}.
As an improvement, recent work~\cite{Wang2018ocnn} utilizes local planar patches to approximate local geometry in leaf nodes. However, planar patches still have limited capability of describing local geometry, especially for complex local shapes.  The patches are in general not smooth or connected, and require further processing, which might degrade the quality of generated shapes.

\noindent
\paragraph{Multi-view images}
To exploit image-like structures while avoiding the high cost of voxels, projecting shapes to multiple 2D views is a feasible approach.
Su et al.~\shortcite{Su2015ICCV} project 3D shapes to multi-view images, along with a novel pooling operation for 3D shape recognition. This representation is regular and efficient. However, it does not contain the full 3D shape information. So, although it can be directly used for recognition, additional efforts and processing are needed to reconstruct 3D shapes~\cite{3DVAE}. It also may not fully recover geometric details due to the incomplete information in multi-view images.

\noindent
\paragraph{Point clouds}
Point clouds have been widely used to represent 3D shapes, since they are flexible and can easily represent the raw data obtained from 3D scanners. The major challenge for deep learning on point clouds is their irregular structure. Qi et al.~\shortcite{Qi2017cvpr,Qi2017nips} propose PointNet and PointNet++ for 3D classification and segmentation, utilizing pooling operations that are order independent.
{Yang et al.~\shortcite{yang2017view} exploit an interactive system for segmenting point clouds of indoor scenes.}
Fan et al.~\shortcite{fan2016point} use point clouds to reconstruct 3D objects from a given image. 
Achlioptas et al.~\shortcite{achlioptas18a} introduce a deep autoencoder network for shape representation and generation.
However, learning from irregular point clouds is still challenging and their method is only able to produce relatively coarse geometry.

\noindent
\paragraph{Meshes and multi-chart representations}
Deformable modeling of a shape collection, especially of human models~\cite{anguelov2005scape,pons2015dyna}, operates on meshes with the same connectivity while altering the mesh vertex positions; the shape collection can be viewed as deformations of a template model. For high quality shape generation, especially with large deformations, a manually crafted deformation representation~\cite{Gao2017} is employed by~\cite{meshvae2017,Gao2018}. Although these methods can represent and generate shapes with fine details, they require meshes to have the same connectivity.
Wang et al.~\shortcite{wang2018pixel2mesh} reconstruct a mesh-based 3D shape from an RGB image by deforming a sphere-like genus-zero mesh model. Dominic et al.~\shortcite{Dominic2018} use a CNN to infer the parameters of free-form deformation (FFD) to deform a template mesh, guided by a target RGB image. Both methods require an image as input to provide guidance, and thus cannot be used for general shape generation tasks without guidance. Moreover, deforming a single mesh limits the topological and geometric complexity of generated shapes.

Multi-chart representations attempt to overcome the above restriction by generating multiple patches that cover a 3D shape. \YL{Zhou et al.~\shortcite{Zhou2004} create texture atlases with less stretches for texture mapping.}
Hamu et al.~\shortcite{Hamu2018} generate a 3D shape as a collection of conformal toric charts~\cite{Maron2017TOG}, each of which provides a cover of the shape with low distortion. Since toric covers are restricted to genus-zero shapes, their multi-chart method still has the same limitation. AtlasNet~\cite{AtlasNet2018} generates a shape as a collection of patches, each of which is parameterized to a 2D domain as an atlas. While the patches together cover the shape well, visible seams can often be observed.
In general, neither the atlases nor the toric charts correspond to meaningful shape parts; the collection is optimized to approximate a shape, but is otherwise unstructured. In contrast, SDM-NET produces structured deformable meshes.

\noindent
\paragraph{Implicit representations}
Several very recent works~\cite{chen2019-IMNET,park2019-DeepSDF,mescheder2019-Occupancy} show great promise of generative shape modeling using implicit representations. These deep networks learn an implicit function which defines the inside/outside statuses of points with respect to a shape or a signed distance function. The generative models can be applied to various applications including shape autoencoding, generation, interpolation, completion, and single-view reconstruction, demonstrating superior visual quality over methods based on \YL{voxels, point clouds}, as well as patch-based representations. However, none of these works generate structured or deformable shapes.

\noindent
\paragraph{Shape structures}
Man-made shapes are highly structured, which motivates structure-aware shape processing~\cite{Mitra2013}. Works on learning generative models of 3D shape structures can be roughly divided into two categories~\cite{chaudhuri2019}: probabilistic graphical models and deep neural networks.

Huang et al.~\shortcite{Huang2015CGF} propose a probabilistic model which computes part templates, shape correspondences, and segmentations from clustered shape collections, and 
their points in each part are influenced by their correspondence in the template.  Similar to \cite{Huang2015CGF}, ShapeVAE~\cite{nash2017shape} generates 
\YL{point}
coordinates \YL{and normals}
based on different parts, %
but uses a deep neural network instead of a probabilistic model. Compared to the above two works, our method does not require point-wise correspondences, which \YL{can be difficult or expensive to obtain reliably}.
Moreover, our method encodes both global spatial structure like support relations, and local geometric deformation, producing shapes with reasonable structures and fine details.

Li et al.~\shortcite{li_sig17} introduce GRASS, a generative recursive autoencoder for shape structures, based on Recursive Neural Networks (RvNNs). Like SDM-NET, GRASS also decouples structure and geometry representations. However, a key difference is that SDM-NET {\em jointly} encodes global shape structure and part geometry, while GRASS trains {\em independent\/} networks for structure and part geometry generations. In terms of outputs, GRASS generates a {\em hierarchical\/} organization of bounding boxes, and then fills them with voxel parts. SDM-NET produces a {\em set\/} of shape \YLN{parts}, each of which is a deformable mesh \hongbo{to} 
better capture finer surface details. \YL{Lastly}, the structural autoencoder of GRASS requires symmetry hierarchies for training while SDM-NET only requires \rz{consistent semantic %
segmentation} and employs the support information to produce shapes with support stability.

\rz{Concurrent to SDM-NET, Mo et al.~\shortcite{mo2019structurenet} develop \emph{StructureNet}, which learns a generative autoencoder of shape structures based on graph neural networks. StructureNet shares much commonality with GRASS but extends it in two important ways. First, unlike GRASS, which is limited to encoding binary trees, StructureNet can directly encode shapes represented as $n$-ary graphs, aimed to facilitate a consistent hierarchical representation of shapes within the same category. Second, StructureNet also accounts for horizontal inter-part relationships between siblings. The outputs from StructureNet are either box structures or point cloud shapes.} \GL{\hongbo{Our SDM-Net analyzes and encodes  shape structures by not only using the consistent representation across the same shape families but also with support stability. In addition, it is expected our mesh-based representation with deformable parts is able to capture more geometry details than box and point cloud based representations adopted by StructureNet.}}

\section{Methodology}
\label{sec:method}
\textbf{Overview.} Given a collection of shapes of the same category with part-level labels, our method represents them using a {structured} set of deformable boxes, each corresponding to a part. The pioneering works~\cite{Ovsjanikov2011,Kim2013TOG} have shown the representation power of using a collection of boxes to analyze and explore shape collections. However, it is highly non-trivial to extend their techniques to shape generation, since boxes are generally of a coarse representation. We tackle this challenge by allowing individual boxes to be flexibly deformable and propose a two-level VAE architecture called SDM-NET, including PartVAE for encoding the geometry of deformable boxes, and SP-VAE for joint encoding of part geometry and global structure such as symmetry and support.
Moreover, to ensure that decoded shapes are physically plausible and stable, we introduce an optimization based on multiple constraints including support stability, which can be compactly formulated and efficiently optimized.
Our SDM-NET model 
allows easy generation of plausible meshes with flexible structures and fine details.

We first introduce the encoding of each part, including both the geometry and structure information. %
We then introduce our SDM-NET involving VAEs at both the local part geometry level (PartVAE), and global joint embedding of structure and geometry (SP-VAE). Then we briefly describe how the support relationships are extracted, and finally present our optimization for generating plausible and well supported shapes.

\subsection{Encoding of a Shape Part}
\label{sec:shape:encoding}

\begin{figure}[tb]
	\centering
	\includegraphics[width=.99\columnwidth]{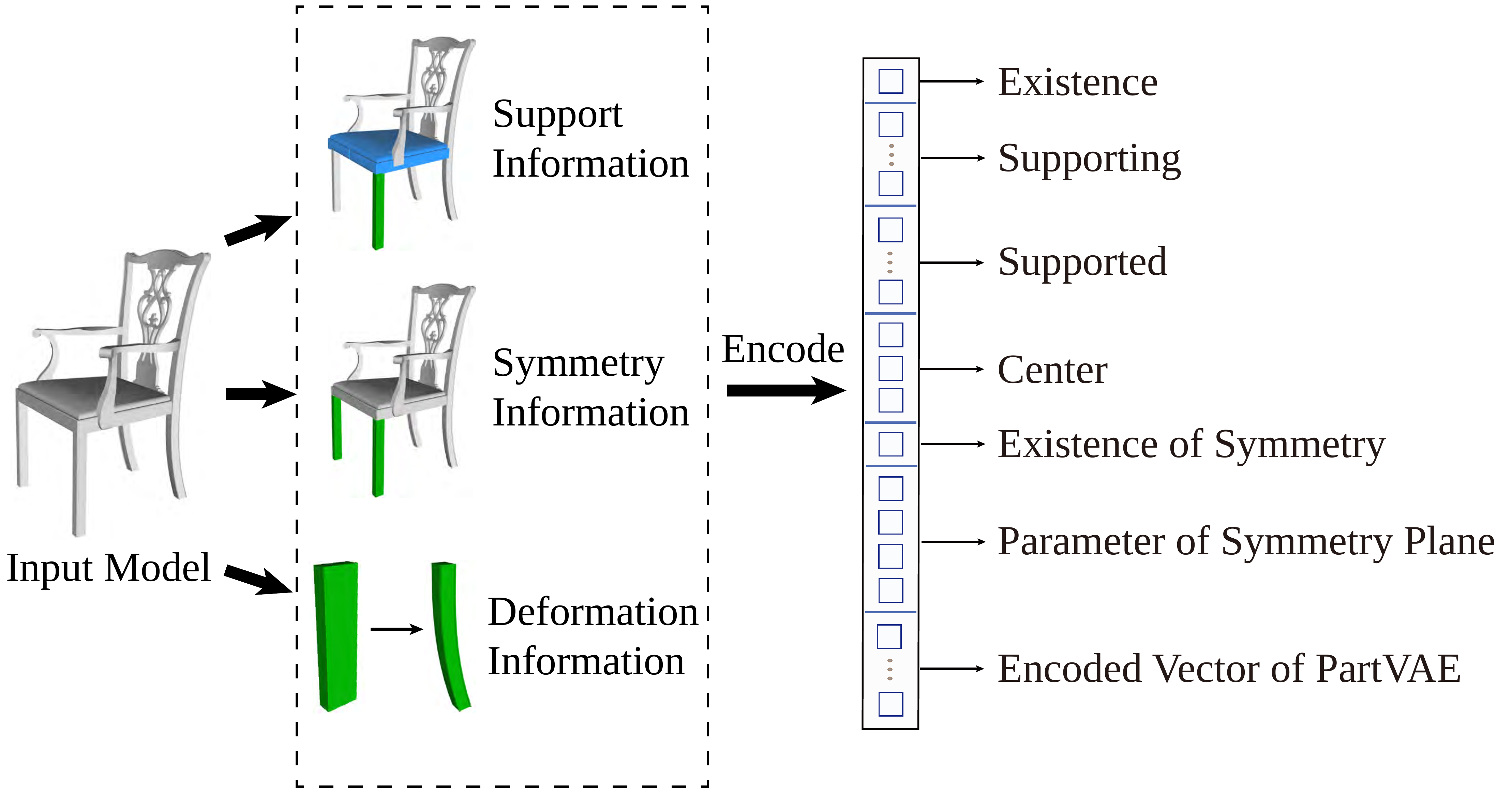}	
	\caption{Encoding of each shape part, including its structure and geometry information. The former includes both the support and symmetry information, and the latter describes the deformation of the bounding box, compactly represented as the latent vector of PartVAE. Detailed explanation of each \YL{entry} of the code is presented in Section~\ref{sec:shape:encoding}.}
	\label{fig:code}
\end{figure}

\begin{figure}[tb]
	{
		\centering
		{
		    {\includegraphics[width=0.99\linewidth]{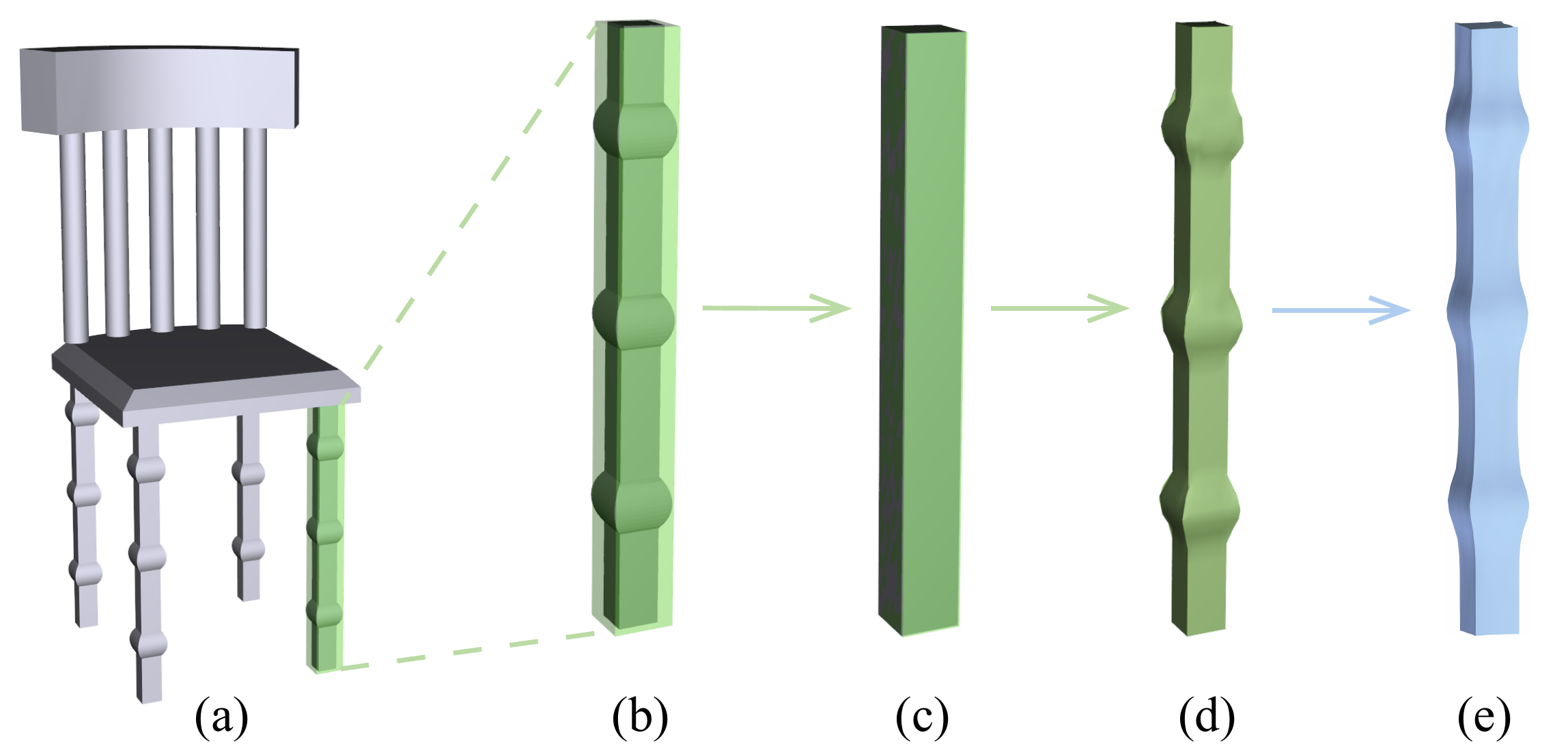}}			
			\caption{%
		    An example of representing a chair leg with detailed geometry by a deformable bounding box. (a) a chair with one of its leg parts highlighted, (b) the targeted part in (a) with the bounding box overlaid, (c) the bounding box used as the template, (d) deformed bounding box after non-rigid registration, (e) recovered shape using PartVAE.} \label{fig:registration}
		}
	}
\end{figure}

Based on semantic part-level labels, a shape is decomposed into a set of parts. %
Each part is represented using a deformable bounding box, as illustrated in Figure~\ref{fig:registration}. Let $n$ be the total %
number of part labels that appear across different shapes in the specified object category. For a given shape, it may contain a fewer number of parts as some parts may not be present.
To make analysis and relationship representation easier, we assume the initial bounding box (before deformation) \YL{of} each part is axis aligned. This is sufficient in practice, since each bounding box is allowed to have substantial deformation to fit the target geometry. 
\rv{%
The initial bounding primitive being a box does not prevent the internal part geometry from being complex, since geometric details can be captured and preserved through non-rigid registration (see Section~\ref{sec:cvae} for details).
Without loss of generality, the bounding boxes are used in our framework.}%

The geometry and associated relationships of each part are encoded by a representation vector $\mathbf{rv}$, as illustrated in Figure~\ref{fig:code}.  %
The detailed definition of this vector is given as follows.
\begin{itemize}
\item {$\mathbf{rv}_1 \in \{0,1\}$ indicates the existence of this part.}
\item {$\mathbf{rv}_2 \in \{0,1\}^n$ is a vector with $n$ dimensions to indicate which parts are supported by this part.}
\item {$\mathbf{rv}_3 \in \{0,1\}^n$ is a vector with $n$ dimensions to indicate which parts support the current part.}
\item $\mathbf{rv}_4 \in \mathbb{R}^3$ is the 3D position of the bounding box center.
\item {$\mathbf{rv}_5 \in \{0,1\}$ indicates the existence of a symmetric part.} %
\item {$\mathbf{rv}_6 \in \mathbb{R}^4$ records the parameters $a$-$d$ of the symmetry plane represented in an implicit form, i.e., $ax+by+cz+d=0$.}
\item $\mathbf{rv}_7$ is the encoded vector from the PartVAE described in Section~\ref{sec:cvae}, which encodes its geometry. By default, $\mathbf{rv}_7\in \mathbb{R}^{64}$.
\end{itemize}

{The ID of each part, used in $\mathbf{rv}_2$ and $\mathbf{rv}_3$, is pre-determined and stored in advance for the dataset.}
Each value in $\mathbf{rv}_1$, $\mathbf{rv}_2$, $\mathbf{rv}_3$ and $\mathbf{rv}_5$ is $1$ if exists and $0$ otherwise.
For generated vectors, we treat a value above $0.5$ as true and below as false. 
The length of this vector is \YL{$2n+73$} and between 77 and 101
for all the examples in this paper. Note that other information such as the label of the part that is symmetric to the current one (if exists) is fixed for a collection (e.g. the right armrest of a chair is symmetric to the left armrest of the chair) and therefore not encoded in the vector. 
In our current implementation, we only consider reflection symmetry and keep one symmetric component (if any) for each part. Although this is somewhat restrictive,  
it is very common and sufficient to cope with most cases. In practice, we first perform global reflection symmetry detection~\cite{podolak2006planar} to identify components that are symmetric to each other w.r.t. a symmetry plane. This is then supplemented by local reflection symmetry detection by checking if pairs of parts have reflective symmetry.

\subsection{{PartVAE for Encoding Part Geometry}}\label{sec:cvae}
For each part, the axis-aligned bounding box (AABB) is first calculated. %
The bounding box of the same part type provides a uniform domain across different shapes, and the geometry variations are viewed as different deformation functions applied to the same domain. We take a common template, namely a unit cube {mesh $box_0$ with $19.2K$ triangles}, 
to represent each part. We first translate and scale it to fit the bounding box of the part. Denote by $b_{i,j}$ the bounding box transformed from $box_0$ for the $j^{\rm th}$ part $c_{i,j}$ on the $i^{\rm th}$ shape $s_i$. We treat it as initialization, and apply 
\YL{non-rigid coarse-to-fine registration~\cite{Zollhofer2014},}
which deforms $b_{i,j}$ to $b'_{i,j}$, as illustrated in Figure~\ref{fig:registration} (d).
$b^{'}_{i,j}$ shares the geometry details with the part $c_{i,j}$ %
and has the same mesh connectivity as the unit cube box $box_0$.

\begin{figure*}[t]
	\centering
	{
		\includegraphics[width=.8\linewidth]{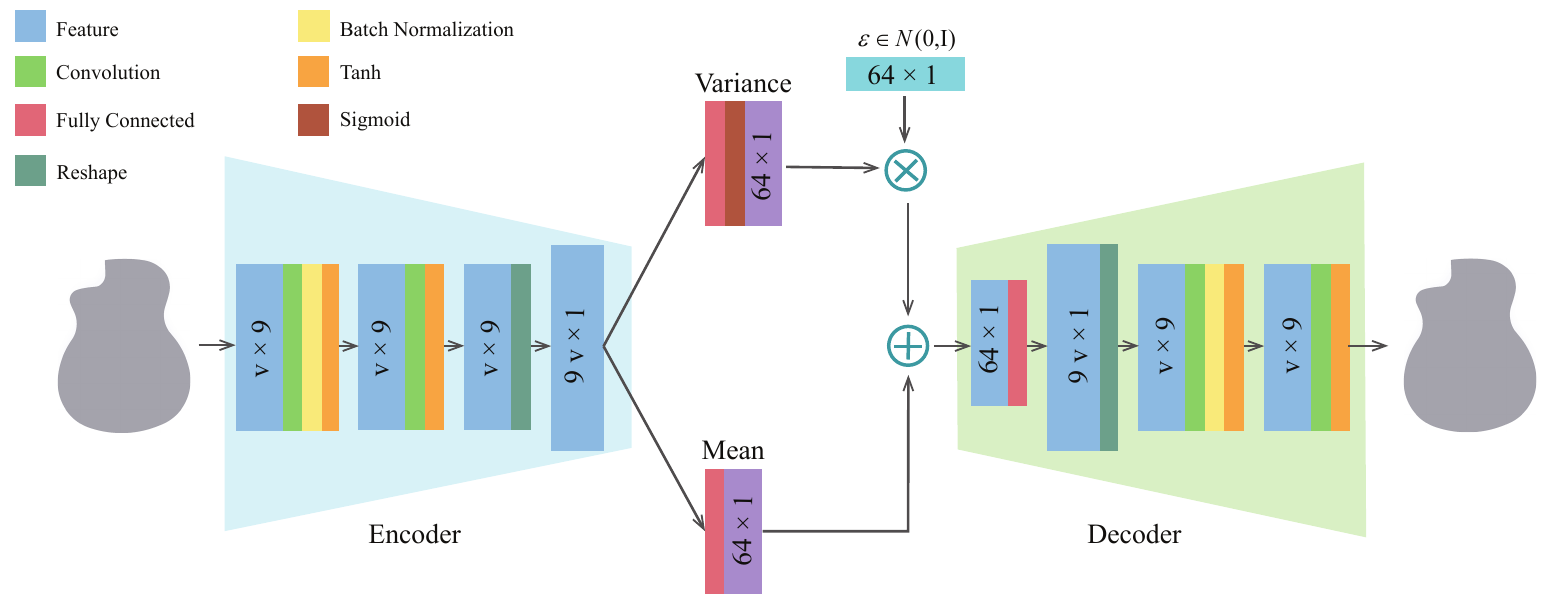}		
	}
	\caption{%
	Architecture of PartVAE for encoding the geometry details of a part represented as the deformation of the associated template box. $V$ is the number of vertices in the template box. $N(\mathbf{0}, \mathbf{I})$ is the Gaussian distribution with $\mathbf{0}$ mean and identity covariance.}
	\label{fig:compvae}
\end{figure*}
{The variational autoencoder has been used to encode the geometric priors for point cloud segmentation~\cite{meng2018vv} and mesh generation~\cite{meshvae2017}. Similar to~\cite{meshvae2017},
using meshes of the same connectivity makes it feasible to build a variational autoencoder to represent the deformation of $b'_{i,j}$. }
The convolutional VAE architecture in~\cite{Gao2018} is employed for compactly representing plausible deformation of each part, allowing new variations to be synthesized.
The architecture is shown in Figure~\ref{fig:compvae}. The input is a $V\times 9$ dimensional matrix, where $V$ is the number of vertices for the template bounding box mesh. Each row of the matrix is a 9-dimensional vector that characterizes the local deformation of 1-ring neighborhood of each vertex including the rotation axis, rotation angle and scaling factor. It passes through two convolutional layers followed by a fully connected layer to obtain the mean and variance. \rv{The decoder mirrors the structure of the encoder to recover the deformation representation, but with different trainable weights.} 
Since each part type has its own characteristics, we train a PartVAE for all the parts with the same part type across different shapes.

\begin{figure}[t]
	\centering
	{
	   {\includegraphics[width=0.99\linewidth]{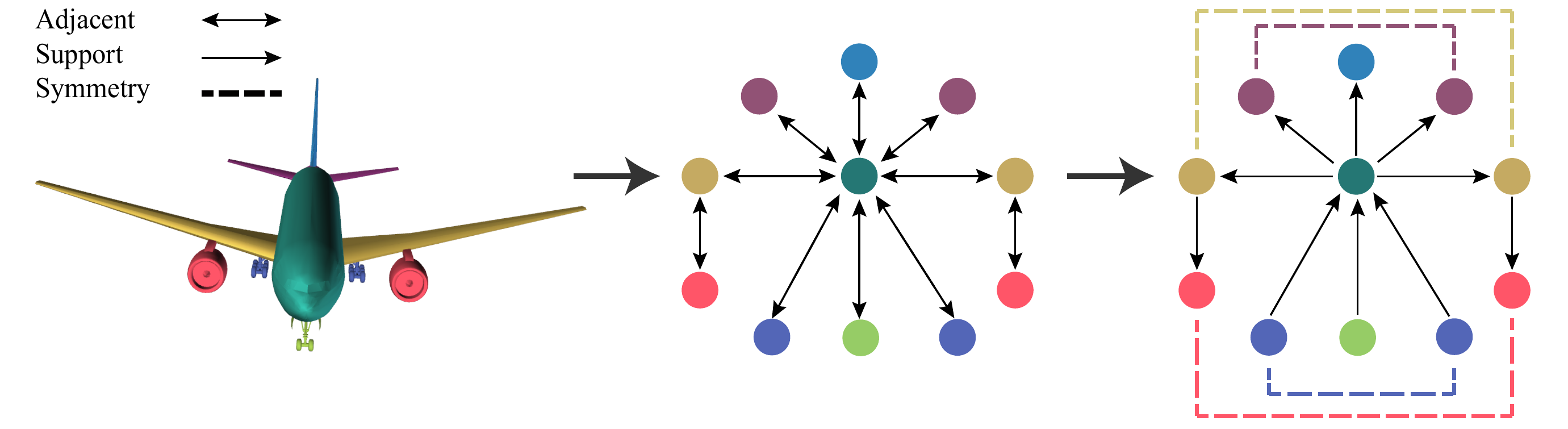}}
	}\caption{Illustration of {support and symmetry relations between parts of an airplane model. The support relations, detected by the approach in ~\cite{Huang2016TVCG}, turn an undirected adjacency graph (Middle) to a directed support graph (Right).} %
	}\label{fig:planesupport}
\end{figure}

\begin{figure*}[!t]
	\centering
	\includegraphics[width=.99\linewidth]{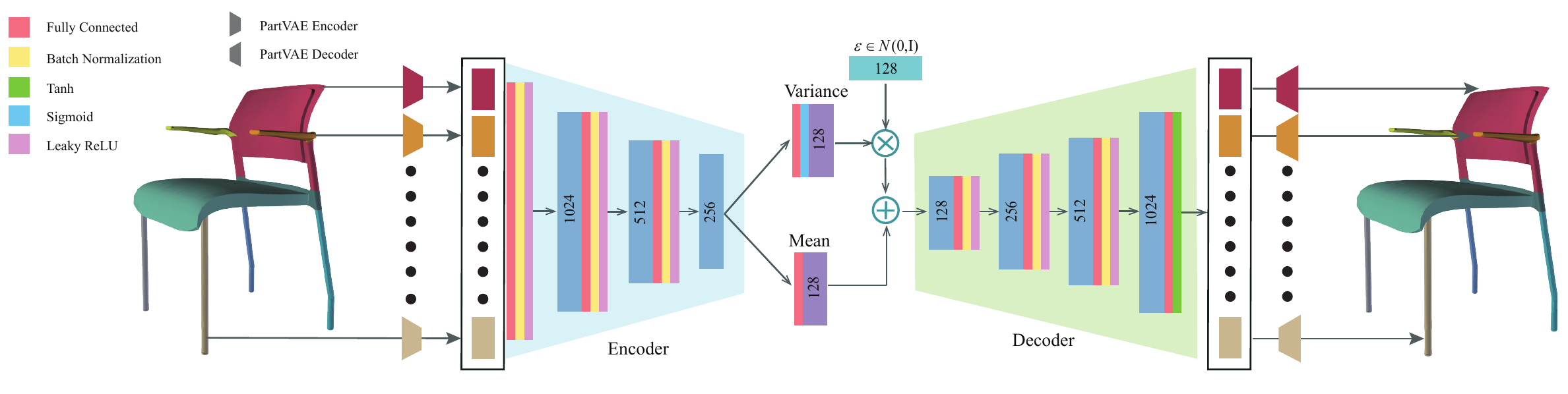}
	\caption{Architecture of  Structured Parts VAE (SP-VAE). The detailed geometry is encoded by  PartVAE. The {support- and symmetry-induced} structure and the associated latent code of PartVAE are encoded by the Structured Parts VAE.} %
	\label{fig:pipeline}
\end{figure*}

\subsection{Supporting Structure Analysis}
Structure captures the relationships between parts, and proper encoding of shape structure is crucial to generate plausible shapes. Symmetry as one of the structural relationships has been well explored, for \YLN{example} effectively used in GRASS \cite{li_sig17}. Besides symmetry, support relationships have been demonstrated useful for structural analysis to synthesize physically plausible new structures with support and stability~\cite{Huang2016TVCG}. Compared to symmetry, support relationships provide a more direct mechanism to model the relations between adjacent parts. We thus use both symmetry and support to encode the structural relationships between parts (Section~\ref{sec:shape:encoding}). 
Note that our work is the first attempt to encode support-based shape structures in  deep neural networks. 

Following~\cite{Huang2016TVCG}, we detect the support relations between adjacent parts as one of three support sub-structures, namely, ``support from below'', ``support from above'', and ``support from side''. As illustrated in Figure \ref{fig:planesupport}, the detected support relations turn an undirected adjacency graph to a directed support graph. For each detected support relation of a part, we encode the labels of its supported and supporting parts in our part feature vector (Section~\ref{sec:shape:encoding}). Our feature coding is flexible to represent the cases including one part being supported by multiple parts, as well as multiple parts being supported by one part. 
Since for all the cases in our shape dataset, the sub-structure type for each support relation between two adjacent parts is fixed, the support sub-structure types are kept in a look-up table but not encoded in our part feature vector. Given the supporting and supported part labels from a decoded part feature vector, we can efficiently obtain the corresponding sub-structures from this look-up table.

\subsection{SP-VAE for Structured Deformable Mesh Encoding}\label{sec:SPVAE}

We build SP-VAE to jointly encode the structure of a shape represented as the layout of boxes, and the geometry of its parts. By analyzing their joint distribution, it helps ensure that the geometry of the generated shape is coherent with the structure and the geometries of individual parts are consistent (i.e., of compatible styles).
Our SP-VAE takes the concatenation of representation vectors for all the parts as input (see Section~\ref{sec:shape:encoding}). \YL{It encodes parts in a consistent order during encoding and decoding.} This concatenated vector covers both the geometric details of individual parts encoded using PartVAE, and the relationships between them. 
The SP-VAE uses multiple fully connected layers, and the architecture is illustrated in Figure~\ref{fig:pipeline}.

Let $Enc_S(\cdot)$ and $Dec_S(\cdot)$ denote the encoder and decoder of our SP-VAE network, respectively. $\mathbf{x}$ represents the input concatenated feature vector of a shape, $\tilde{\mathbf{x}}=Enc_S(\mathbf{x})$ is the encoded latent vector, and $\mathbf{x'}=Dec_S(\tilde{\mathbf{x}})$ is the reconstructed feature vector.
Our SP-VAE minimizes the following loss:
\begin{equation}\label{eq:Lvae}
L_{SP-VAE}=\lambda_{1}L_{recon}+\lambda_{2}L_{KL}+L_{RegVAE},
\end{equation}
where $\lambda_1$ and $\lambda_2$ are the weights of different loss terms, and
\begin{equation}
L_{recon}=\frac{1}{N}\underset{\mathbf{x}\in \mathcal{S}}{\sum}||\mathbf{x}-\mathbf{x'}||^{2}_2
\end{equation}
denotes the MSE (mean squared error) reconstruction loss to ensure better reconstruction. Here $\mathcal{S}$ is the training dataset and $N = |\mathcal{S}|$ is the number of shapes in the training set. %
\begin{equation}
L_{KL}=D_{KL}(\hat{q}(\tilde{\mathbf{x}}|\mathbf{x})|\hat{p}(\tilde{\mathbf{x}}))
\end{equation}
is the KL divergence to promote Gaussian distribution in the latent space, where $\hat{q}(\tilde{\mathbf{x}}|\mathbf{x})$ is the posterior distribution given feature vector $\mathbf{x}$, and $\hat{p}(\tilde{\mathbf{x}})$ is the Gaussian prior distribution. $L_{RegVAE}$ is the squared $\ell_2$ norm regularization term of the network parameters used to avoid overfitting. The Gaussian distribution makes it effective to generate new shapes by sampling in the latent space, which is used for random generation and interpolation.

\subsection{Shape Generation and Refinement}\label{sec:shapegen}

The latent space of SP-VAE provides a meaningful space for shape generation and interpolation. Extensive experimental results are shown in Section \ref{sec:exp}. Random sampling in the latent space can generate novel shapes. However, \YL{although the desired geometry and structure from the decoded feature vector are generally reasonable, they may not satisfy supporting and physical constraints exactly, resulting in shapes which may include parts not exactly in contact, or may be unstable.}
Inspired by~\cite{Averkiou2014EG}, we propose to use an effective global optimization to {refine the spatial relations between parts by mainly using the associated symmetry and support information.}

\begin{figure}
	\centering
    \subfigure[]{\includegraphics[width=0.19\linewidth]{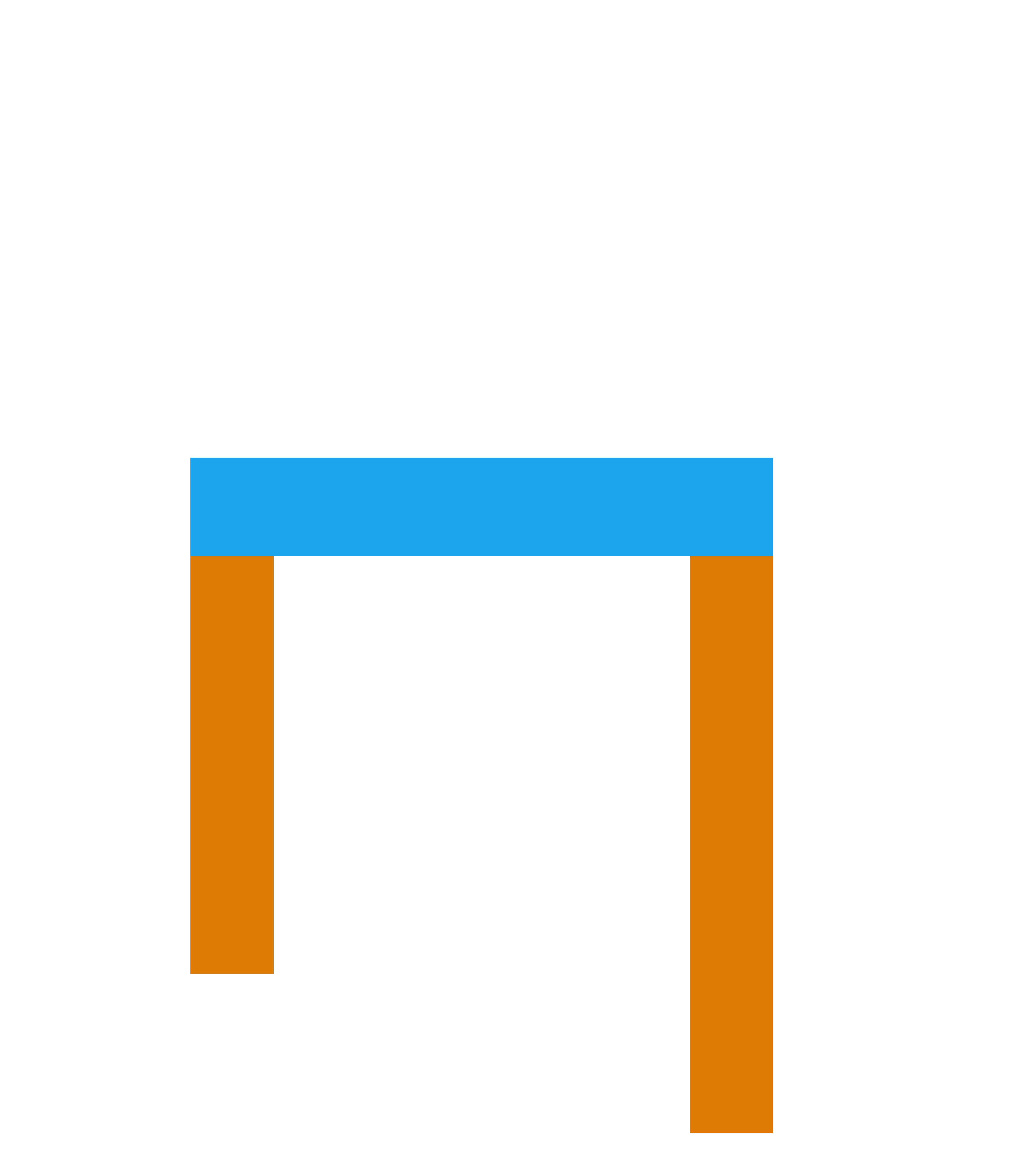}}
    \subfigure[]{\includegraphics[width=0.19\linewidth]{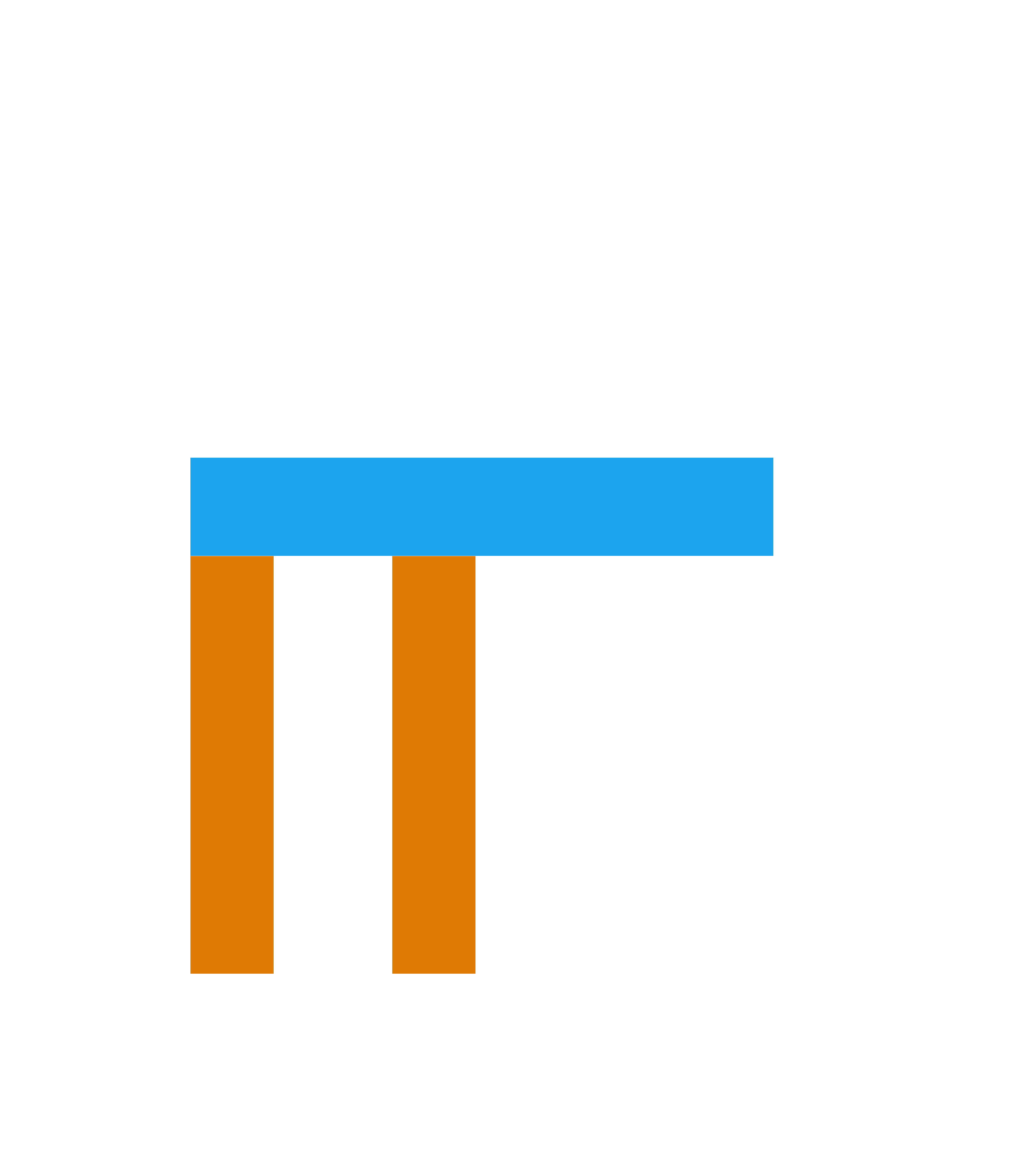}}
    \subfigure[]{\includegraphics[width=0.19\linewidth]{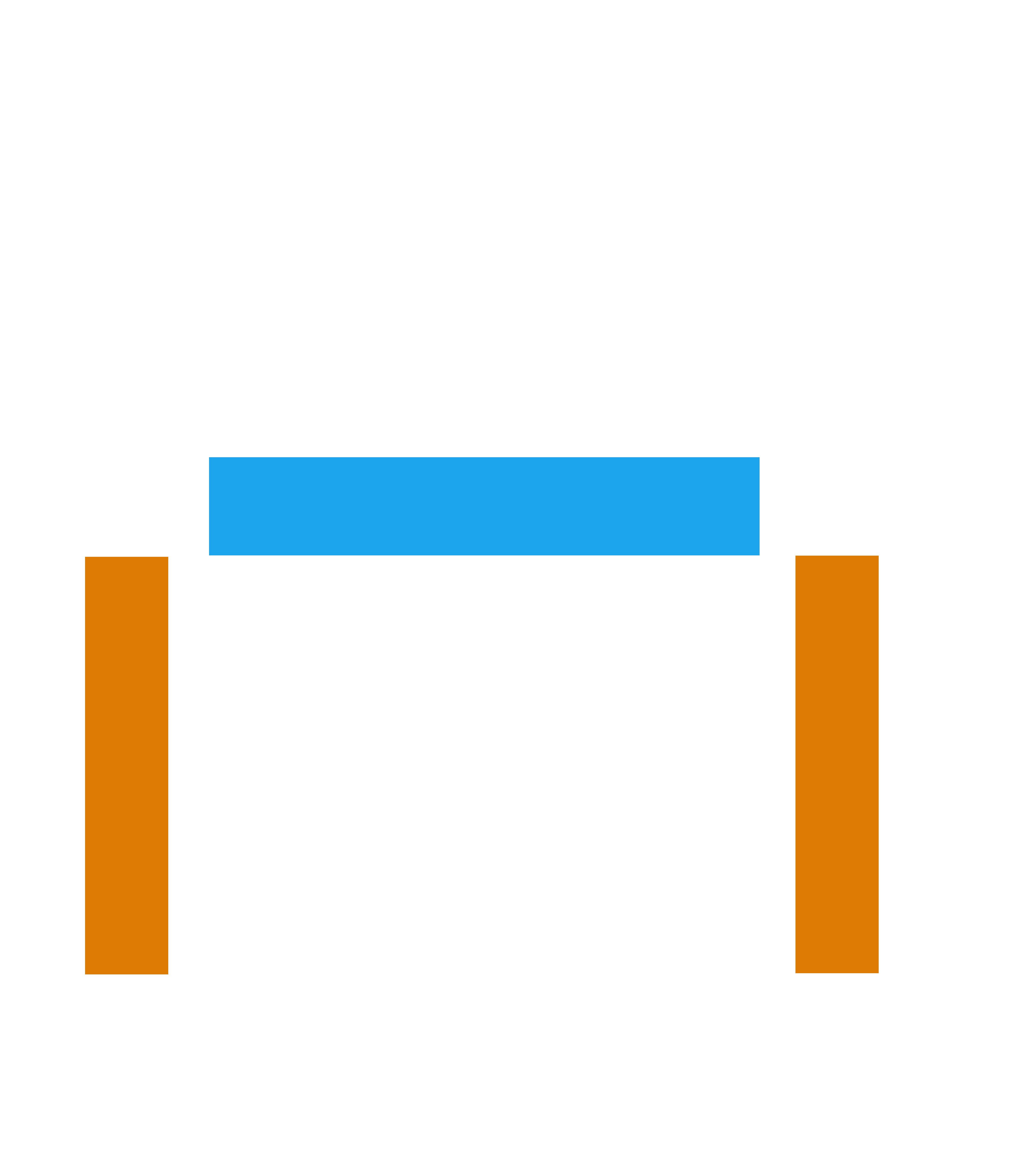}}
    \subfigure[]{\includegraphics[width=0.19\linewidth]{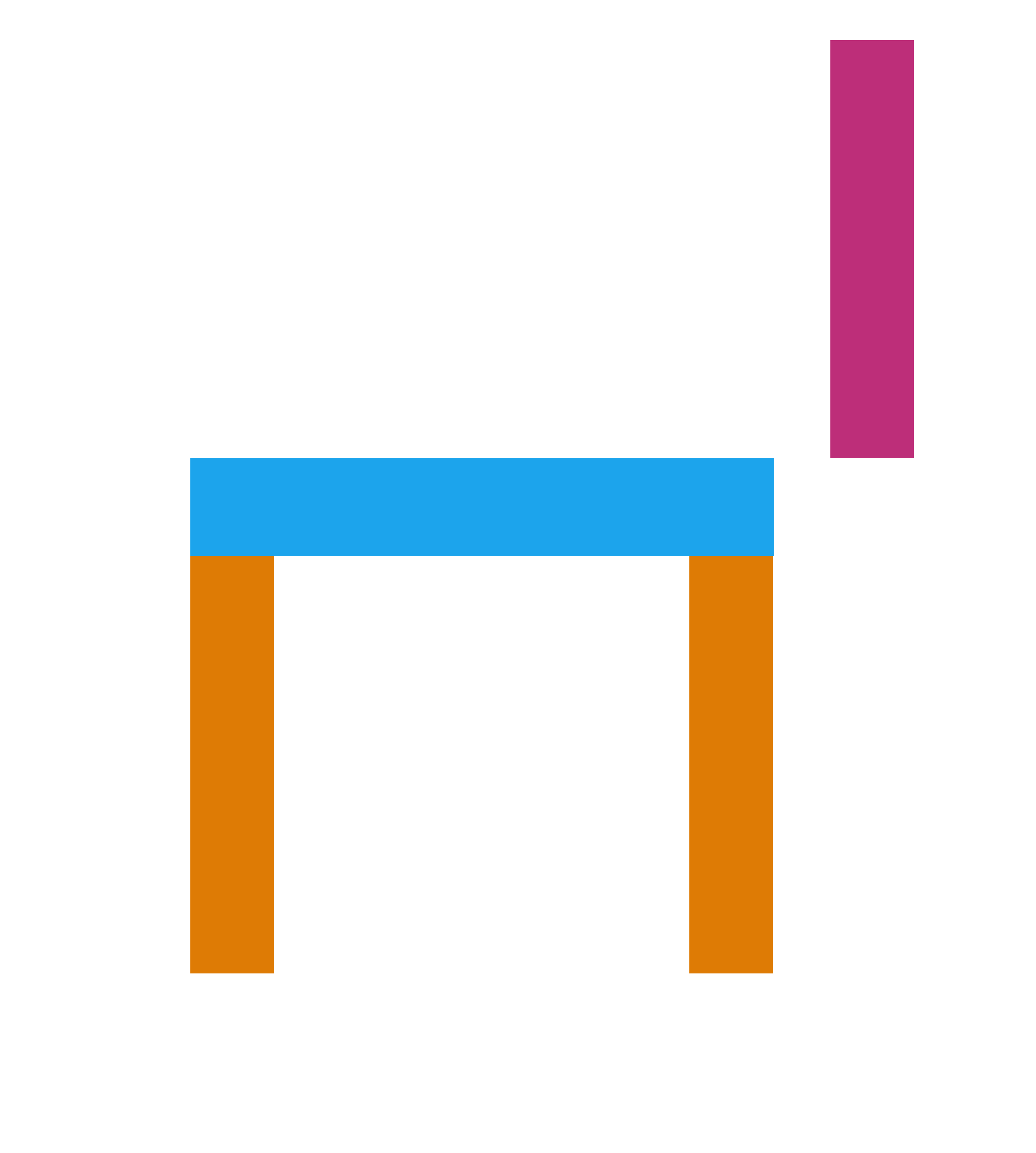}}
    \subfigure[]{\includegraphics[width=0.19\linewidth]{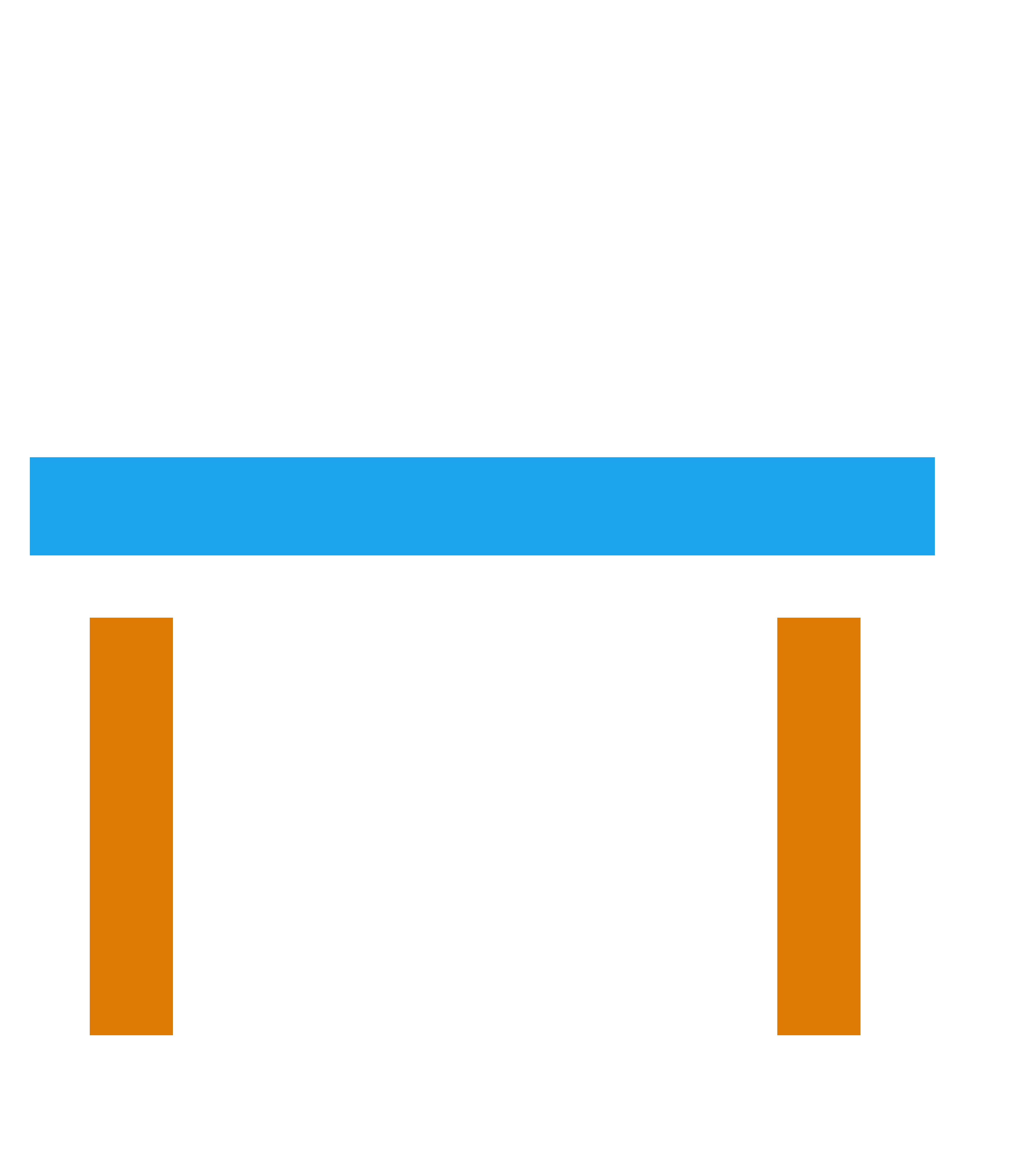}}
	\caption{Illustration of different cases addressed by  refinement optimization: (a) equal length constraint, (b) stable support constraint, (c-e) support constraint.}
	\label{fig:refinement}
\end{figure}

Denote the center position and size (half of the length in each dimension) of the $i^{\rm th}$ part 
\YL{as $\mathbf{p}_i$ and $\mathbf{q}_i$, each being a 3-dimensional vector corresponding to $x$, $y$ and $z$ axes, where $\mathbf{p}_i$ is directly obtained from the representation vector, and $\mathbf{q}_i$ is determined by the bounding box after recovering the part geometry.}
Denote by $\mathbf{p}'_i$ and $\mathbf{q}'_i$ the position and size of the $i^{\rm th}$ part after global optimization.
The objective of this optimization is to minimize the changes between the optimized position/scale and the original position/scale
\begin{equation}
{\sum_i \|	{\mathbf{p}^{'}_{i}}-{\mathbf{p}_{i}}
	\|}^2 + \alpha {\|{\mathbf{q}^{'}_{i}}-{\mathbf{q}_{i}}
	\|}^2,
\end{equation}
while ensuring the following constraints are satisfied. $\alpha$ is a weight to balance the two terms, and is fixed to $10$ in our experiments.
The symmetry and equal length constraints are from~\cite{Averkiou2014EG}, though we use the support relationships to help identify equal length constraints more reliably. The remaining constraints are unique in our approach. {Figure~\ref{fig:refinement} illustrates typical problematic cases which are handled by our refinement optimization.}

\textbf{Symmetry Constraint}: 
If the generated $i^{\rm th}$ part has the symmetry indicator flagged in its representation vector, its symmetry part (denoted using index $j$) also exists. Let $\mathbf{n}_i$ and $d_i$ denote the normal and the intercept of the symmetry plane, respectively. Enforcing the symmetry constraint leads to the following constraints to be satisfied: \YLN{$\frac{(\mathbf{p}^{'}_{i}-\mathbf{p}^{'}_{j})}{2}\cdot \mathbf{n}_i+d_i=0$, $(\mathbf{p}^{'}_{i}-\mathbf{p}^{'}_{j}) \times \mathbf{n}_i = 0$.} %
The symmetry of two parts is viewed as an undirectional relationship. If the symmetry indicator ($\mathbf{rv}_5$ of the representation vector) of either part $i$ or $j$ is $1$, we consider these two parts as symmetric.

\textbf{Equal Length Constraint}: 
A set of parts which are simultaneously supported by a common part, and simultaneously support another common part, are considered as a group to have the same length along the supporting direction. For this purpose, the ground is considered as a virtual part.
For example, the four legs of a table supporting the same table top part and at the same time being supported by the ground should have the same height. These can be easily detected by traversing the support structure. An example that violates this constraint is illustrated in Figure~\ref{fig:refinement} (a). 
The equal length constraints can be formulated as
$\mathbf{q}^{'}_{i}[t]=\mathbf{q}^{'}_{k}[t], k \in g_i$, where $g_i$ is an equal-length group containing the $i^{\rm th}$ part, and $t$ is the supporting direction. $t = 0, 1, 2$ respectively represents $x$, $y$ and $z$ directions where $y$ is the upright direction. 

\textbf{Support Relationship Constraint}: In order for a supporting part to well support a part being supported, two requirements are needed: 1) in the supporting direction, the bounding box of the supporting part should have tangential relation (or a small amount of intersection in practice) with the bounding box of the part being supported (see Figure~\ref{fig:refinement} (e) for a problematic case violating this constraint).
If the $i^{\rm th}$ part supports the $j^{\rm th}$ part, the following inequality should be satisfied 
along the supporting direction. $\mathbf{p}^{'}_{j}[t]-\mathbf{q}^{'}_{j}[t] + \varepsilon \mathbf{q}^{'}_{j}[t]\leq \mathbf{p}^{'}_{i}[t]+\mathbf{q}^{'}_{i}[t] \leq \mathbf{p}^{'}_{j}[t]-\mathbf{q}^{'}_{j}[t]+\YLN{2 \varepsilon} \mathbf{q}^{'}_{j}[t]$, %
where $t$ is the supporting direction, and $\varepsilon$ \YLN{controls} the amount of overlap allowed and is set to $\varepsilon=0.1$ in our experiments.
2) assuming $\tilde{b}_i$ and $\tilde{b}_j$ are the bounding boxes of parts $i$ and $j$ projected onto the plane orthogonal to the supporting direction $t$,
it should satisfy that either $\tilde{b}_i \subseteq \tilde{b}_j$ or $\tilde{b}_j \subseteq \tilde{b}_i$ (see \YLN{Figure}~\ref{fig:refinement} (c-d) for examples). This constraint can be formulated as an integer programming problem and solved efficiently during the optimization. The detailed formulation is provided in the Appendix.

\textbf{Stable Support Constraint}: For the ``support above'' relation ($t=1$), the center of a supported part should be located in the supporting bounding box (the bounding box that covers all the bounding boxes of the supporting parts) for stable support (see Figure~\ref{fig:refinement} (b) for an example violating this constraint). For a single supported part, the following constraints should be followed. $\mathbf{p}^{'}_{i}[l]-\mathbf{q}^{'}_{i}[l] \leq \mathbf{p}^{'}_{j}[l] \leq \mathbf{p}^{'}_{i}[l]+\mathbf{q}^{'}_{i}[l]$, $l\in\{0,2\}$. 
For multiple supporting parts %
\YL{(e.g. four legs supporting the table top), }
the lower bound and upper bound of the $x$ and $z$ directions will be chosen from the corresponding parts. 

\begin{figure}
	\centering
	{
    {\includegraphics[width=1.0\linewidth]{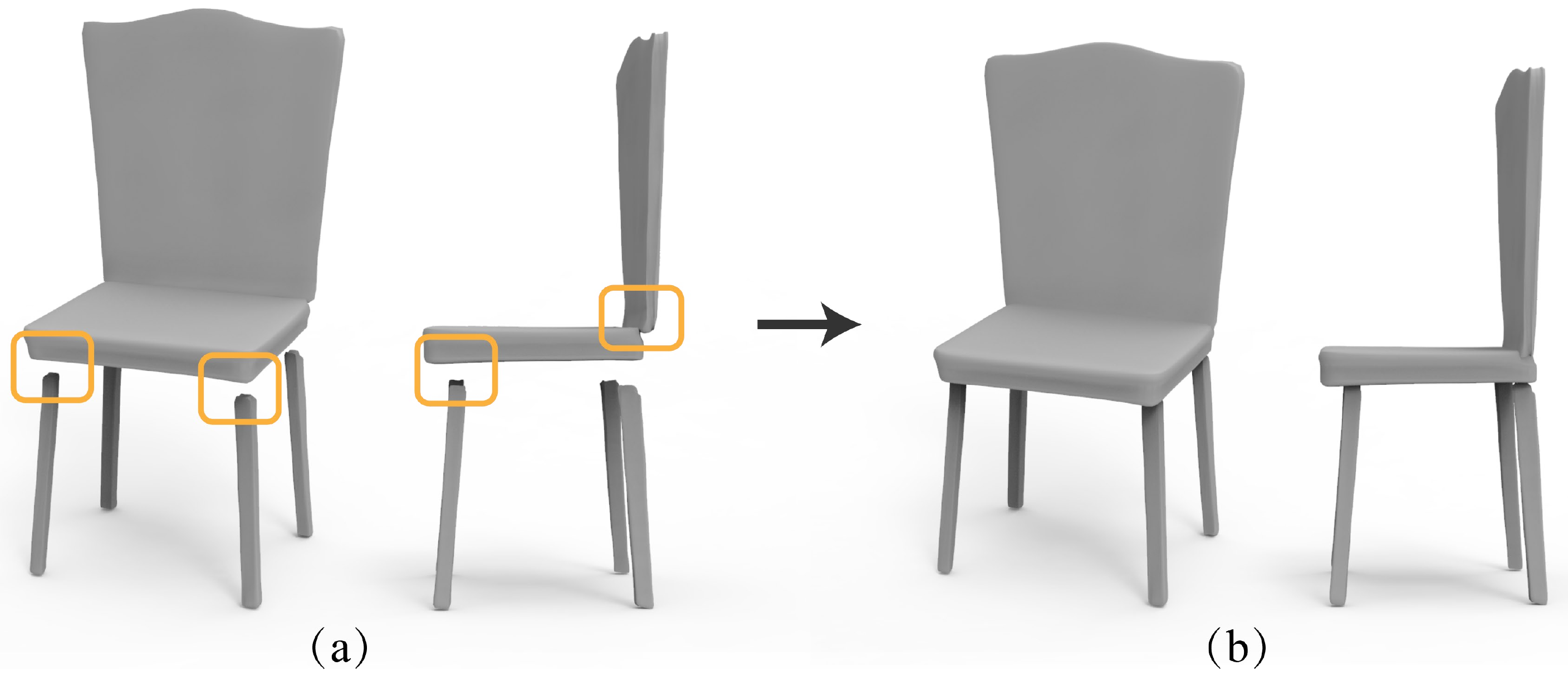}}
	}\caption{An example %
	showing the effect of shape refinement. (a) shape decoded by SDM-NET directly, (b) shape after optimization. The artifacts of gap and unstable support are fixed. 
	\YL{The sub-figures show the same shape from two viewpoints, with problematic areas highlighted.}}\label{fig:opti}
\end{figure}

\YLN{This} quadratic optimization with linear integer programming is solved by TOMLAB\YL{~\cite{holmstrom2004tomlab}} %
efficiently. We show an example of shape refinement in Figure~\ref{fig:opti}.

\begin{figure*}[ht]
	\centering
	{		
		{\includegraphics[width=0.2\linewidth]{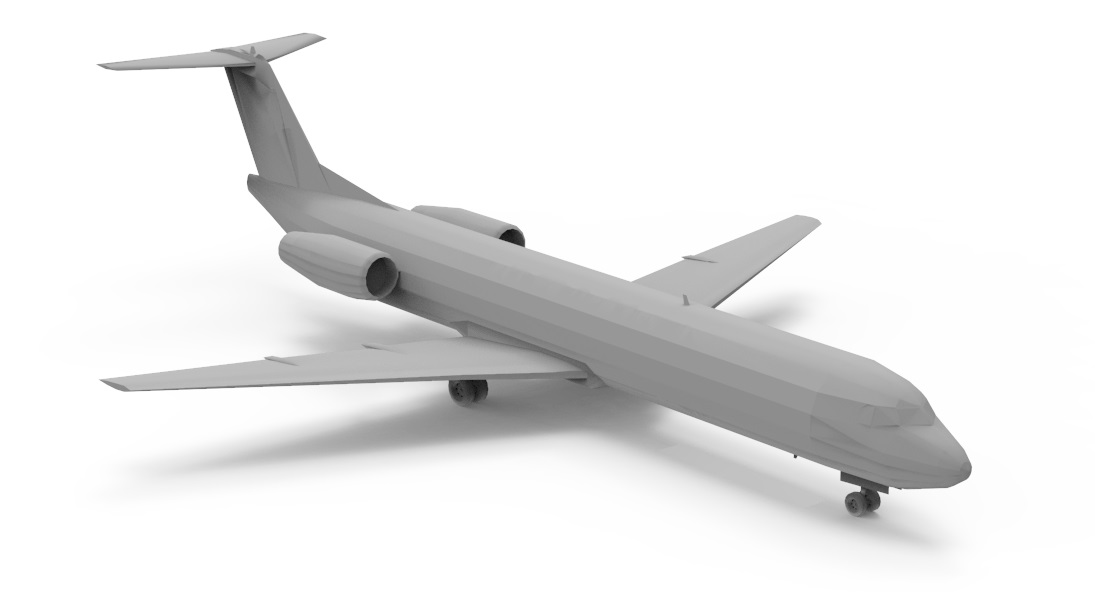}}
		{\includegraphics[width=0.2\linewidth]{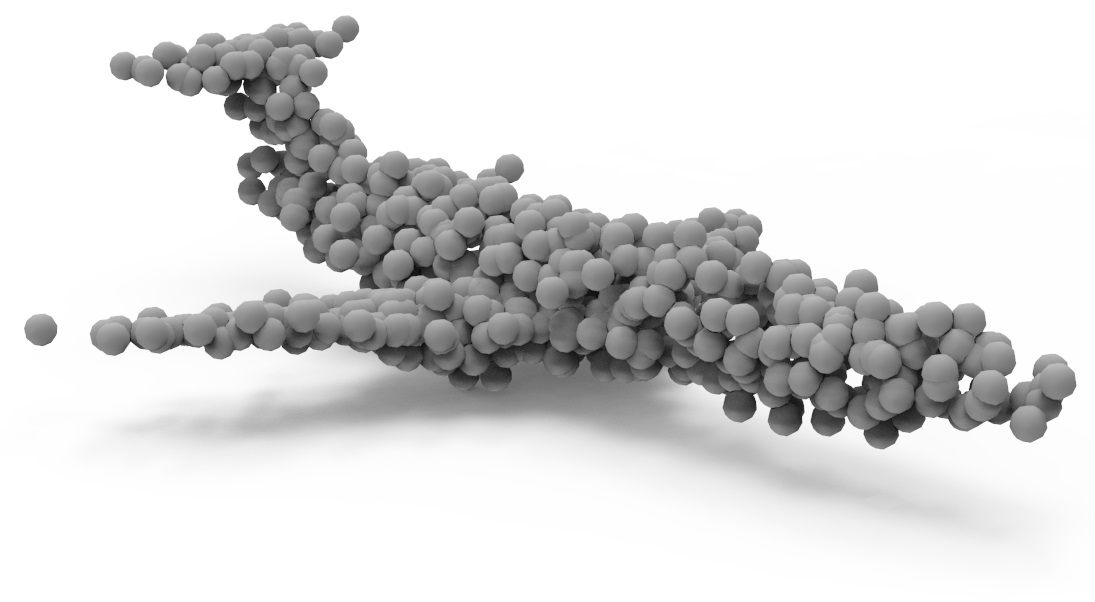}}
		{\includegraphics[width=0.2\linewidth]{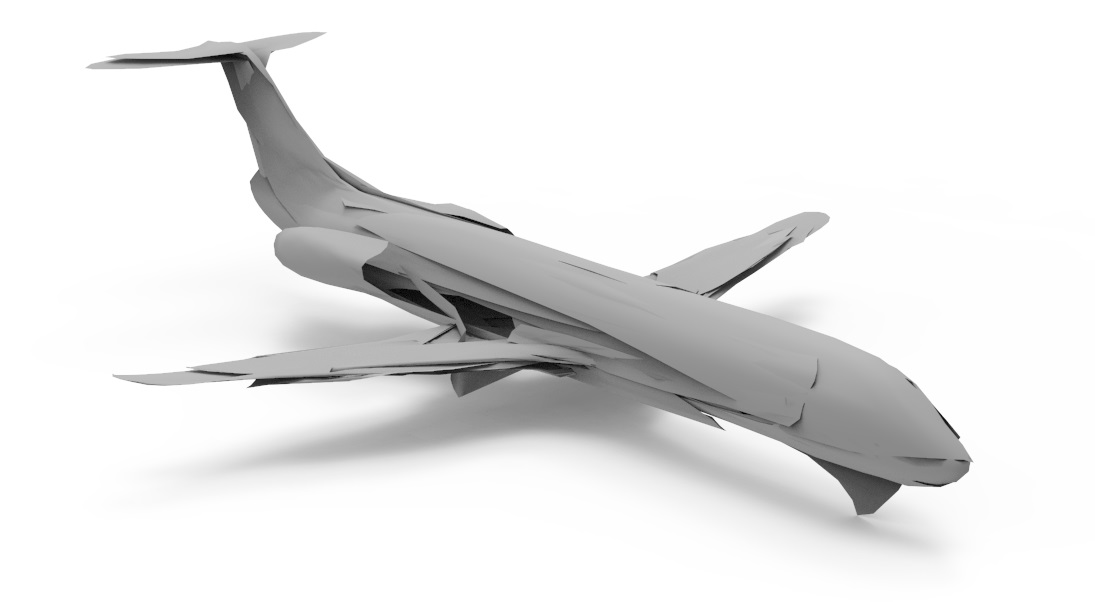}}
		{\includegraphics[width=0.2\linewidth]{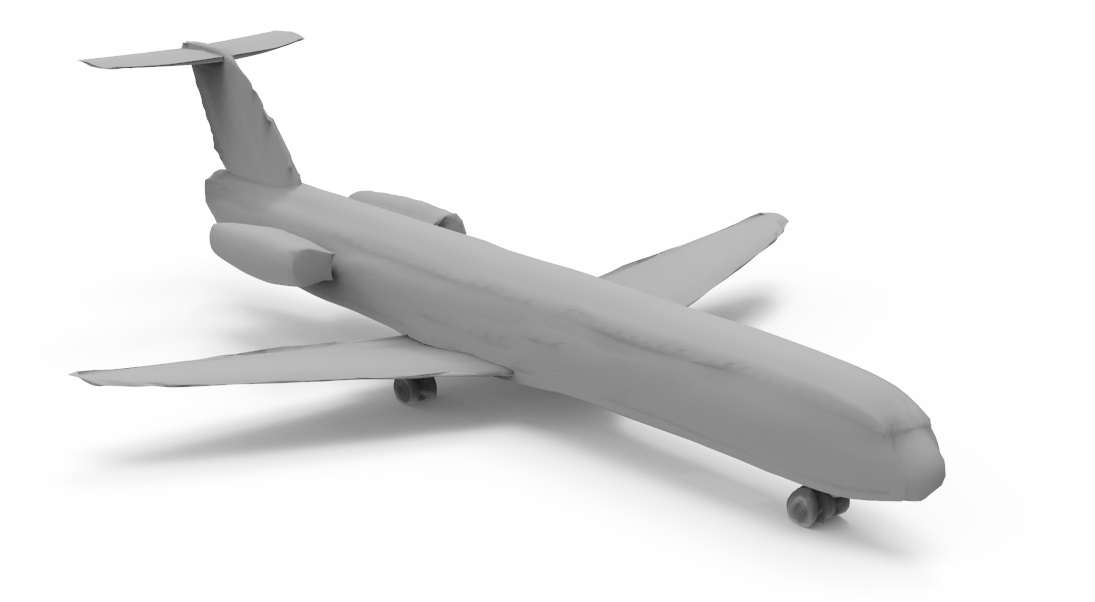}}
		
		{\includegraphics[width=0.2\linewidth]{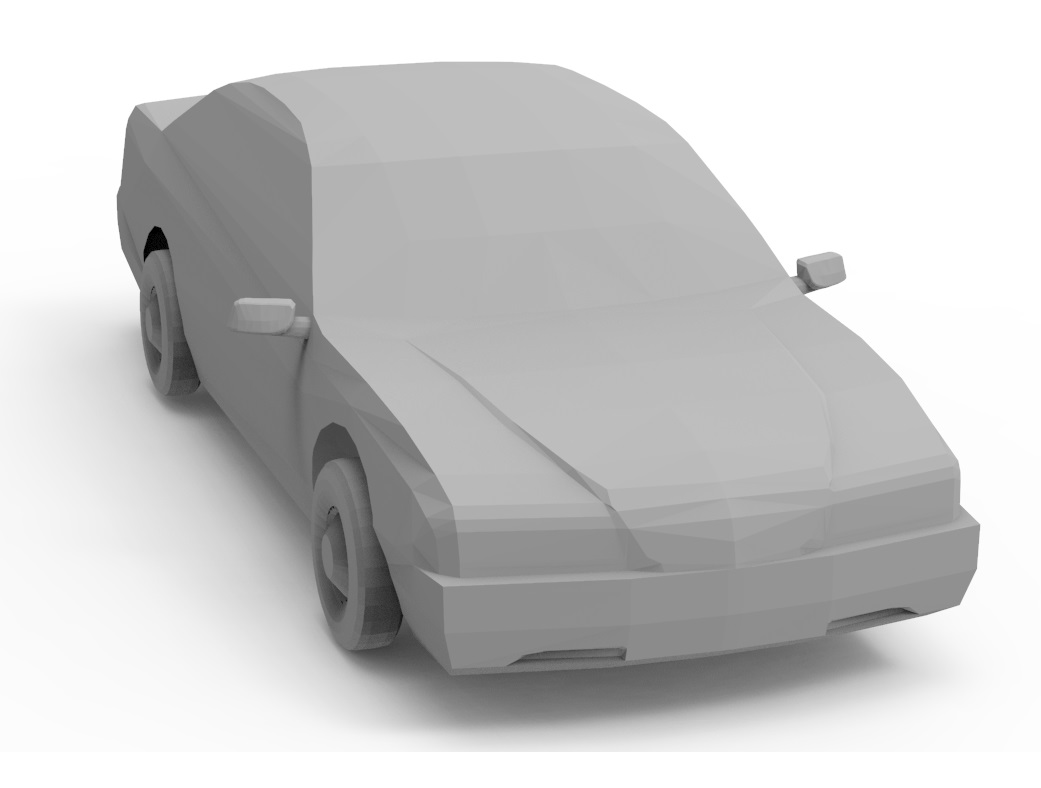}}
		{\includegraphics[width=0.2\linewidth]{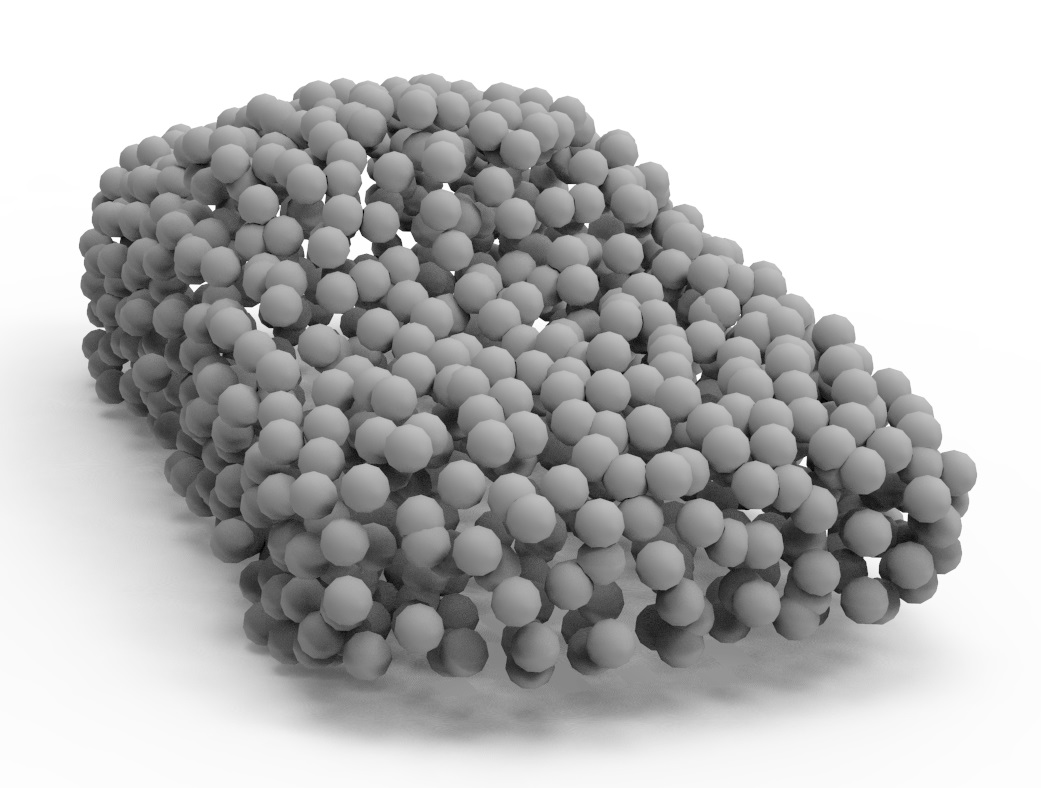}}
		{\includegraphics[width=0.2\linewidth]{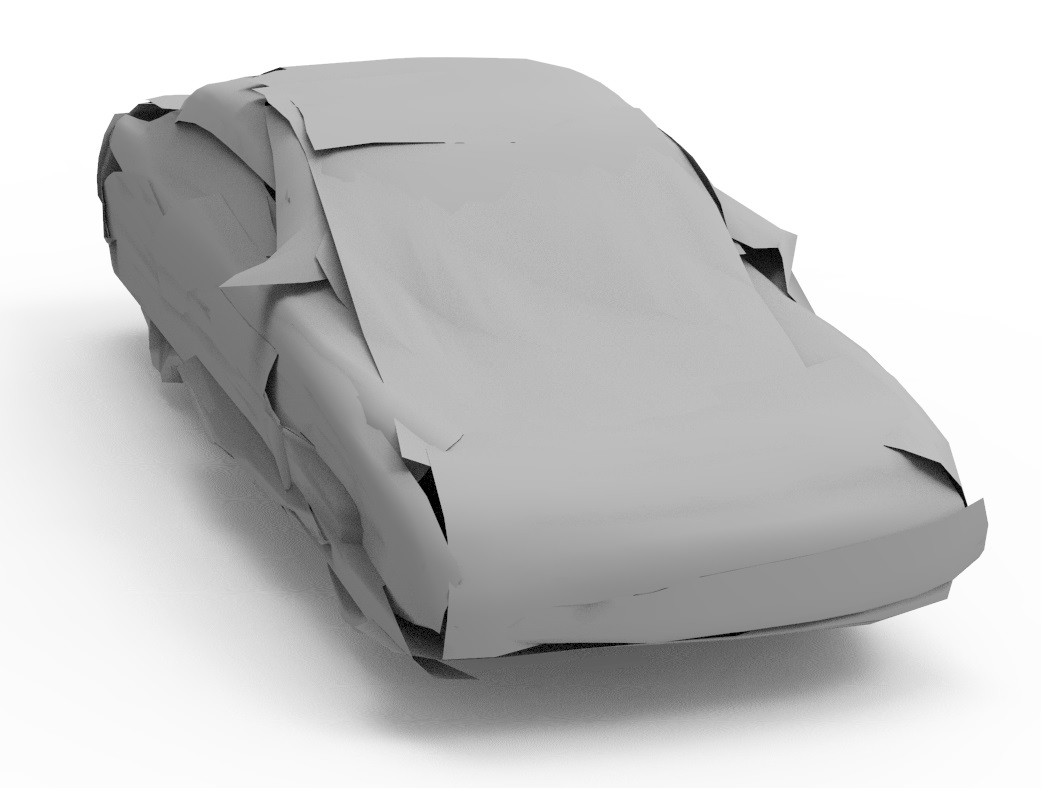}}
		{\includegraphics[width=0.2\linewidth]{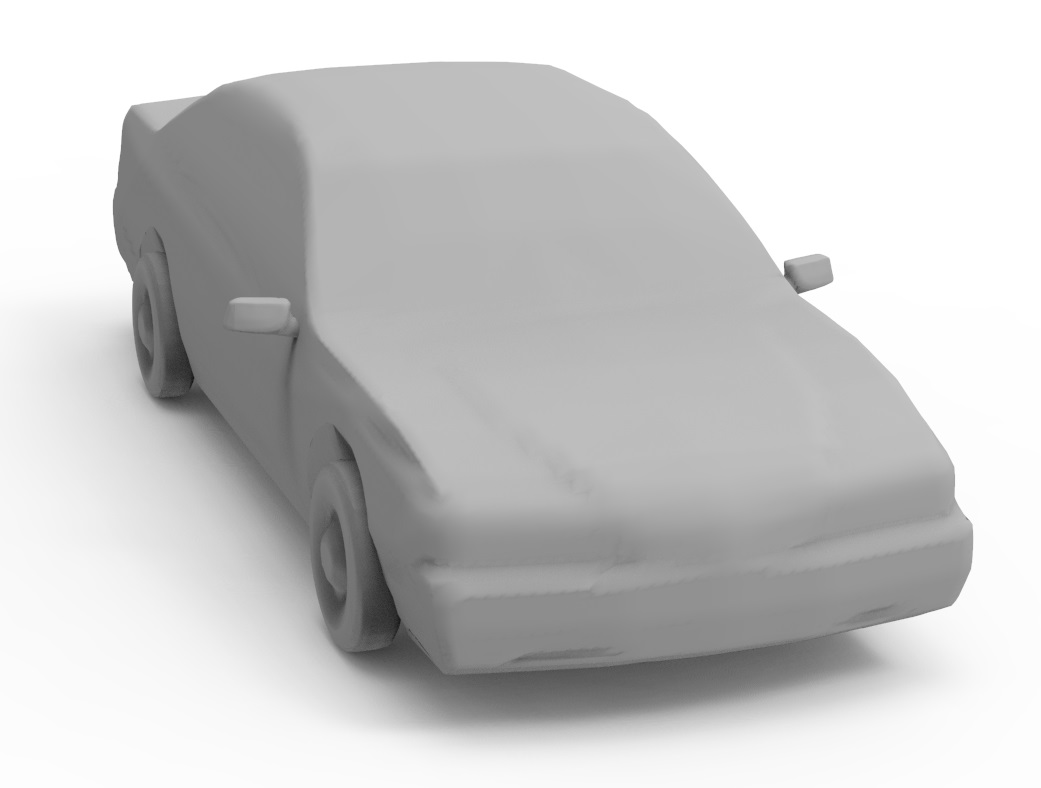}}
		
		\subfigure[Input shape]{\includegraphics[width=0.2\linewidth]{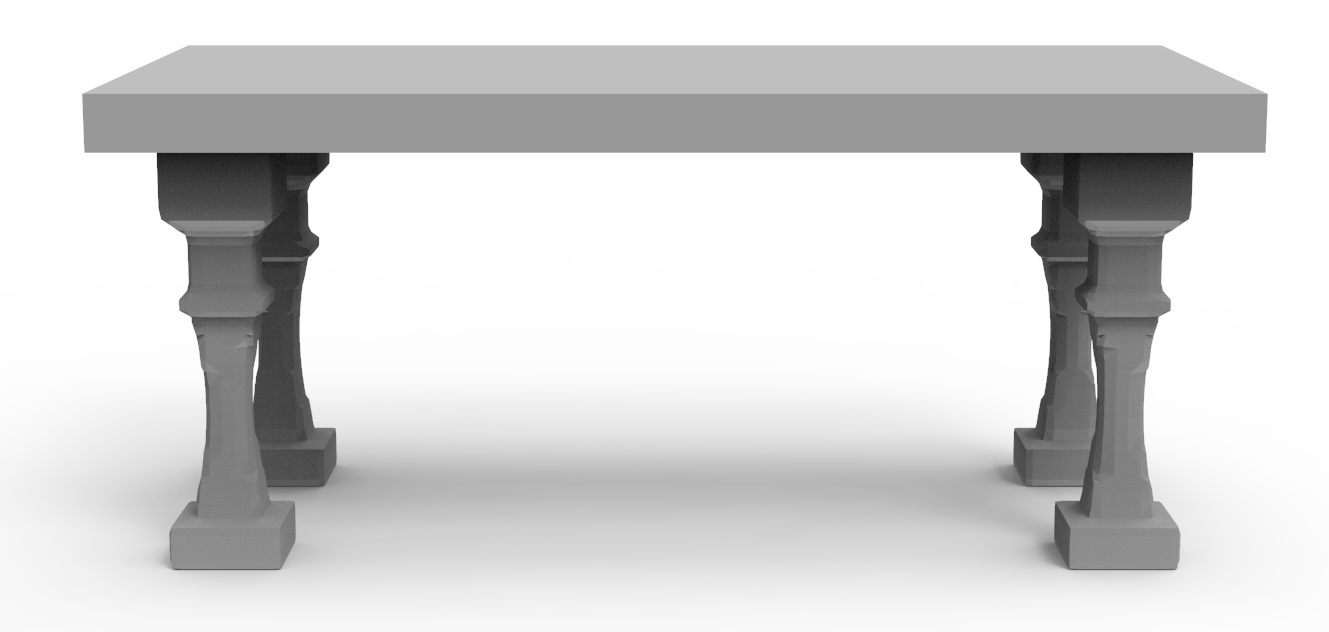}}
		\subfigure[PSG]{\includegraphics[width=0.2\linewidth]{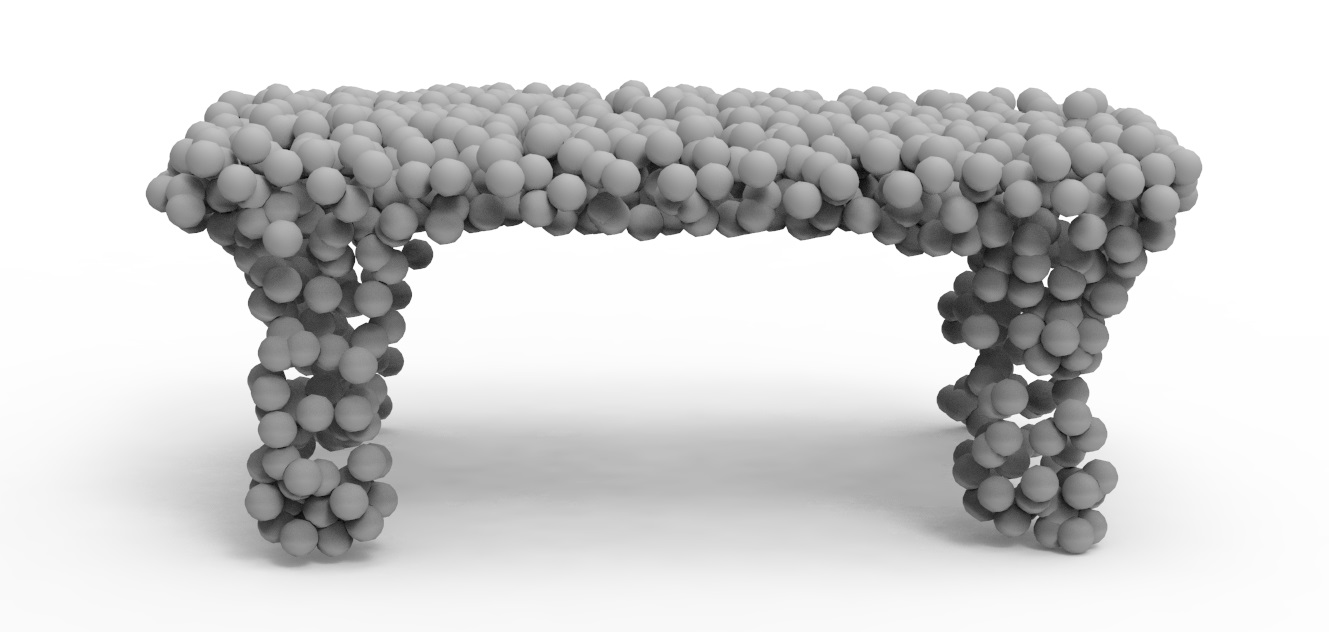}}
		\subfigure[AtlasNet]{\includegraphics[width=0.2\linewidth]{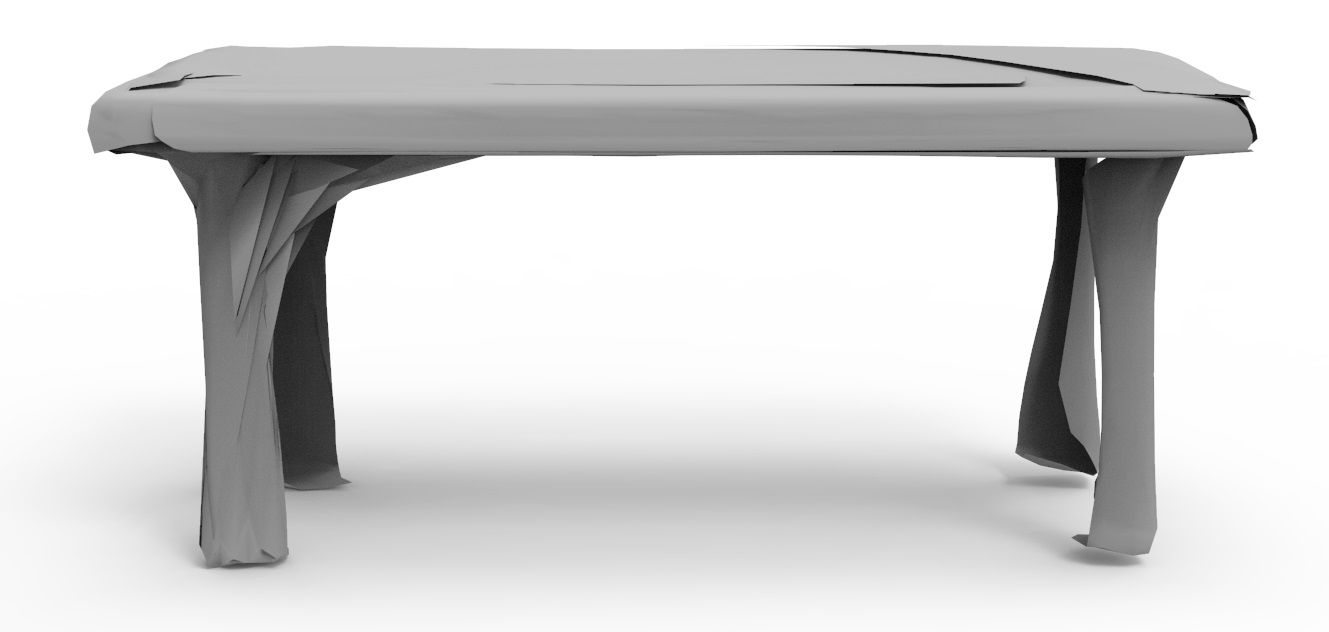}}
		\subfigure[Our Method]{\includegraphics[width=0.2\linewidth]{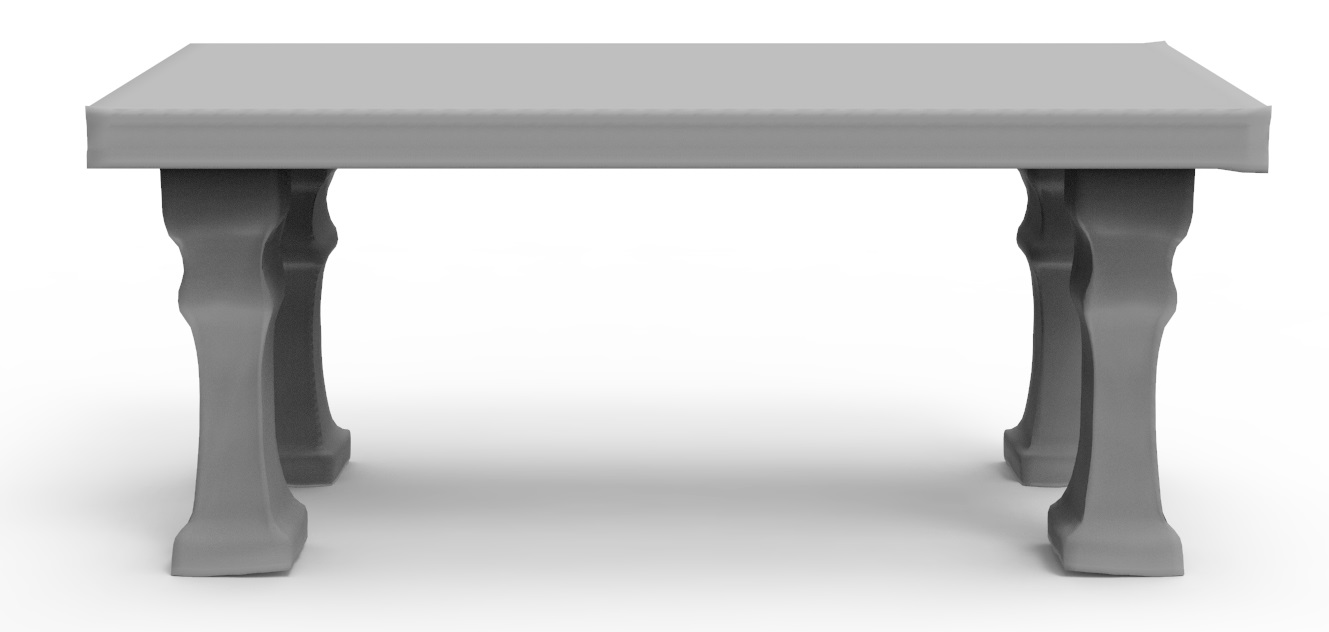}}
	}\caption{{Representative results of decoded shapes with different methods. Compared with PSG~\cite{fan2016point} and AtlasNet~\cite{AtlasNet2018}, our method produces the decoded shapes of higher quality.} PSG results are rather coarse point samples. 
	{The results by AtlasNet exhibit clearly noticeable patch artifacts.}
	}\label{fig:decodecompare}
\end{figure*}

\GL{
}

\section{Dataset and Network Implementation}
\label{sec:impl}

We now give the details of our network architecture and training process. The experiments were carried out on a computer with an i7 6850K CPU, 64GB RAM, and a GTX 1080Ti GPU.

\begin{figure}[ht]
	\centering
	{		
		\subfigure[Input shape]{\includegraphics[width=0.27\linewidth]{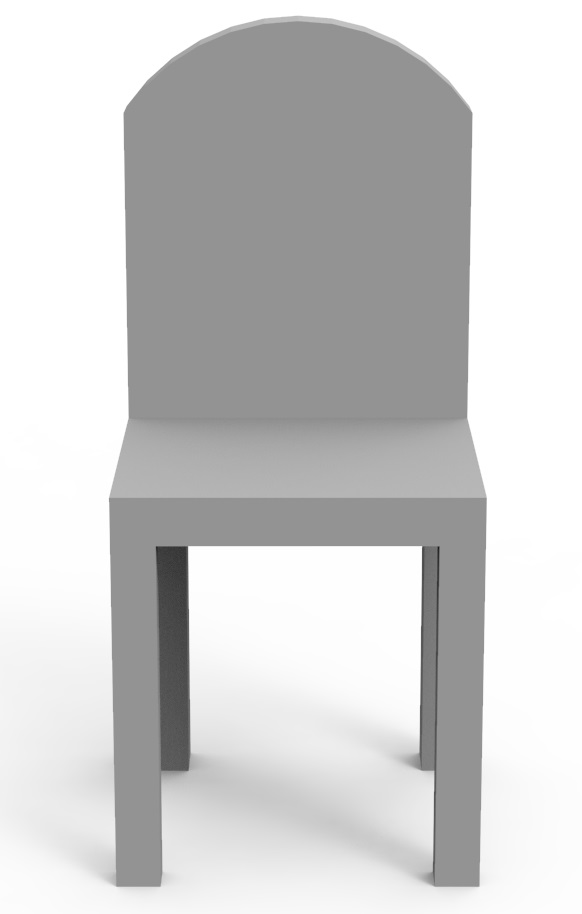}}
		\subfigure[Adaptive O-CNN]{\includegraphics[width=0.27\linewidth]{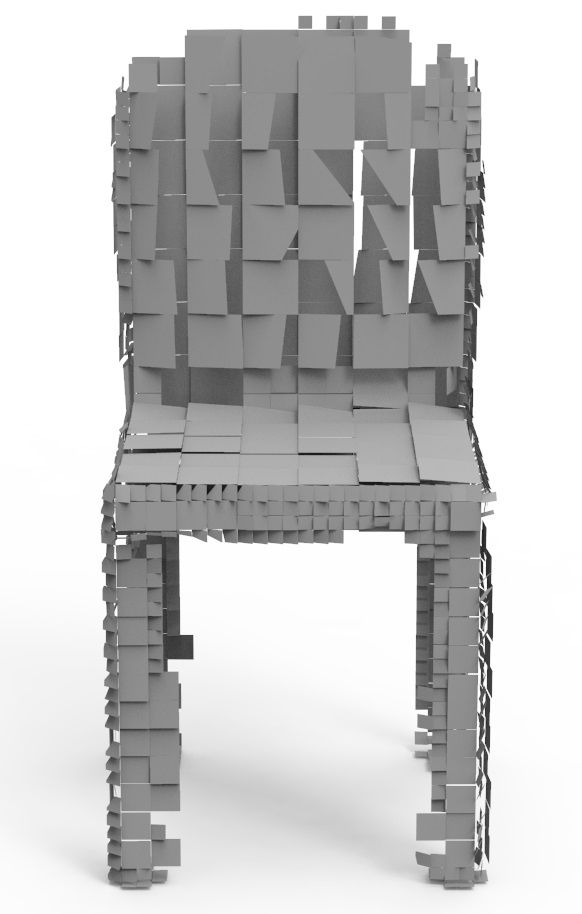}}
		\subfigure[Our Method]{\includegraphics[width=0.27\linewidth]{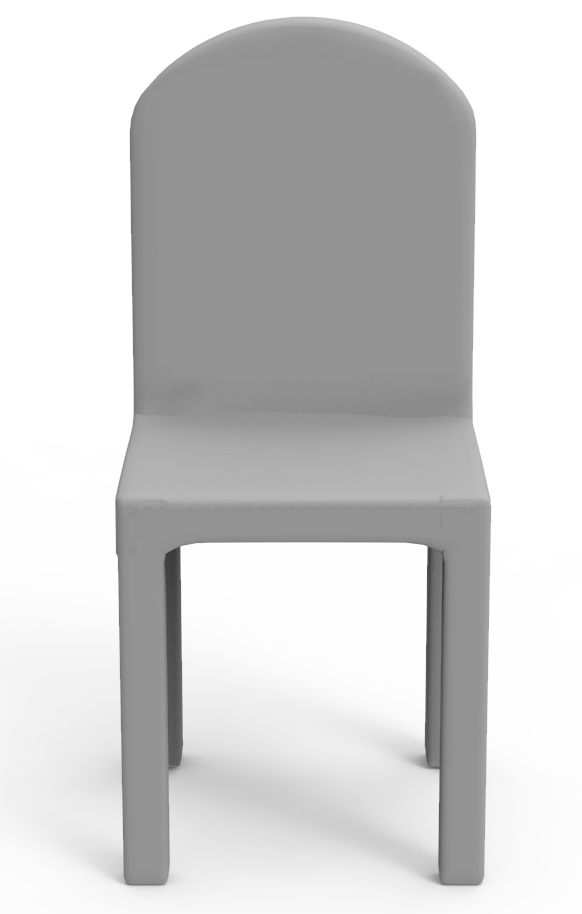}}
	}
	\caption{Visual comparison of the decoded shapes with Adaptive O-CNN~\protect\cite{Wang2018ocnn} and our method. Compared with Adaptive O-CNN, while the planar regions of the chair can be decoded by both methods, the curved regions such as the top of the chair back can be recovered only by our method.}\label{fig:ocnncomp}
\end{figure}

\subsection{Dataset Preparation}
The mesh models used in our paper are from~\cite{Yi16}, including a subset of ShapeNet Core V2 models~\cite{shapenet}, as well as ModelNet~\cite{Wu_2015_CVPR}.
These datasets include pre-aligned models. However, ModelNet does not contain semantic segmentation, and models from~\cite{Yi16} sometimes do not have sufficiently detailed segmentation to describe support structure (e.g. the car body and four wheels are treated as a single segment). 
To facilitate our processing, we use an active learning approach~\cite{Yi16} to perform a refined semantic segmentation. We further
use refined labels to represent individual parts of the same type,
 e.g., to have \emph{left armrest} and \emph{right armrest} labels for two armrest parts. The statistics of the resulting dataset are shown in Table~\ref{tab:dataset}.
 
\rv{\YLN{Our network takes 3D shapes with consistent segmentation as input. \hongbo{The segmentation of test shapes can be obtained by some supervised methods such as~\cite{Qi2017cvpr,Qi2017nips}. Each segmented part is registered from the bounding box by non-rigid deformation. 
} Our method allows each part type to include substantial geometric variations, e.g., the swivel leg and bar stool in Figure~\ref{fig:grassrandom}.}
Within the confines of the consistent segmentation, the SP-VAE is capable of handling variations of part structure and topologies, as well as varying part counts (by marking certain parts as non-existing).}

\begin{table}[h]
\fontsize{9}{12}\selectfont
  \centering
    \begin{tabular}{c||p{0.9cm}<{\centering}p{0.4cm}<{\centering}p{0.5cm}<{\centering}p{0.5cm}<{\centering}p{0.4cm}<{\centering}p{0.8cm}<{\centering}p{0.6cm}<{\centering}}
    \hline
    Category & Airplane & Car & Chair & Table & Mug & Monitor & Guitar \\
    \hline\hline
    \# Meshes & 2690 & 1824 & 3746 & 5266 & 213 & 465 & 787\\
    \# Labels & 14   &  7   &  10   &  9  &   2 & 3 & 3\\
    \hline
    \end{tabular}%
  \caption{The numbers of meshes and \YL{total} part labels for each category in  our  dataset (after label refinement). 
  }\label{tab:dataset}
\end{table}%

\subsection{Network Architecture}
The whole network includes two components, namely \emph{PartVAE} for encoding the deformation of each part of a shape, and \emph{SP-VAE} for jointly encoding the global structure of the shape and the geometric details of each part.

As illustrated in Figure~\ref{fig:compvae}, the structure of the PartVAE has two convolutional layers and one fully connected layer. We use $tanh$ as the activation function, and in the last convolution layer, we use the linear output. The output of the last convolution layer is reshaped to a vector and mapped into a 64-dimensional latent space by the fully connected layer. The decoder has a mirrored structure, {but not} sharing weights with the encoder. We train the PartVAE once for each part type. 

The input of the SP-VAE is the concatenated representation vector of all parts as shown in Figure~\ref{fig:pipeline}. The input is fully connected with dimensions 1024, 512 and 256, respectively, and the latent space dimension is 128. Leaky ReLU is set as the activation function.

\subsection{Parameters}
\YL{We use fixed hyper-parameters in our experiments for different shape categories. In the following, we perform experiments on the \emph{table} data in the ShapeNet Core V2 to demonstrate how the method behaves with changing hyper-parameters.}
The dataset is randomly split into the training data (75\%) and test data (25\%). The generalization of SP-VAE is evaluated with different hyper-parameters in Table~\ref{tab:gen_comp}, where the bidirectional Chamfer distance is used to measure the reconstruction error on the test data (as unseen data). %
We perform such tests for 10 times and report the average errors in Table~\ref{tab:gen_comp}. As can be seen, SP-VAE has the lowest error with the hyper-parameters $\lambda_1=1.0$ and $\lambda_2=0.5$, where $\lambda_1$ and $\lambda_2$ are the weights of the reconstruction error term and KL divergence term, respectively. 
{The hyper-parameters (weights of reconstruction, KL-divergence, and regularization) of PartVAE are set to the same \YLN{numbers} in~\cite{Gao2018}.}
We set the dimension of the latent space of PartVAE to \YL{64}, and the dimension of the latent space of SP-VAE to \YL{128}. These two parameters are evaluated in Tables~\ref{tab:local_latent_comp} and~\ref{tab:global_latent_comp} with the reconstruction error.
When adjusting the dimension of one VAE, \YLN{we} leave the dimension of the other VAE unchanged.

\begin{table}[h]
\fontsize{8}{12}\selectfont
	\centering
	\begin{tabular}{c||cccc}
		\hline
		\tabincell{c}{($\lambda_1,\lambda_2$)}& (0.5, 0.5) & (1.0, 0.5) & (0.5, 1.0)  & (1.0, 1.0) \\
		\hline\hline
		Recons. Error ($\times 10^{-3}$)    & 2.01 & \textbf{1.85} & 2.24 & 1.94\\
		\hline
	\end{tabular}
	\caption{\YL{Comparison of average SP-VAE reconstruction errors (measured in bidirectional Chamfer distance) for unseen data on the \emph{table} dataset w.r.t. changing hyper-parameters.}} %
	\label{tab:gen_comp}
\end{table}

\begin{table}%
\fontsize{9}{12}\selectfont
	\centering
	\begin{tabular}{c||cccc}
		\hline
		\tabincell{c}{PartVAE Embedding Dimension } & 32  & 64 & 128  & 256 \\
		\hline\hline
		\tabincell{c}{PartVAE Recons. Error ($\times10^{-3}$)}  & 1.92 & 1.76  & \textbf{1.74} & 1.82 \\
		\hline
		\tabincell{c}{SP-VAE Recons. Error ($\times10^{-3}$)} & 2.16 & \textbf{1.85}  & 1.91 & 2.03  \\
		\hline
	\end{tabular}
	\caption{Comparison of average reconstruction errors (measured by bidirectional Chamfer distance) for PartVAE and SP-VAE w.r.t. changing dimension of the PartVAE latent space.}
	\label{tab:local_latent_comp}
\end{table}

\begin{table}[h]
\fontsize{9}{12}\selectfont
    \centering
	\begin{tabular}{c||cccc}
		\hline
		\tabincell{c}{SP-VAE Embedding Dimension} 
		& 32 & 64 & 128  & 256 \\
		\hline\hline
		\tabincell{c}{SP-VAE Recons. Error ($\times 10^{-3}$)} & 2.23 & 1.99  & \textbf{1.85} & 1.91  \\
		\hline
	\end{tabular}
	\caption{Comparison of average reconstruction errors (measured by bidirectional Chamfer distance) of SP-VAE w.r.t. changing embedding dimension.}
	\label{tab:global_latent_comp}
\end{table}

\subsection{Training Details}

\YL{Since a PartVAE encodes the geometry of a specific type of parts, it is trained separately. SP-VAE is then trained using PartVAE for encoding part geometry. Training of both VAEs is optimized using the Adam solver~\cite{adamsolver}.}
The PartVAE is trained with 20,000 iterations and SP-VAE with 120,000 iterations by minimizing their loss functions. For both VAEs, we set the batch size as 512 and learning rate starting from 0.001 and decaying every 1000 steps with the decay rate set to 0.8. The training batch is randomly sampled from the training data set.

\YL{For a typical category,} the training of both  PartVAE and SP-VAE takes about 300 minutes. Once the networks are trained, shape generation is very efficient: generating one shape and structure optimization take only about \YL{36 and 100
milliseconds, respectively.}

\section{Results and Evaluation}
\label{sec:exp}

\begin{figure}
	\centering
	{
		\subfigure[]{\includegraphics[width=0.24\linewidth]{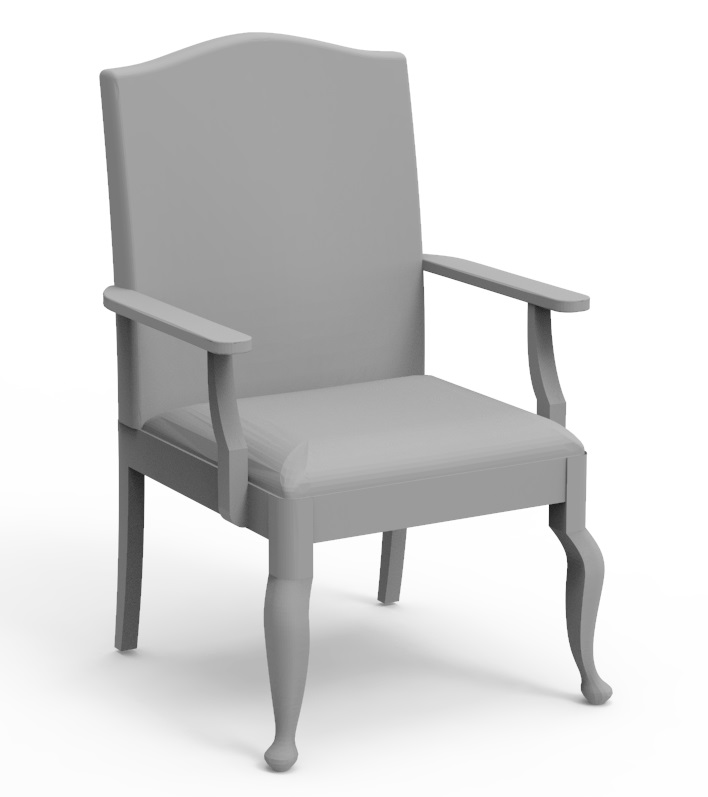}}
		\subfigure[]{\includegraphics[width=0.24\linewidth]{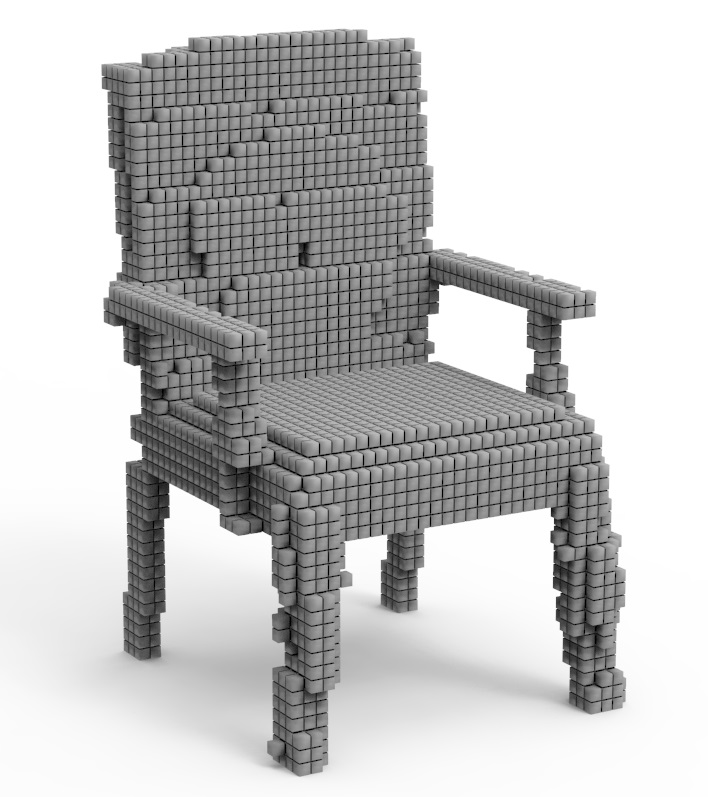}}
		\subfigure[]{\includegraphics[width=0.24\linewidth]{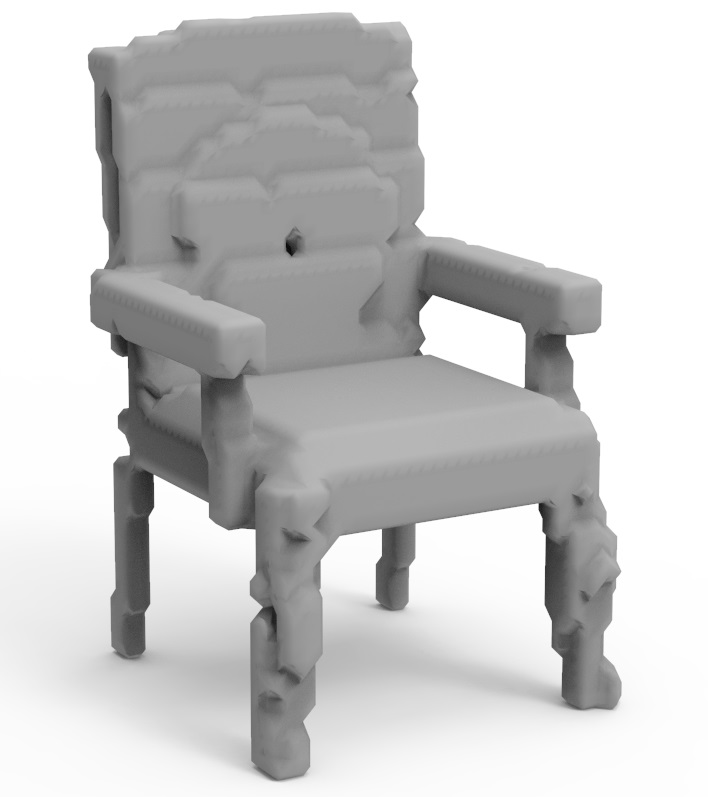}}
		\subfigure[]{\includegraphics[width=0.24\linewidth]{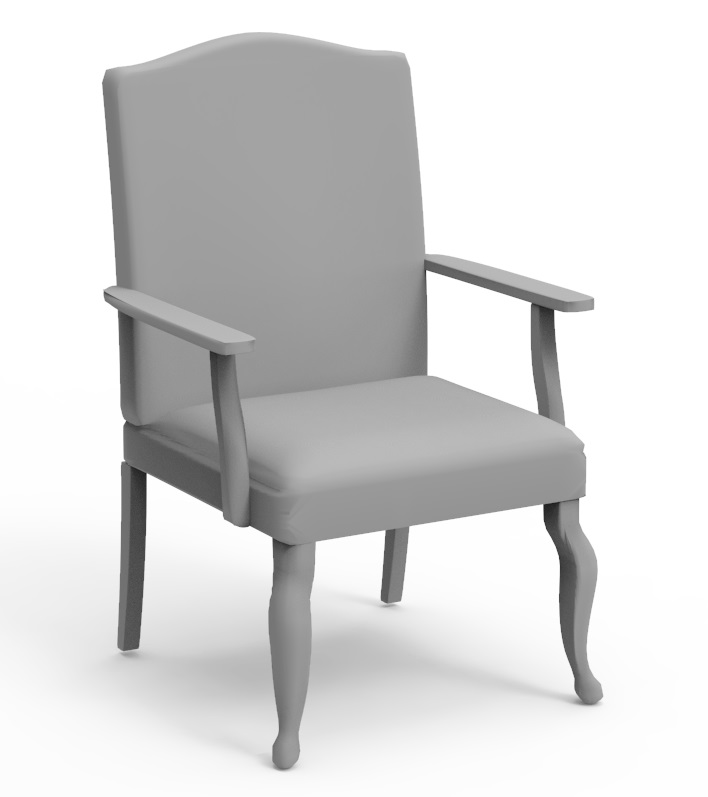}}
	}
	\caption{{Visual comparison of shape reconstruction with  GRASS~\cite{li_sig17} and our technique. (a) input shape, (b)(c) GRASS results in voxels and extracted mesh, (d) Our result.}}
	\label{fig:compwithgrass_voxel}
\end{figure}

\begin{figure}
	\centering
	{
		\subfigure[]{\includegraphics[width=0.35\linewidth]{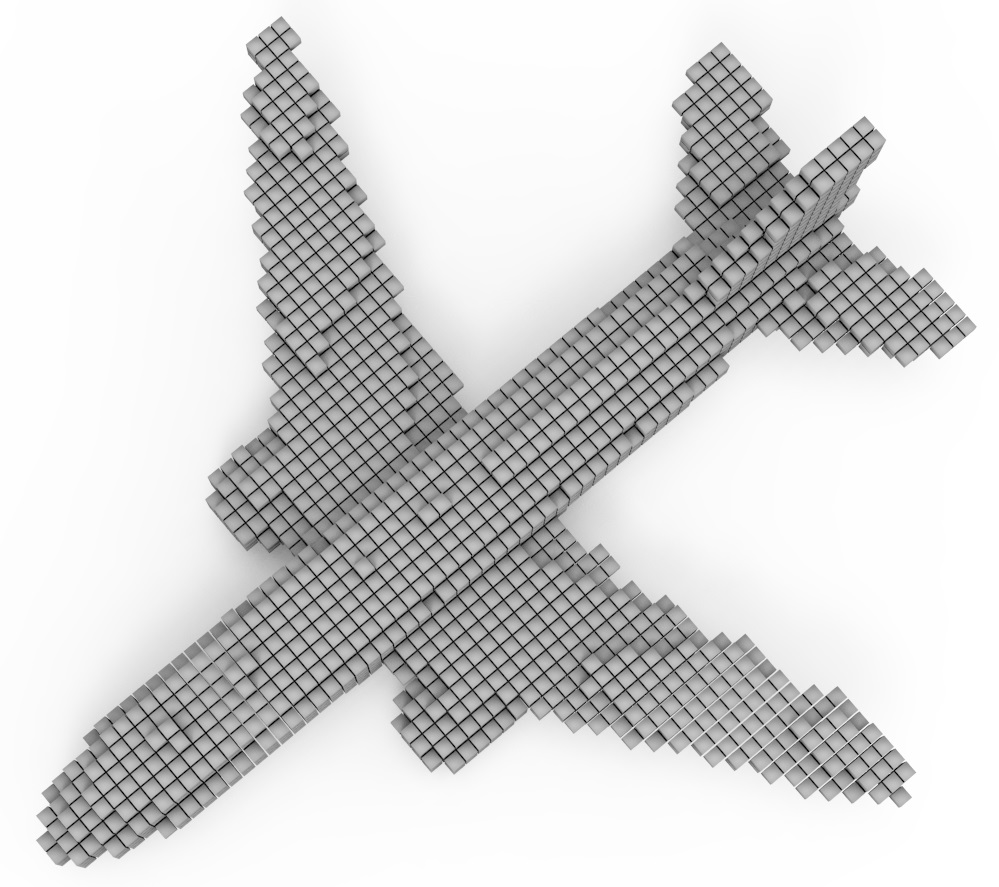}}
		\subfigure[]{\includegraphics[width=0.35\linewidth]{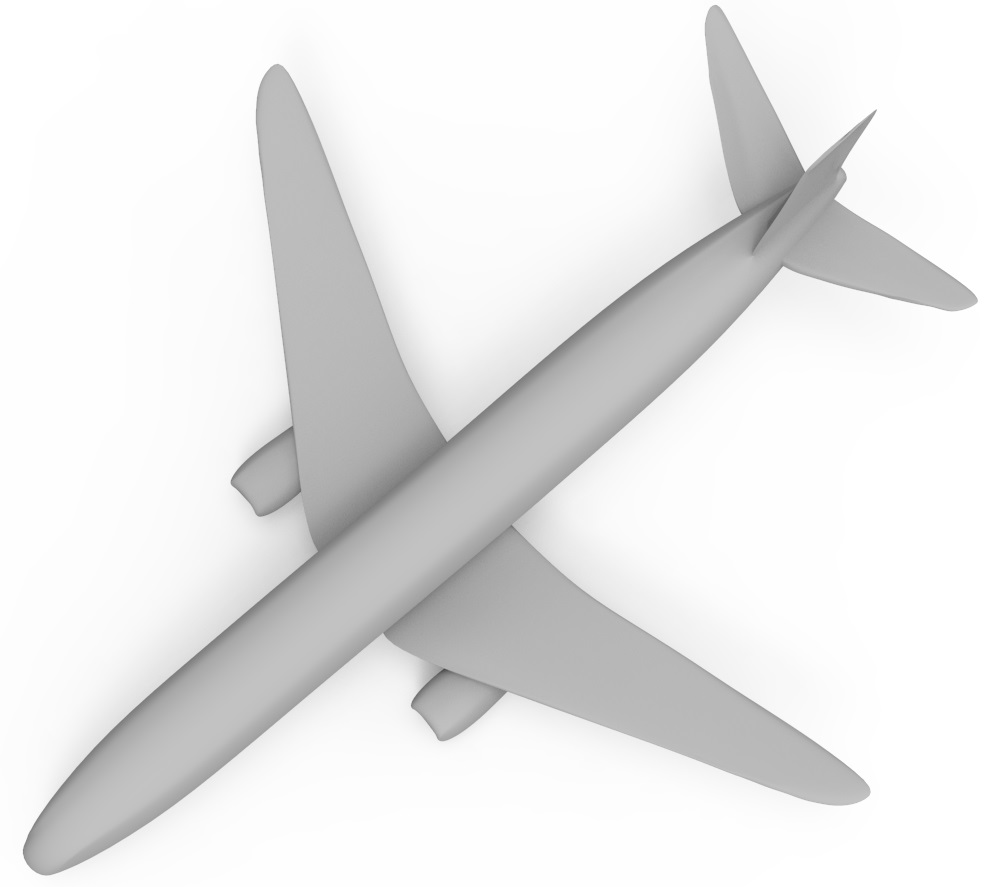}}
	}\caption{Visual comparison between the global-to-local method~\cite{Wang2018TOG} (a) and our technique (b) for shape generation.}\label{fig:g2l}
\end{figure}

\begin{figure}
	\centering
	{
		\subfigure[]{\includegraphics[width=0.25\linewidth]{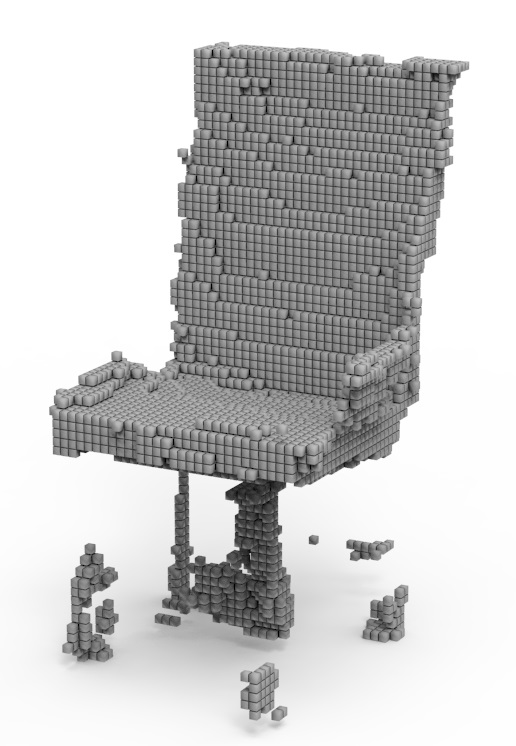}}
		\subfigure[]{\includegraphics[width=0.25\linewidth]{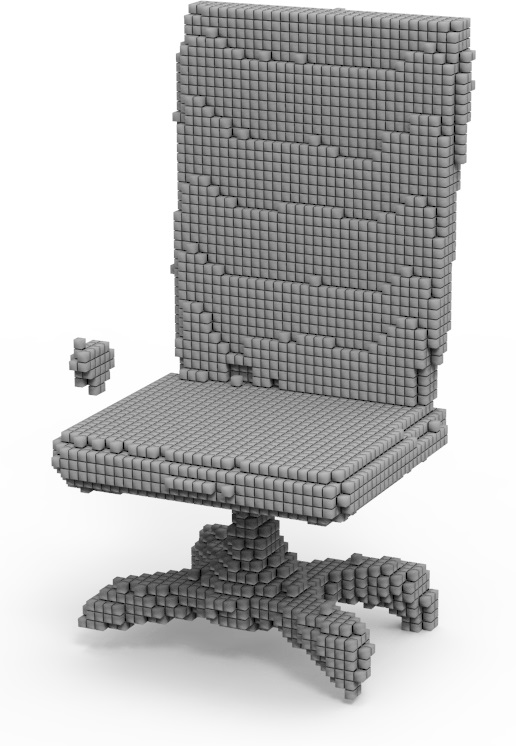}}
		\subfigure[]{\includegraphics[width=0.25\linewidth]{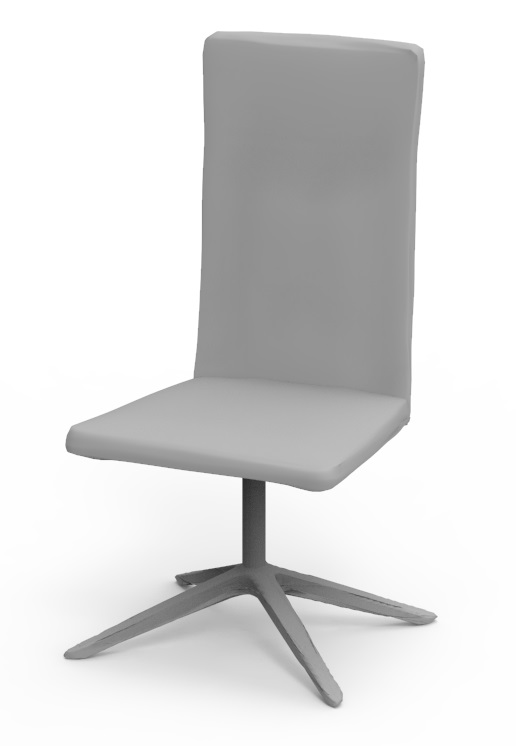}}
	}\caption{\yyja{Visual comparison of shape generation using different methods. (a) 3DGAN~\cite{3dgan2016}, (b) the global-to-local method~\cite{Wang2018TOG}, (c) our technique. }}\label{fig:3dgan}
\end{figure}

We present the results of shape reconstruction, shape generation and shape interpolation to demonstrate the capability of our method, and compare them with those generated by the state-of-the-art methods. We also perform ablation studies to show the advantages of our design. Finally, we present examples to show generalizability (i.e., applying our learned model to new shapes of the same category), editability and limitations of our technique. 

\textbf{Shape Reconstruction.}
 We compare our method with PSG~\cite{fan2016point}, AtlasNet~\cite{AtlasNet2018} and Adaptive O-CNN~\cite{Wang2018ocnn} on the ShapeNet Core V2 dataset. %
In this experiment, we choose four \YL{representative} 
categories \YL{commonly used in the literature} to perform both qualitative and quantitative comparisons. Each dataset is randomly split into the training set ($75\%$) and  test set ($25\%$). 
{For fair comparison, we train PSG, AtlasNet, and Adaptive O-CNN for individual shape categories, similar to ours. To prepare the input for PSG, we use rendered images under different viewpoints.}
 Given the same input models in the test set, we compare the decoded shapes by different methods.
Figures~\ref{fig:decodecompare} and~\ref{fig:ocnncomp} show the visual comparison of representative results on several test shapes. It can be easily seen that the decoded shapes by PSG, Adaptive O-CNN and AtlasNet cannot capture the shapes faithfully. AtlasNet and Adaptive O-CNN are able to produce more details than PSG, but suffer from clearly noticeable patch artifacts. 
In contrast, SDM-NET recovers shapes with higher quality and finer-detailed geometry. Note that we compare the existing methods with SP-VAE followed by structure optimization instead of SP-VAE alone, since structure optimization, which is dependent on the output of SP-VAE, is a unique and essential component in our system, and cannot be directly used with the methods being compared \YL{due to their lack of structure information}.

\begin{figure}
	\centering
	{
		{\includegraphics[width=0.18\linewidth]{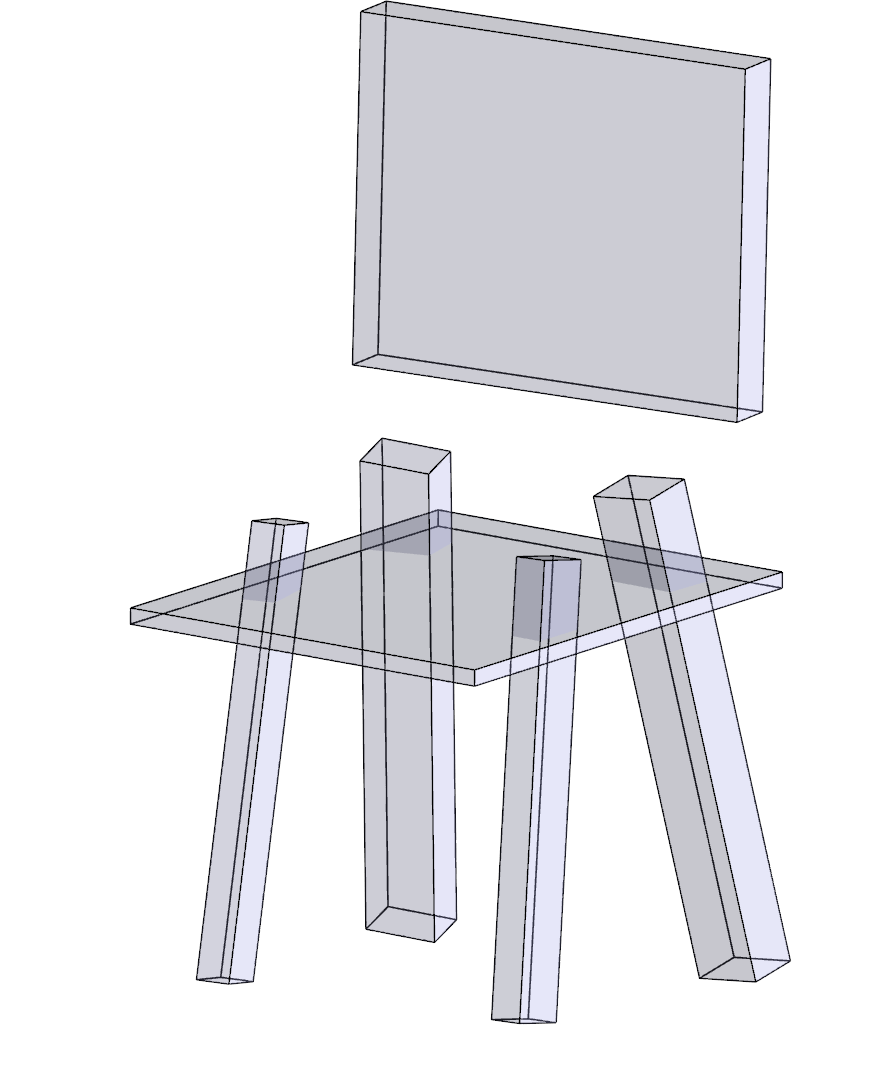}}
		{\includegraphics[width=0.18\linewidth]{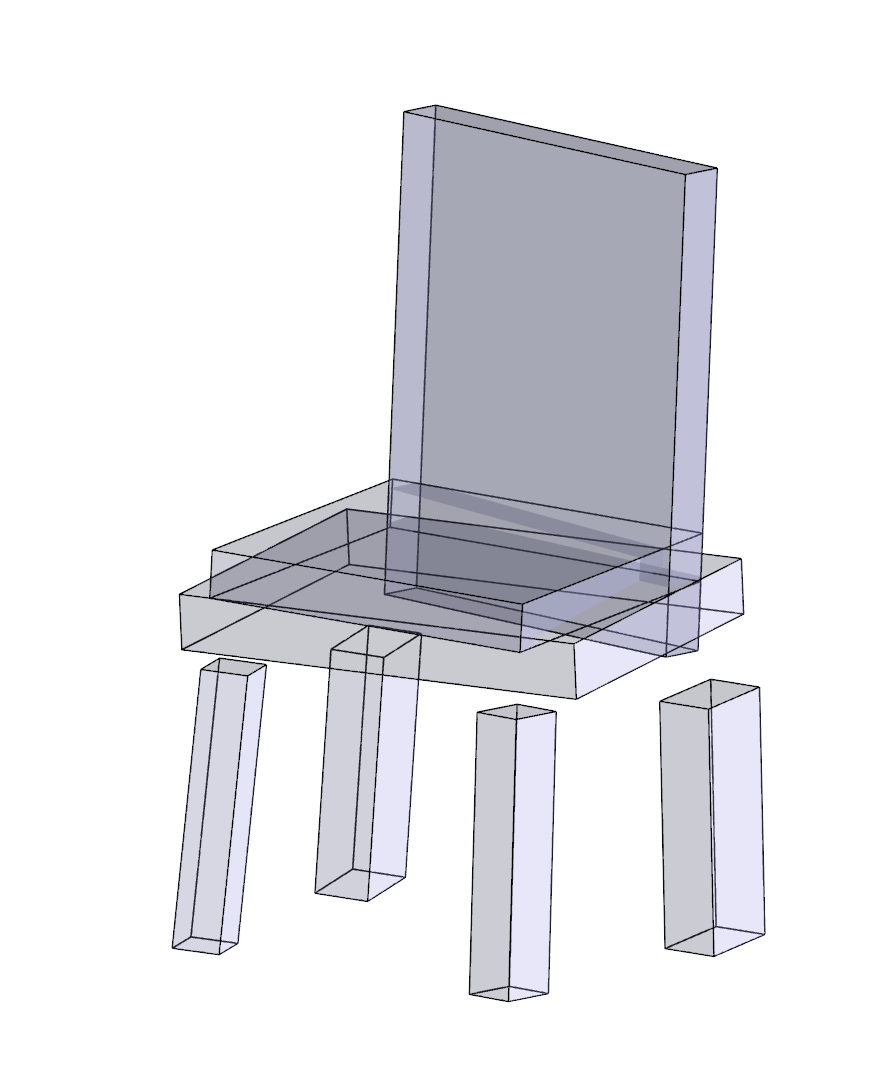}}
		{\includegraphics[width=0.18\linewidth]{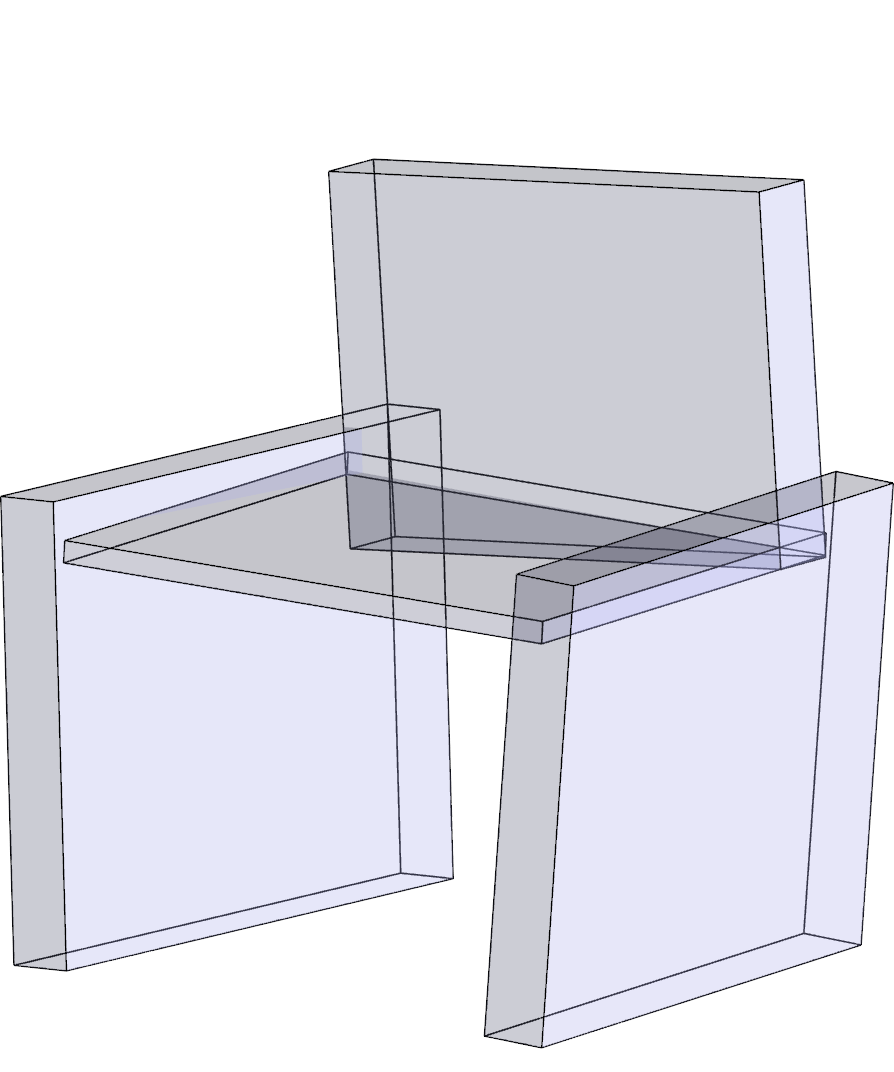}}
		{\includegraphics[width=0.18\linewidth]{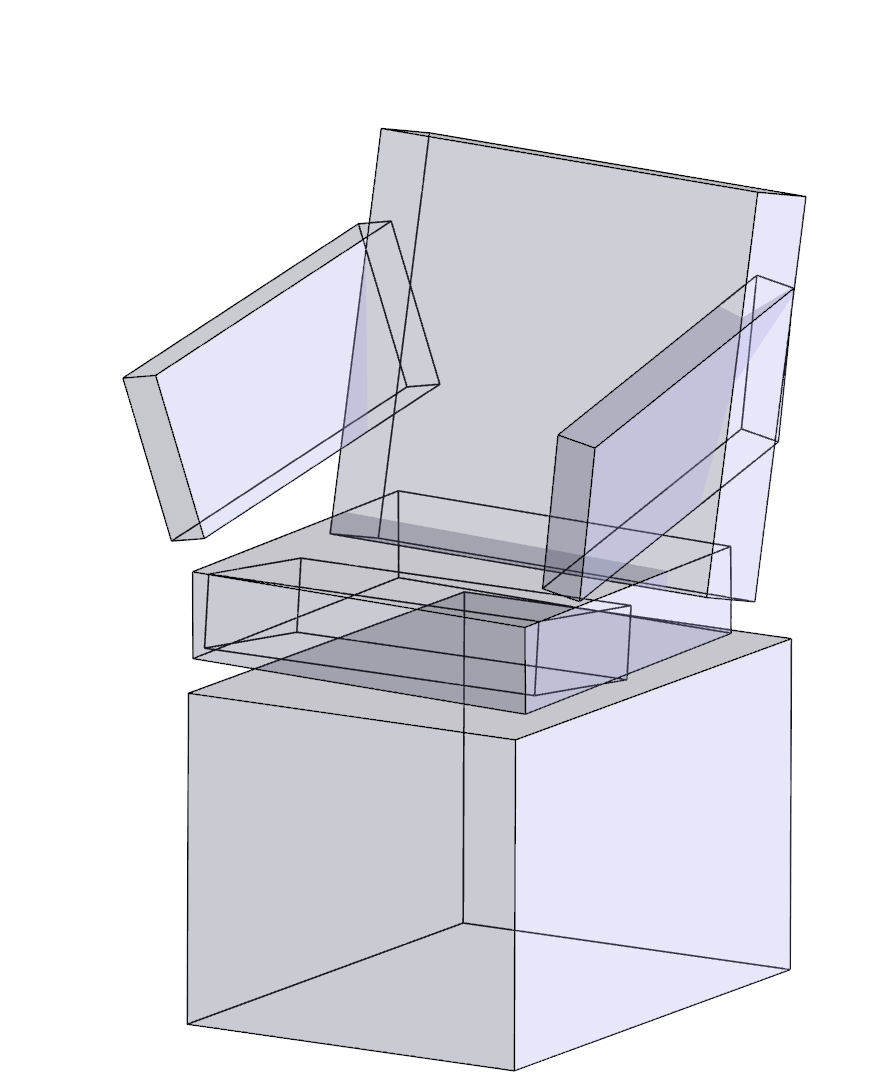}}
		{\includegraphics[width=0.18\linewidth]{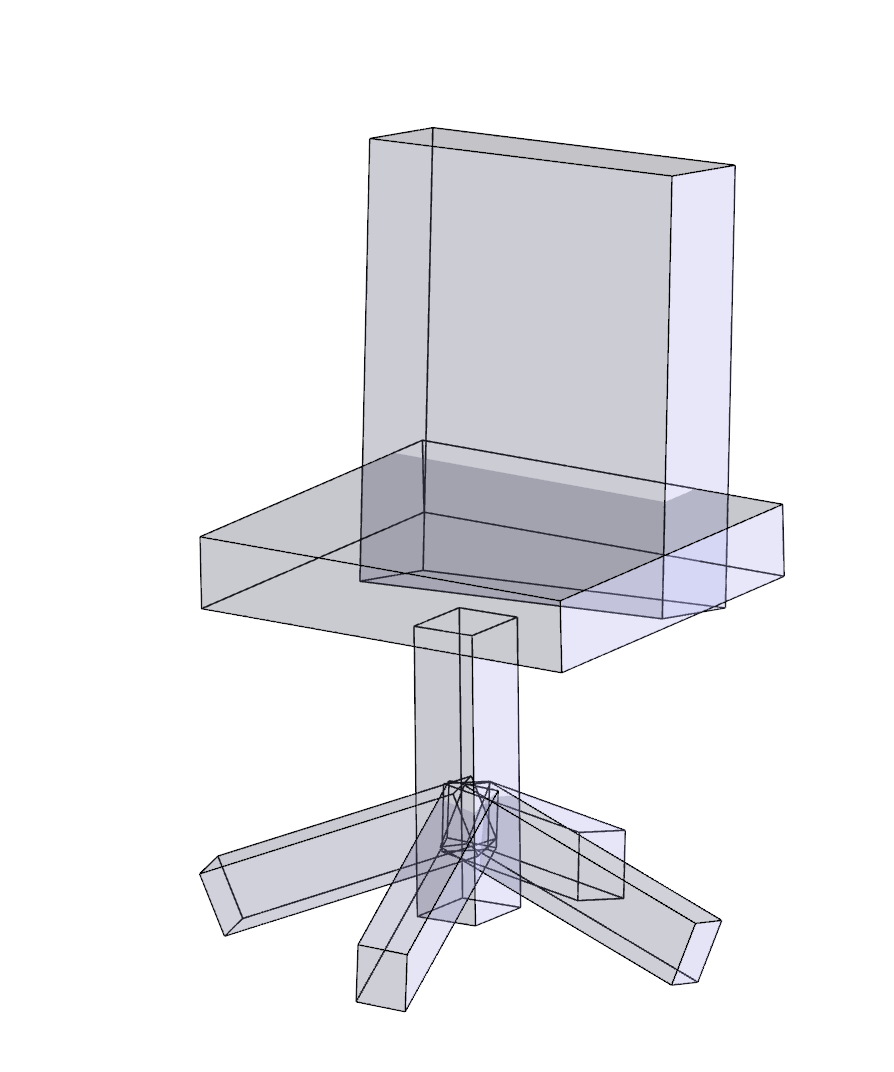}}
\\
		{\includegraphics[width=0.18\linewidth]{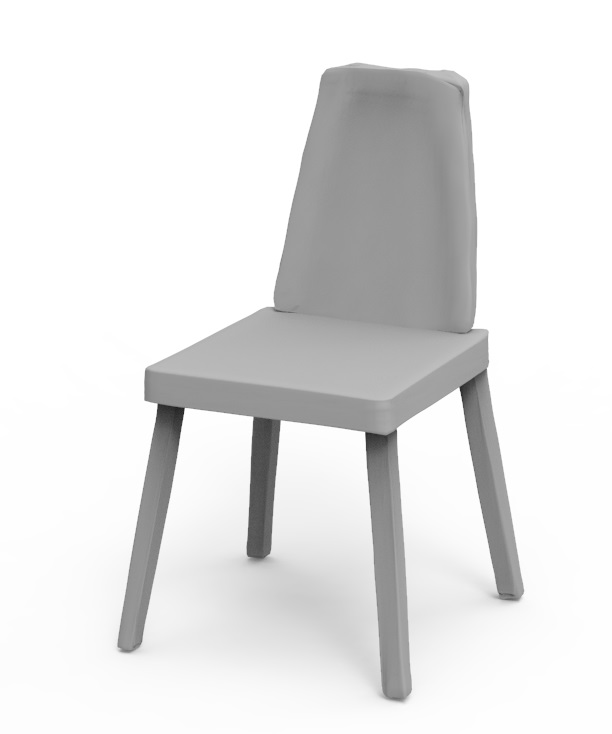}}
		{\includegraphics[width=0.18\linewidth]{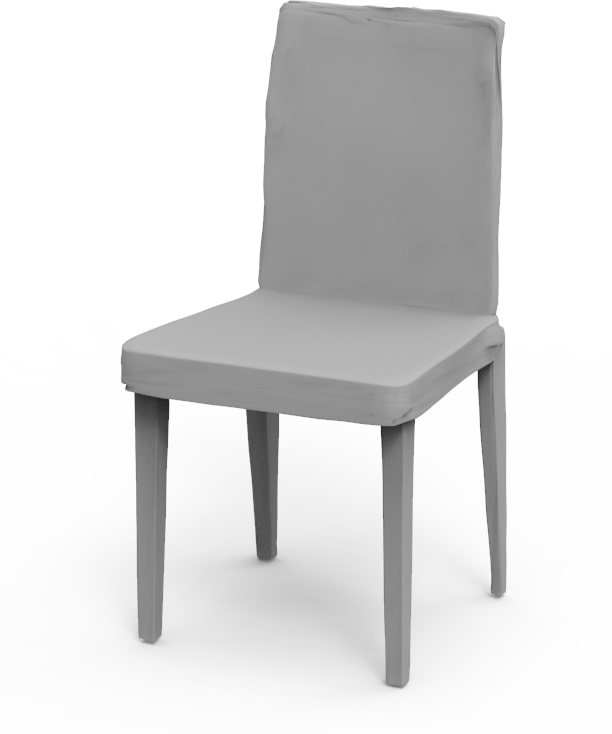}}
		{\includegraphics[width=0.18\linewidth]{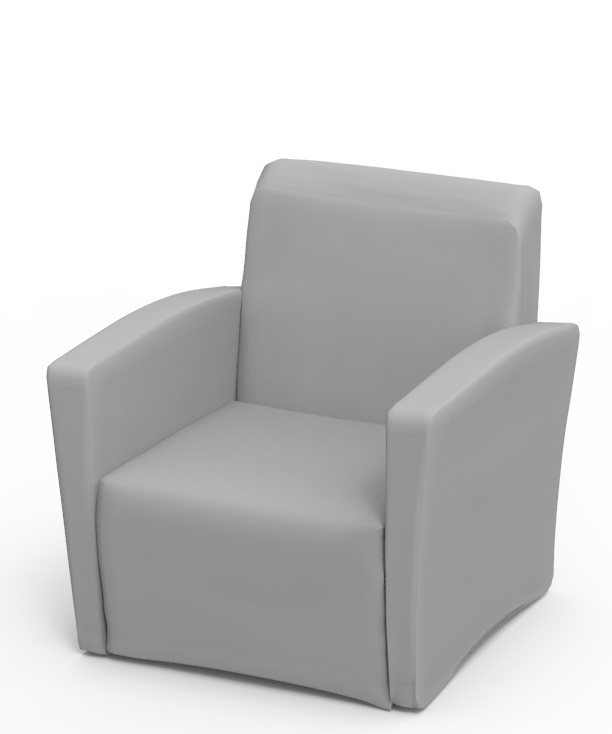}}
		{\includegraphics[width=0.18\linewidth]{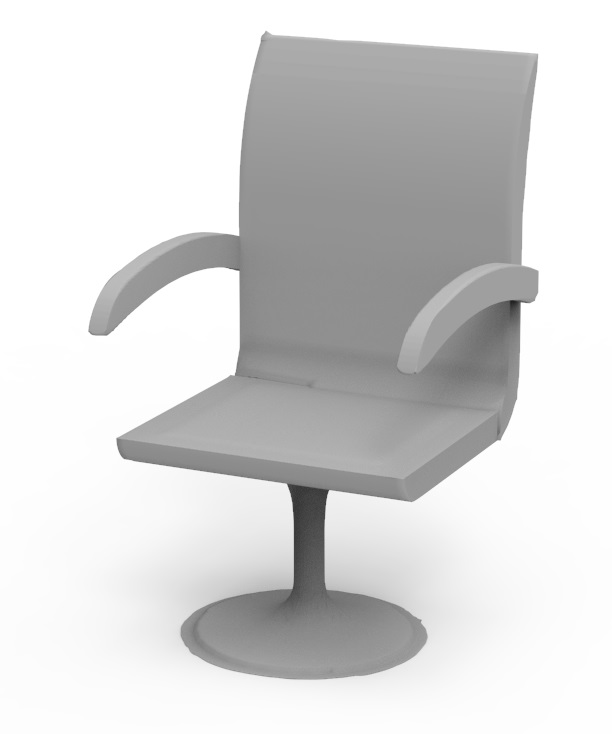}}
		{\includegraphics[width=0.18\linewidth]{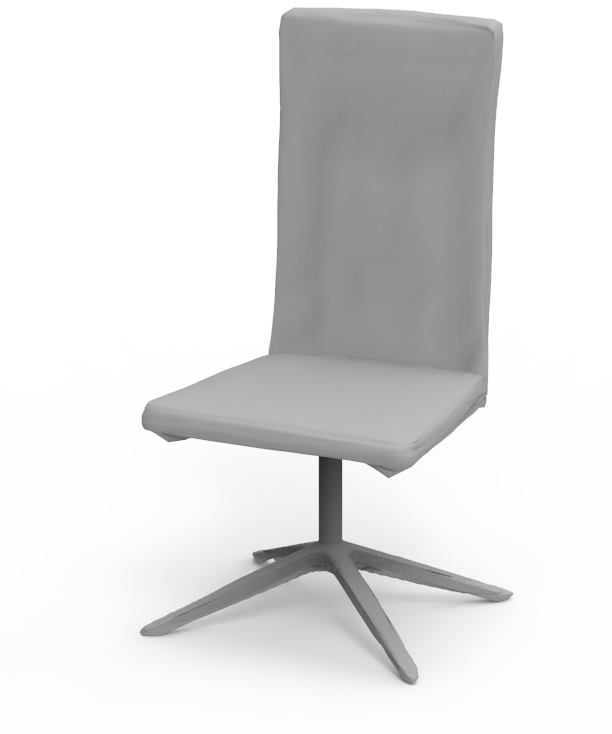}}
	}
	\caption{Comparison between GRASS~\cite{li_sig17} and our method for random generation. We visualize the structures generated by GRASS and shapes generated by our method. }
	\label{fig:grassrandom}
\end{figure}

\begin{figure}
	\centering
	{
		\includegraphics[width=0.24\linewidth]{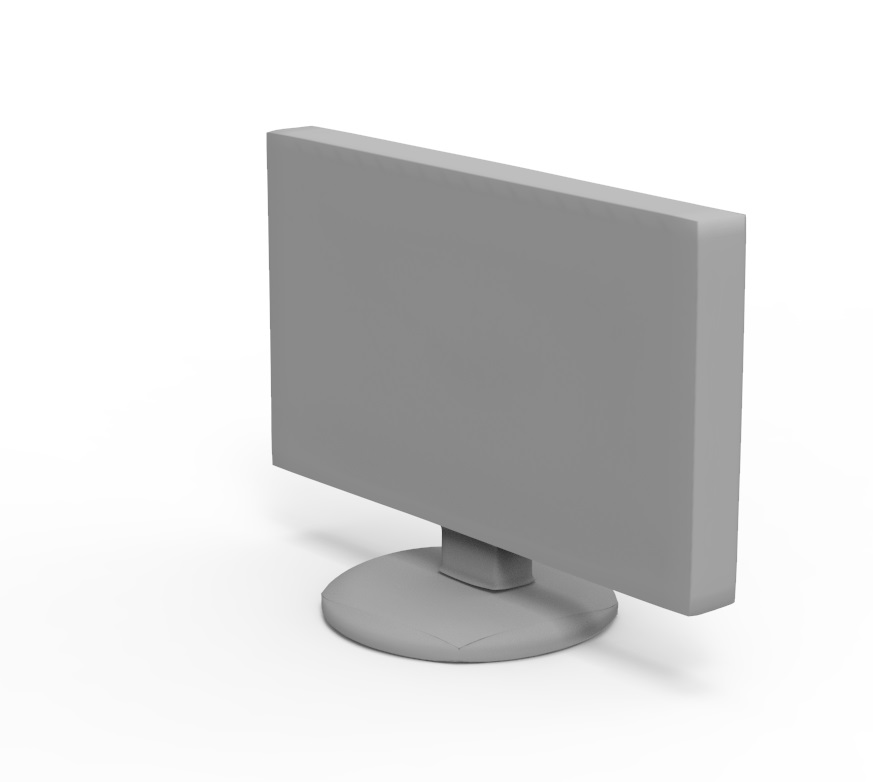}
		\includegraphics[width=0.24\linewidth]{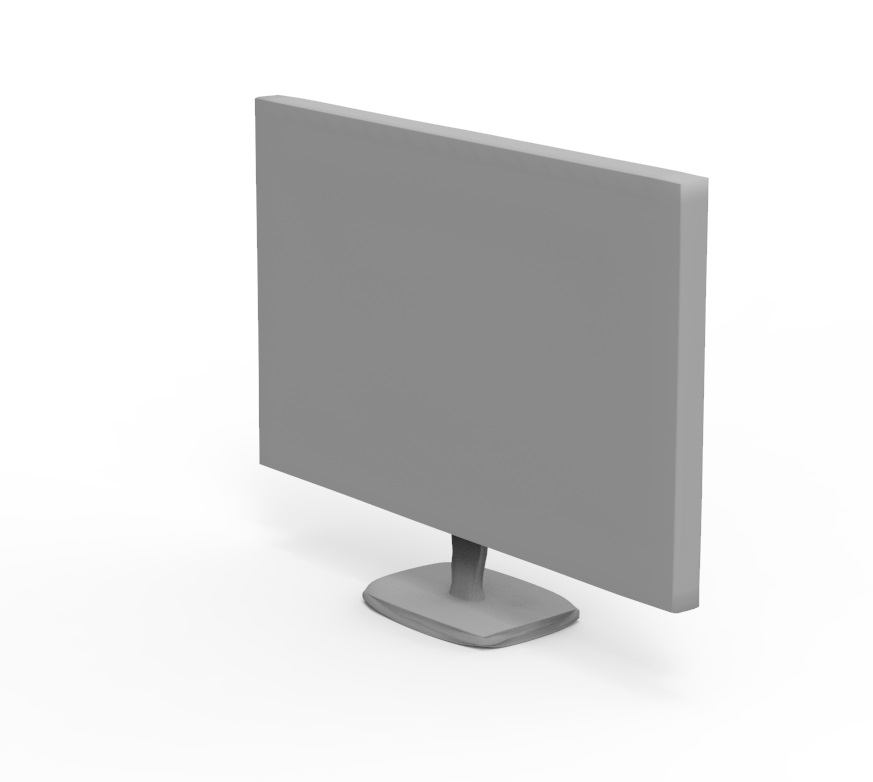}
		\includegraphics[width=0.24\linewidth]{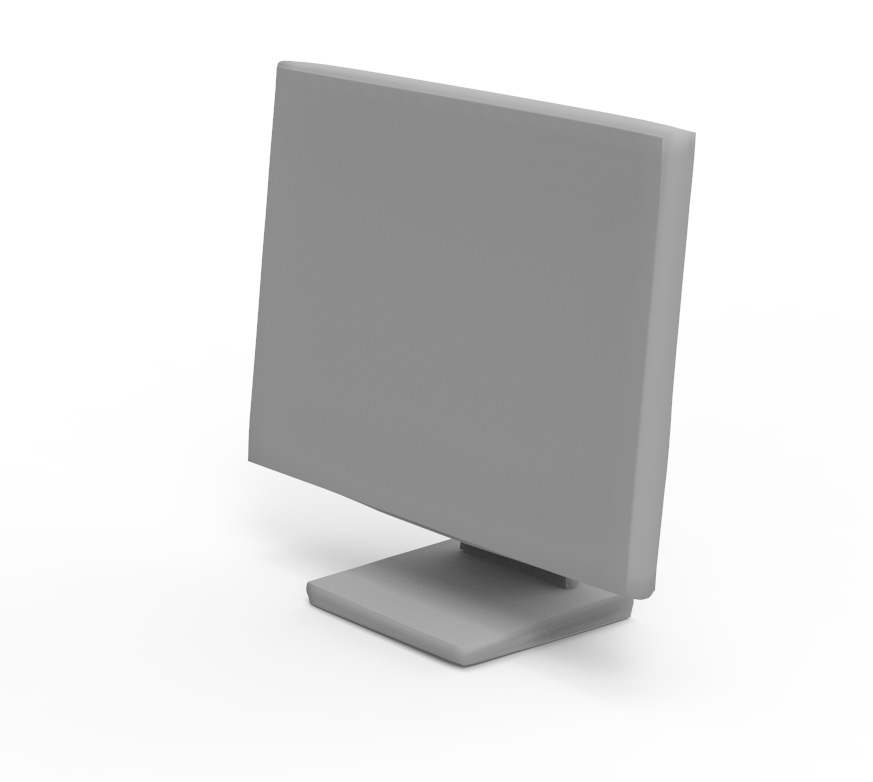}
		\includegraphics[width=0.24\linewidth]{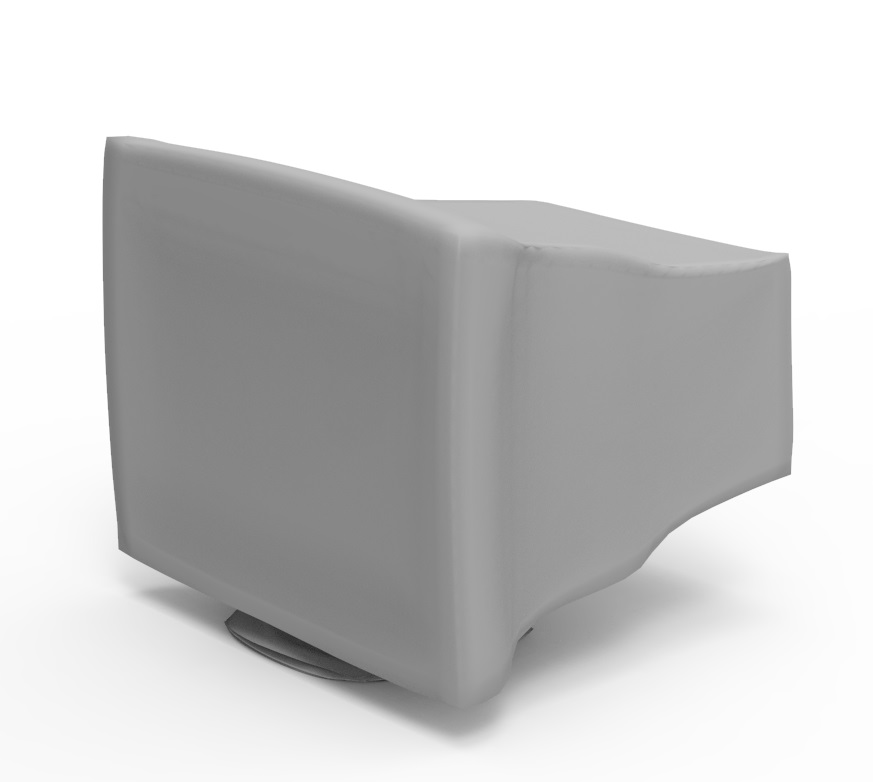}
	}\caption{Random generation of monitor shapes using our method, where the training data is from ModelNet~\cite{Wu_2015_CVPR}.}\label{fig:random_monitor}
\end{figure}

Moreover, we quantitatively compare our method with the existing methods using common metrics for 3D shape \YL{sets}, including Jensen-Shannon Divergence (JSD), Coverage (COV) and Minimum Matching Distance (MMD)~\cite{achlioptas18a}. The latter two metrics are calculated using both the Chamfer Distance (CD) and Earth Mover's Distance (EMD) \YL{for measuring the distance between shapes}. For JSD and MMD, the smaller the better, while for COV, the larger the better.
The \YLN{average} results for different methods on these datasets are shown in Table~\ref{tab:more_metrics}. It can be seen that our method achieves the best performance for nearly all the metrics.

\begin{table*}[htbp]
\fontsize{9}{12.5}\selectfont
  \centering
    \begin{tabular}{ccccccc}
    \toprule
    \multirow{2}[4]{*}{Dataset} & \multirow{2}[4]{*}{Methods} & \multicolumn{5}{c}{Metrics} \\
\cmidrule{3-7}          &       & JSD   & MMD-CD & MMD-EMD & COV-CD & COV-EMD \\
    \midrule\midrule
    \multirow{4}[8]{*}{Airplane} & AOCNN & 0.0665 & 0.0167 & 0.0157 & 84.3  & \textbf{95.5} \\
\cmidrule{2-7}          & AtlasNet & 0.0379 & 0.0147 & 0.0132 & 79.6  & 82.1 \\
\cmidrule{2-7}          & PSG   & 0.0681 & 0.0244 & 0.0172 & 33.5  & 38.9 \\
\cmidrule{2-7}          & Our   & \textbf{0.0192} & \textbf{0.00462} & \textbf{0.00762} & \textbf{87.2}  & 90.6 \\
    \midrule\midrule
    \multirow{4}[8]{*}{Car} & AOCNN & 0.0649 & 0.0264 & 0.0223 & 60.6  & 60.8 \\
\cmidrule{2-7}          & AtlasNet & 0.0393 & 0.0228 & 0.0137 & 75.4   & 81.9 \\
\cmidrule{2-7}          & PSG   & 0.0665 & 0.0365 & 0.0247 & 49.8  & 59.4 \\
\cmidrule{2-7}          & Our   & \textbf{0.0280} & \textbf{0.00247} & \textbf{0.00101} & \textbf{87.2}  & \textbf{88.5} \\
    \midrule\midrule
    \multirow{4}[8]{*}{Chair} & AOCNN & 0.0384 & 0.0159 & 0.0196 & 43.5  & 39.3 \\
\cmidrule{2-7}          & AtlasNet & 0.0369 & 0.0137 & 0.0124 & 51.1  & 52.6 \\
\cmidrule{2-7}          & PSG   & 0.0391 & 0.0131 & 0.0152 & 42.9  & 49.1 \\
\cmidrule{2-7}          & Our   & \textbf{0.0364} &\textbf{0.00375} & \textbf{0.00764} & \textbf{47.3}  & \textbf{55.3} \\
    \midrule\midrule
    \multirow{4}[8]{*}{Table} & AOCNN & 0.0583 & 0.0393 & 0.0256 & 55.2    & 40.1 \\
\cmidrule{2-7}          & AtlasNet & 0.0324 & 0.0154 & 0.0146 & 59.1  & 63.7 \\
\cmidrule{2-7}          & PSG   & 0.0354 & 0.0271 & 0.0276 & 41.2  & 42.5 \\
\cmidrule{2-7}          & Our   & \textbf{0.0123} & \textbf{0.00183} & \textbf{0.00127} & \textbf{63.3}  & \textbf{76.8} \\
    \bottomrule
    \end{tabular}%
    \vspace{3mm}
  	\caption{Quantitative comparison of reconstruction capabilities of different methods on several metrics. For JSD and MMD, the smaller the better, while for COV, the larger the better.}
	\label{tab:more_metrics}%
\end{table*}%

\begin{table*}[htbp]
\fontsize{9}{12.5}\selectfont
  \centering
    \begin{tabular}{ccccccc}
    \toprule
    \multirow{2}[4]{*}{Dataset} & \multirow{2}[4]{*}{Methods} & \multicolumn{5}{c}{Metrics} \\
\cmidrule{3-7}          &       & JSD   & MMD-CD & MMD-EMD & COV-CD & COV-EMD \\
    \midrule\midrule
    \multirow{4}[8]{*}{Chair} & G2L & 0.0357 & 0.0034  & 0.0682 & 83.7 & 83.4 \\
\cmidrule{2-7}          & GRASS & 0.0374 & 0.0030  & 0.0744 &   46.0   &  44.5 \\
\cmidrule{2-7}          & SAGNet   & 0.0342 & \textbf{0.0024}  & 0.0608  &  75.1  &   74.3 \\
\cmidrule{2-7}          & Our   & \textbf{0.0289} &0.00274 &  \textbf{0.00671}&   \textbf{89.3}& \textbf{84.1}\\
    \bottomrule
    \end{tabular}%
    \vspace{3mm}
  	\caption{\rv{Quantitative comparison of reconstruction capabilities of different methods (G2L~\cite{Wang2018TOG}, GRASS~\cite{li_sig17}, SAGNet~\cite{pageSAGnet19}) on several metrics. For JSD and MMD,  the smaller the better, while for COV, the larger the better.}}
	\label{tab:more_eval_chair}%
\end{table*}%

\rv{\hongbo{In Table~\ref{tab:more_eval_chair}} we compare our method with three recent shape generation methods, \hongbo{namely}, GRASS \cite{li_sig17}, G2L \cite{Wang2018TOG} and SAGNet \cite{pageSAGnet19}. For fair comparison, both of our reconstructed shapes and input shapes are voxelized. Particularly, we make comparisons with GRASS on their chair data since GRASS requires symmetry hierarchies as input for training. The results show that our method outperforms the compared methods in most cases with several metrics.
\rv{We also show a visual comparison result between GRASS and our method in Figure~\ref{fig:compwithgrass_voxel}}.} %
\rv{The GRASS result exhibits some artifacts due to the voxel representation. \hongbo{Even after surface extraction GRASS still} fails to capture fine geometric details compared with our method.}

\begin{figure}[!t]
	\centering
	{
		\includegraphics[width=0.15\linewidth]{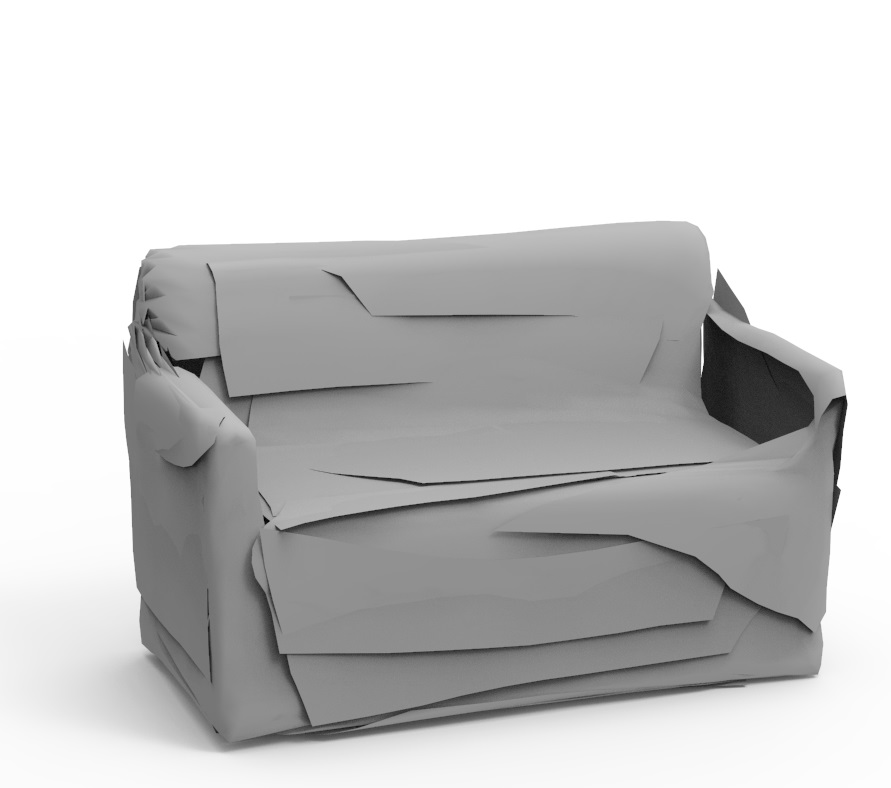}
		\includegraphics[width=0.15\linewidth]{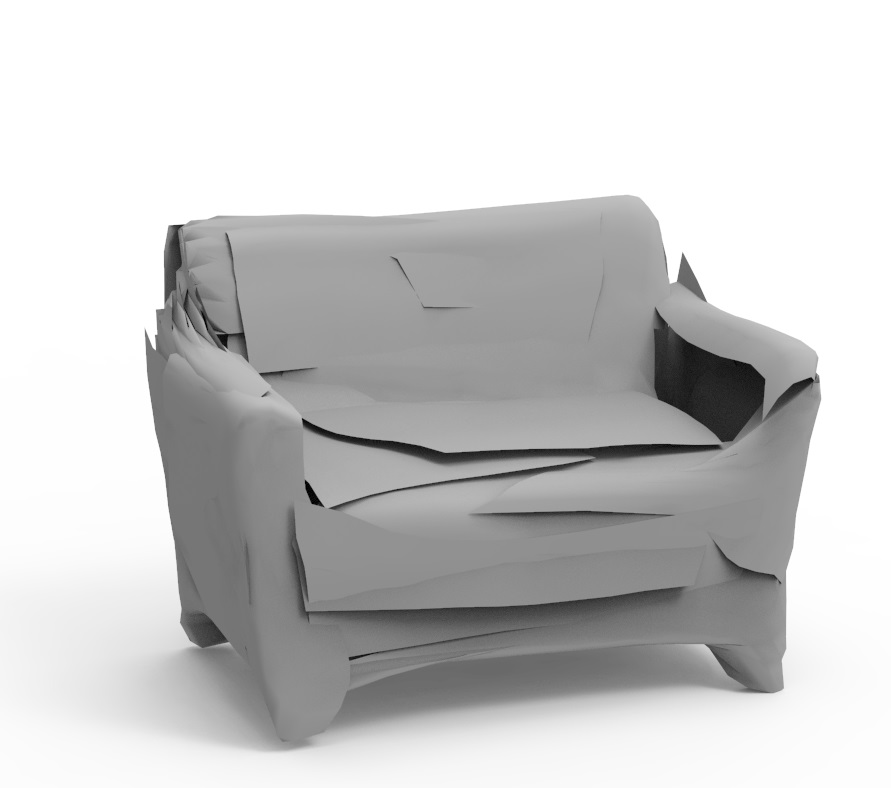}
		\includegraphics[width=0.15\linewidth]{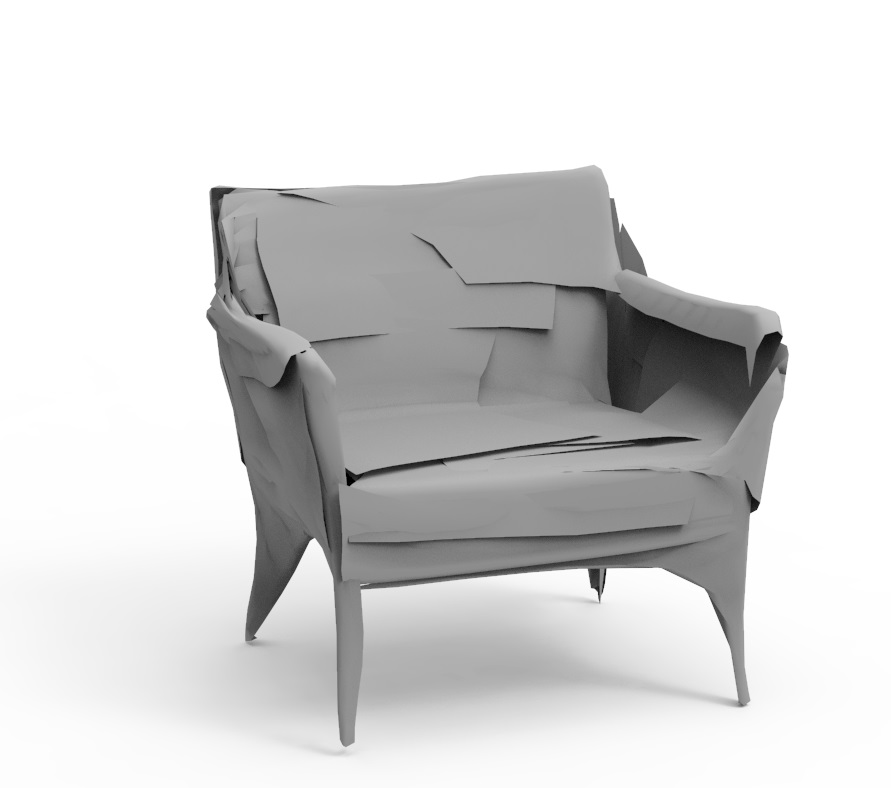}
		\includegraphics[width=0.15\linewidth]{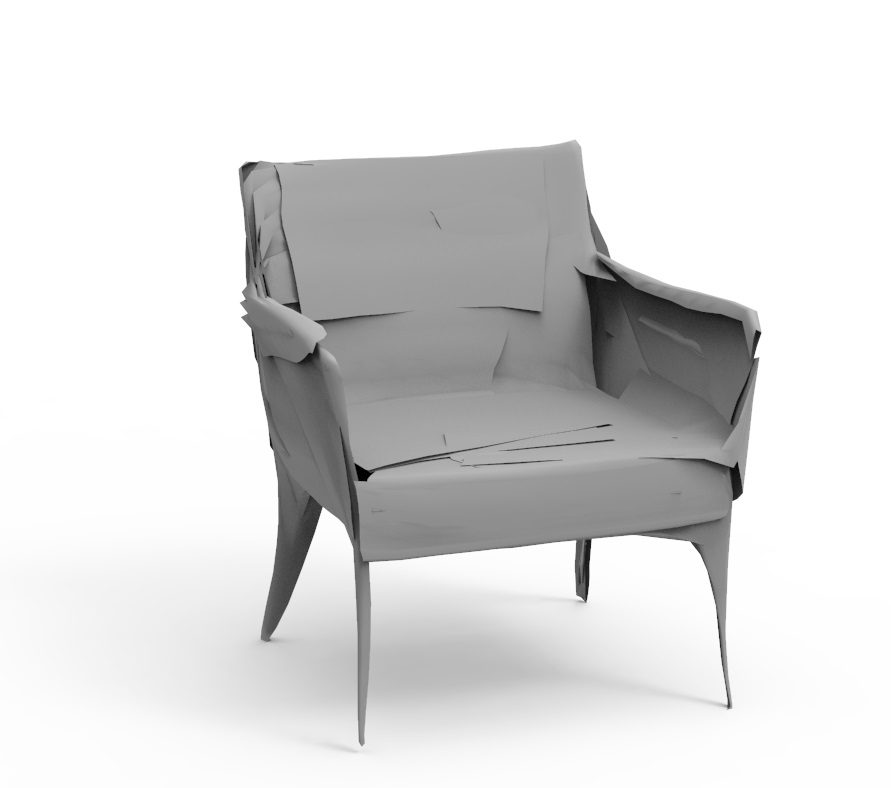}
		\includegraphics[width=0.15\linewidth]{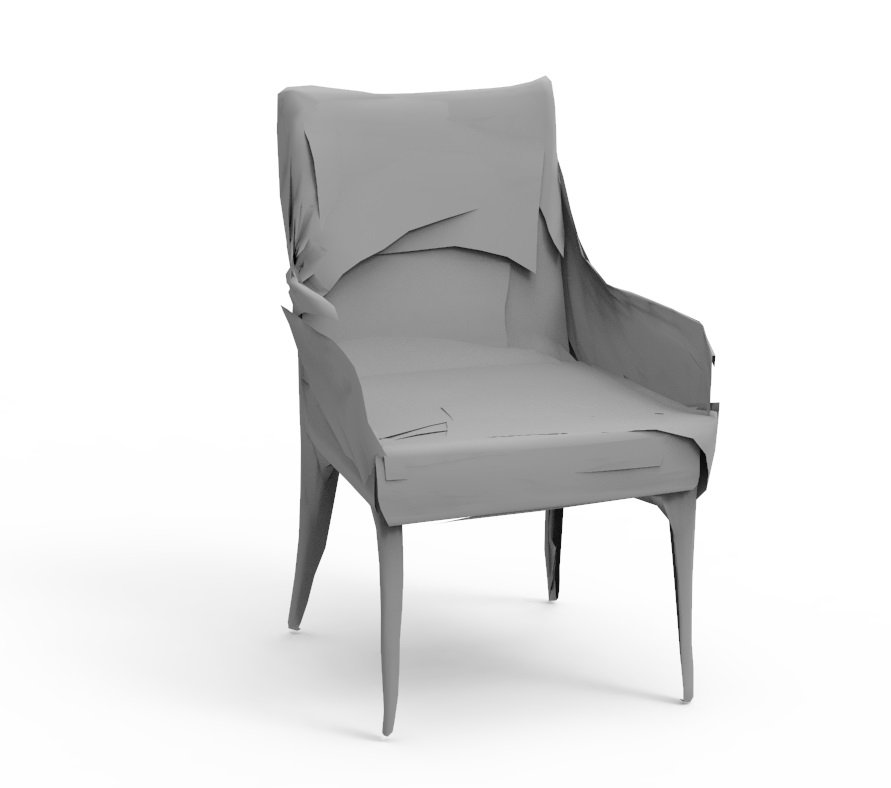}
		\includegraphics[width=0.15\linewidth]{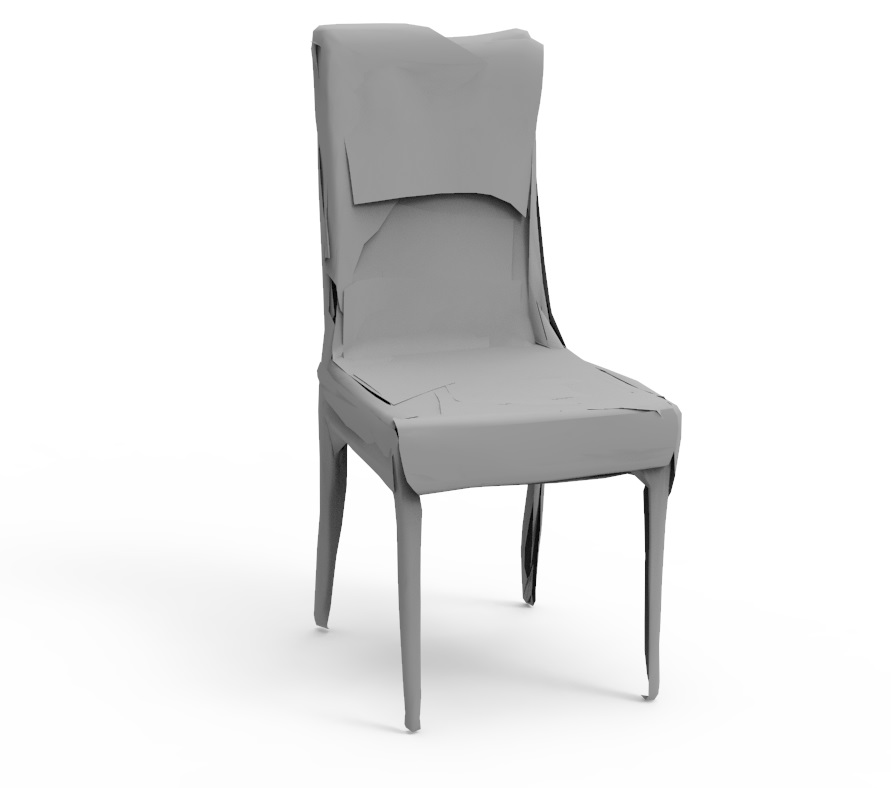}\\
		\includegraphics[width=0.15\linewidth]{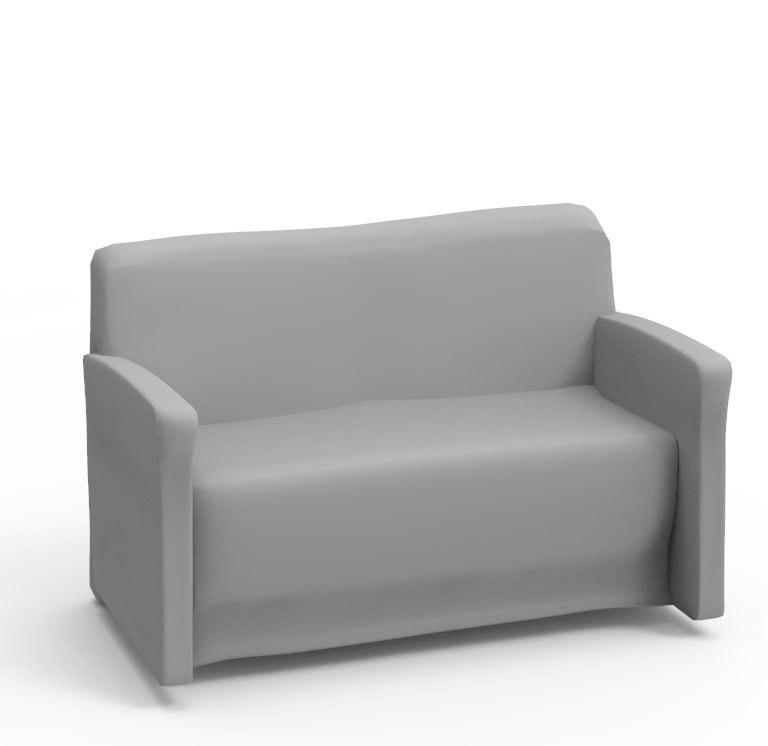}
		\includegraphics[width=0.15\linewidth]{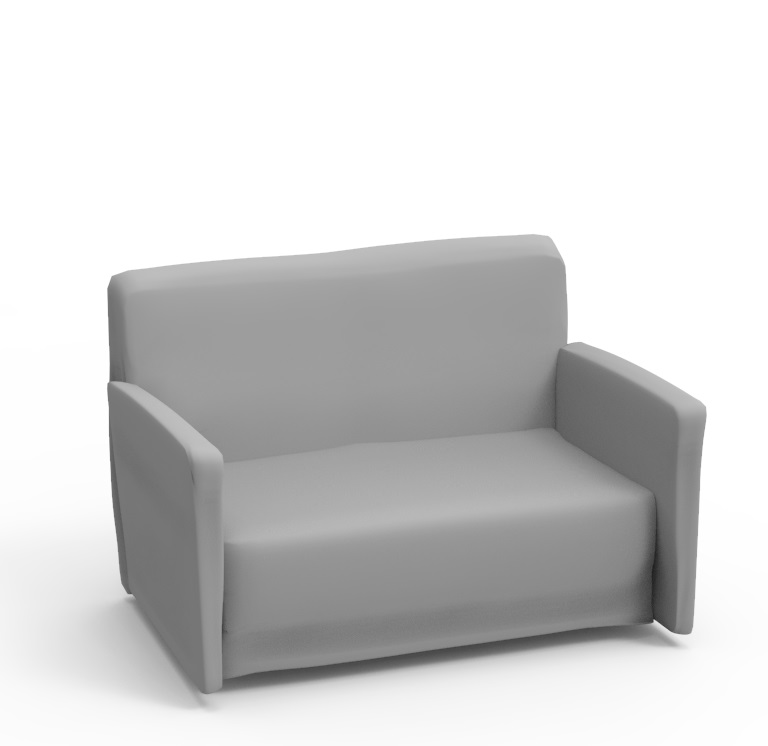}
		\includegraphics[width=0.15\linewidth]{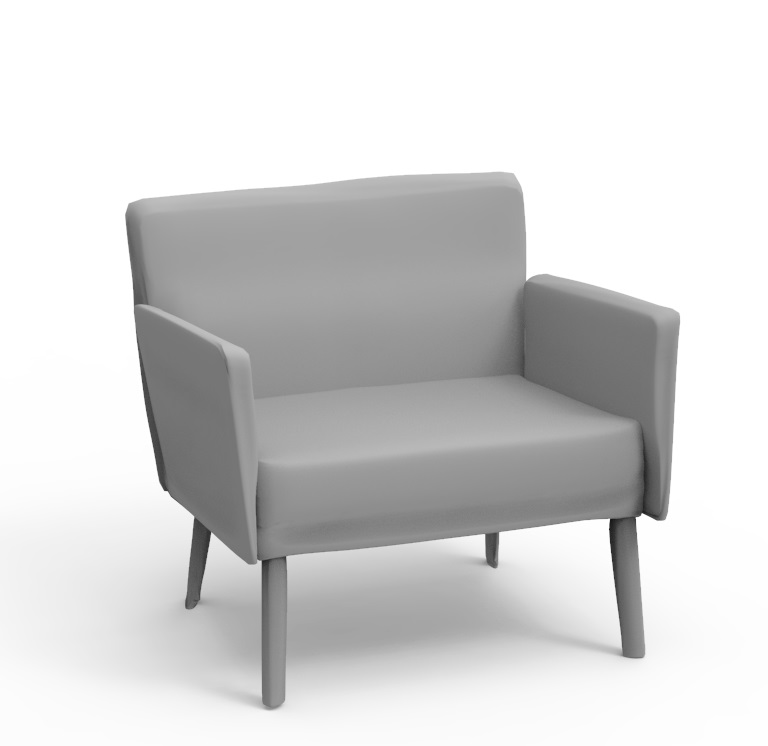}
		\includegraphics[width=0.15\linewidth]{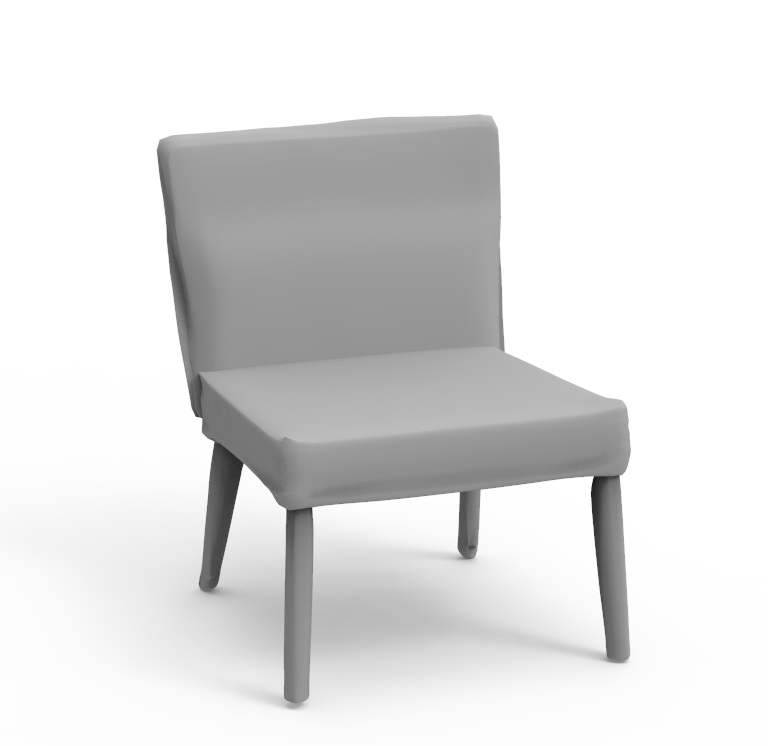}
		\includegraphics[width=0.15\linewidth]{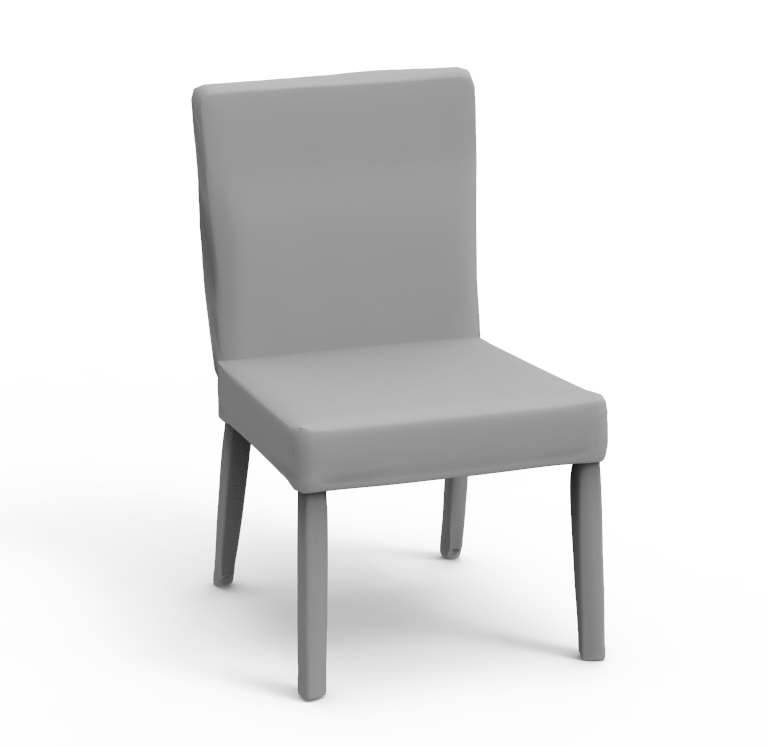}
		\includegraphics[width=0.15\linewidth]{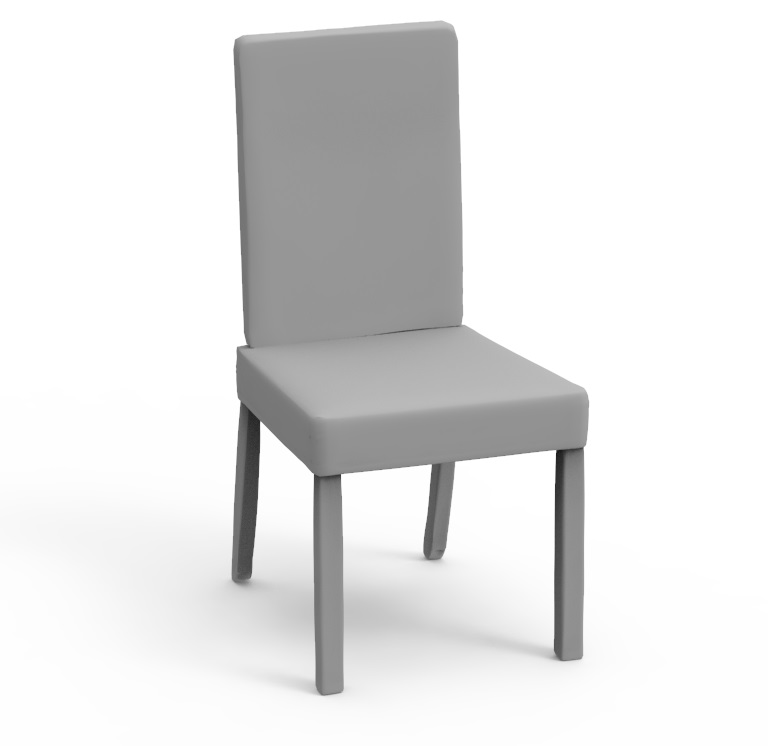}		
	}\caption{Visual comparison of shape interpolation by AtlasNet~\protect\cite{AtlasNet2018} (top) and our technique (bottom). The first and last columns are the two shapes to be interpolated \YL{(which are decoded by respective methods)}. 
	}
	\label{fig:vaecomp}
\end{figure}

\begin{figure}
	\centering
	{
		\includegraphics[width=0.15\linewidth]{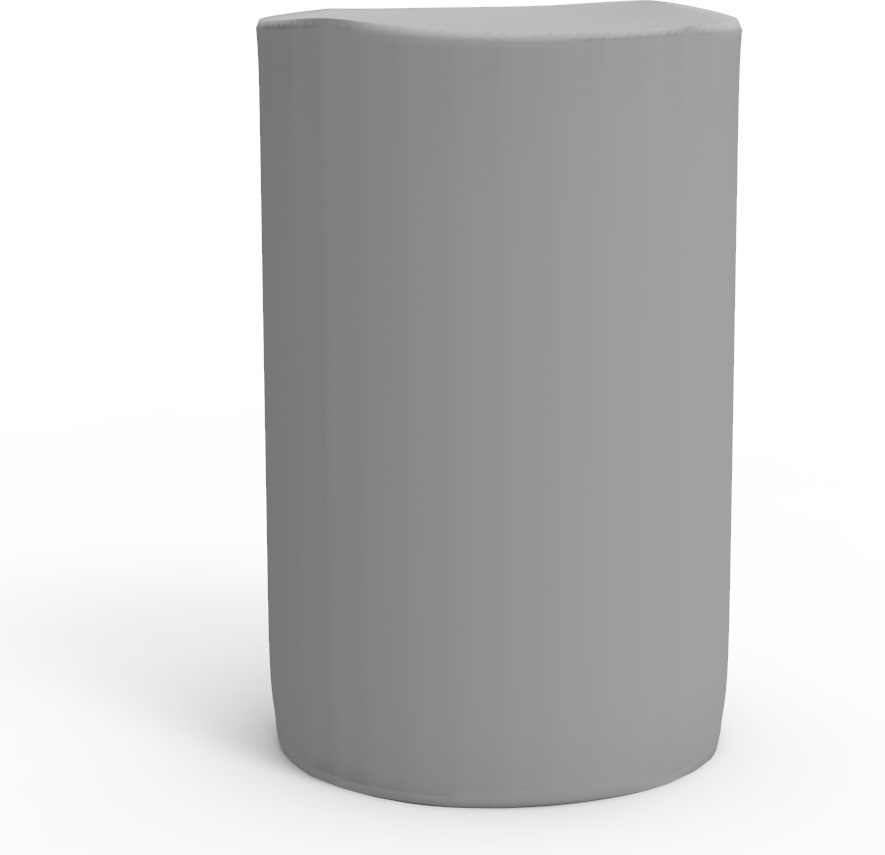}
		\includegraphics[width=0.15\linewidth]{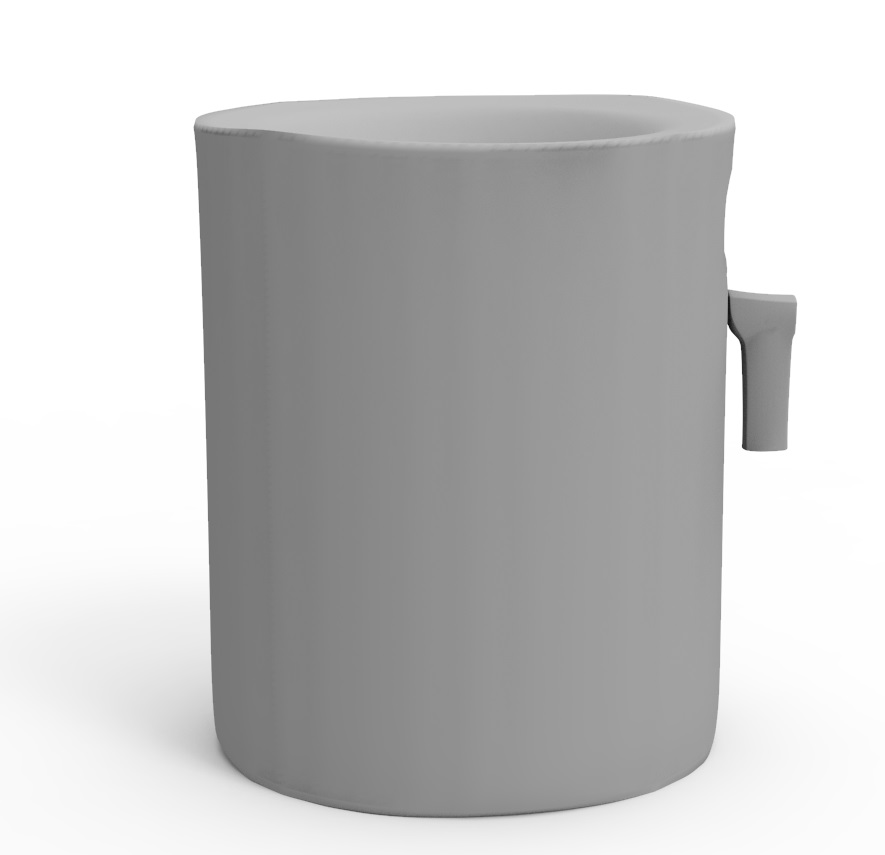}
		\includegraphics[width=0.15\linewidth]{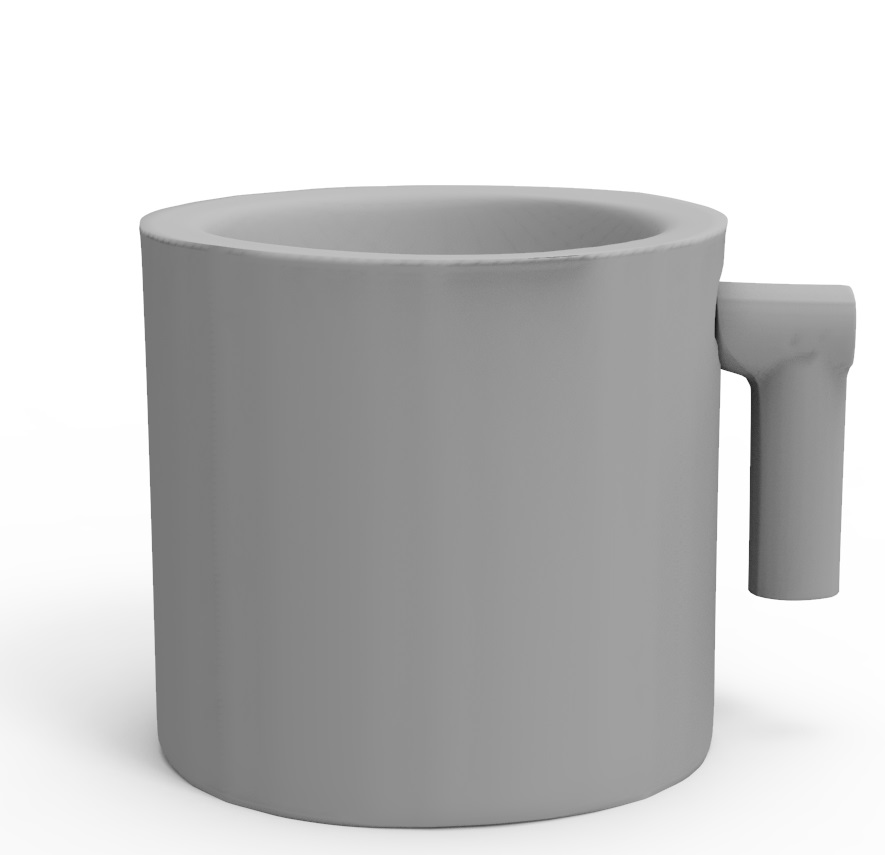}
		\includegraphics[width=0.15\linewidth]{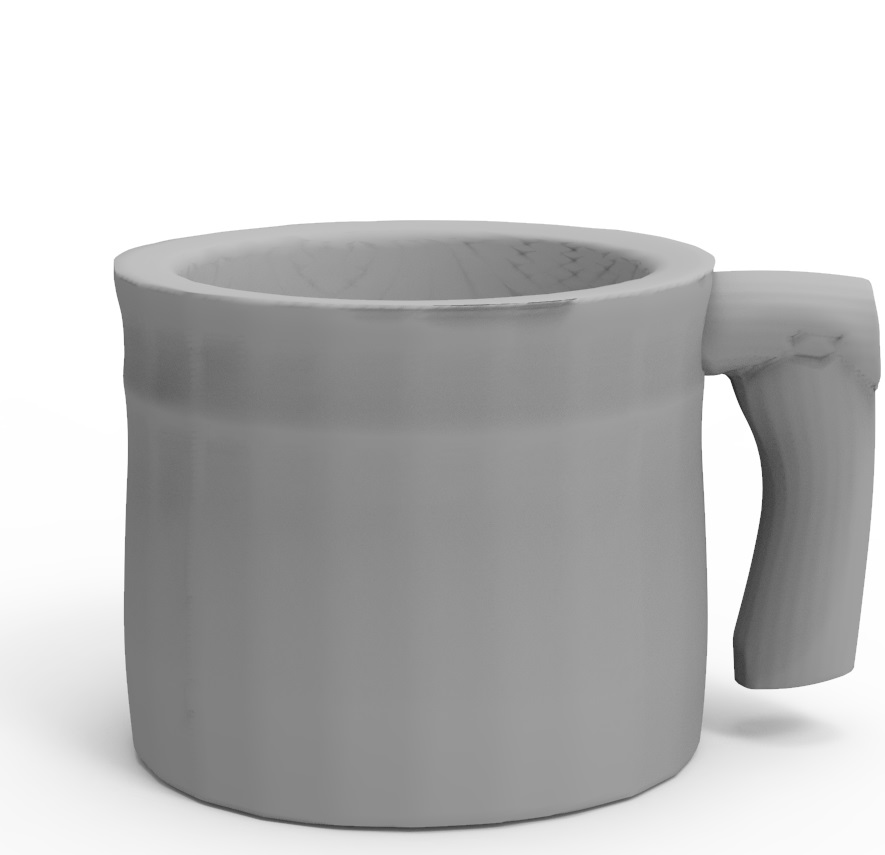}
		\includegraphics[width=0.15\linewidth]{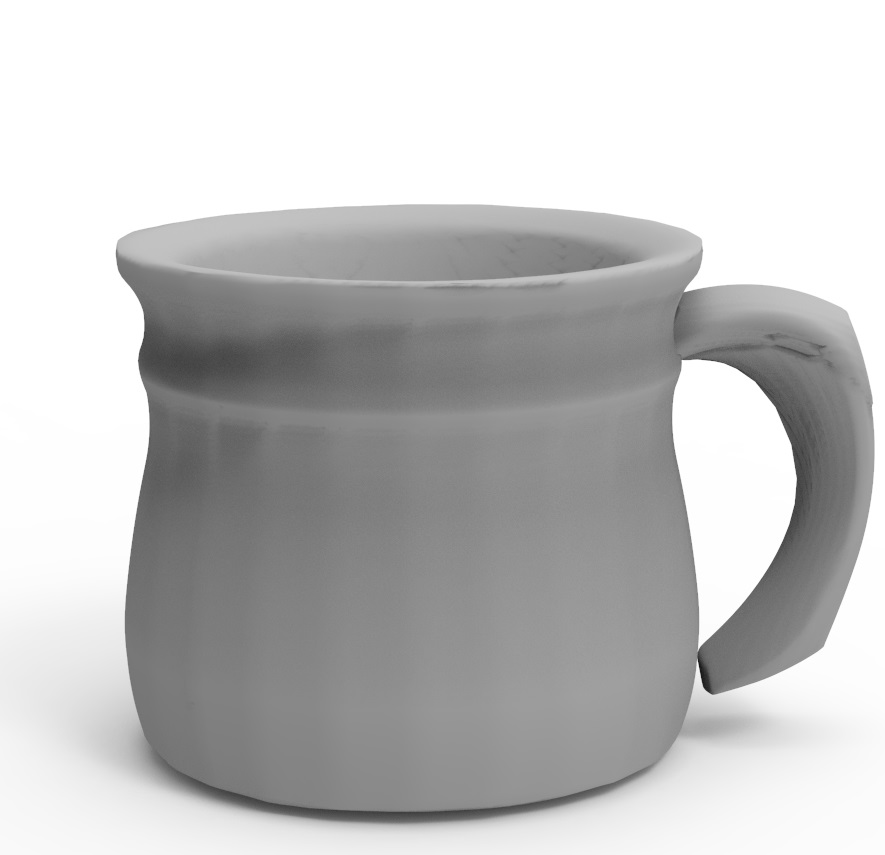}
		\includegraphics[width=0.15\linewidth]{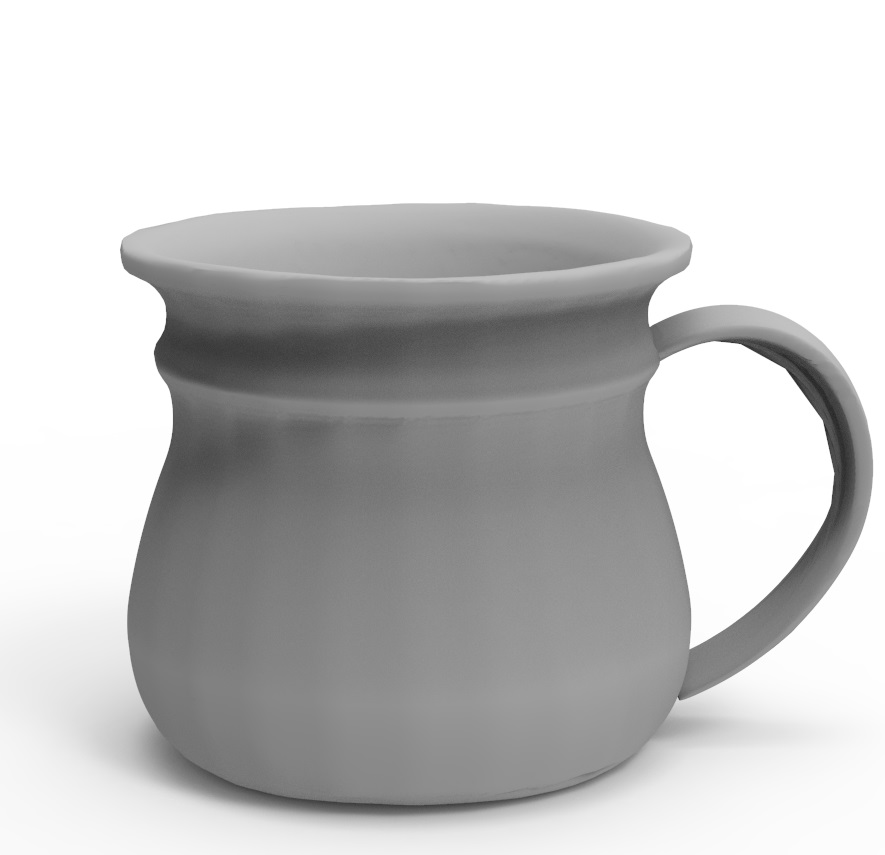}\\
		\includegraphics[width=0.15\linewidth]{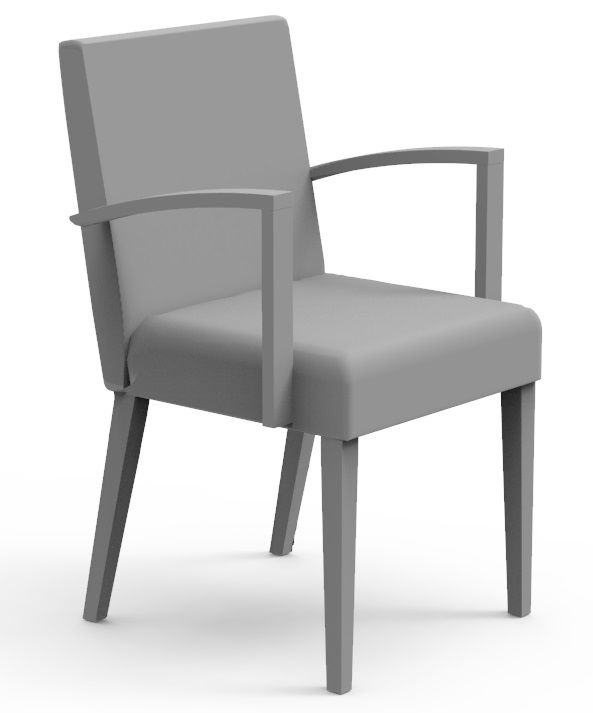}
		\includegraphics[width=0.15\linewidth]{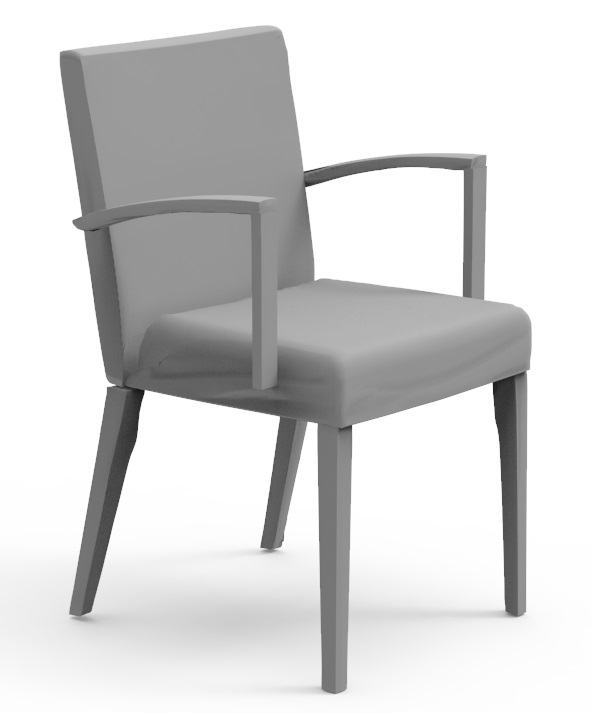}
		\includegraphics[width=0.15\linewidth]{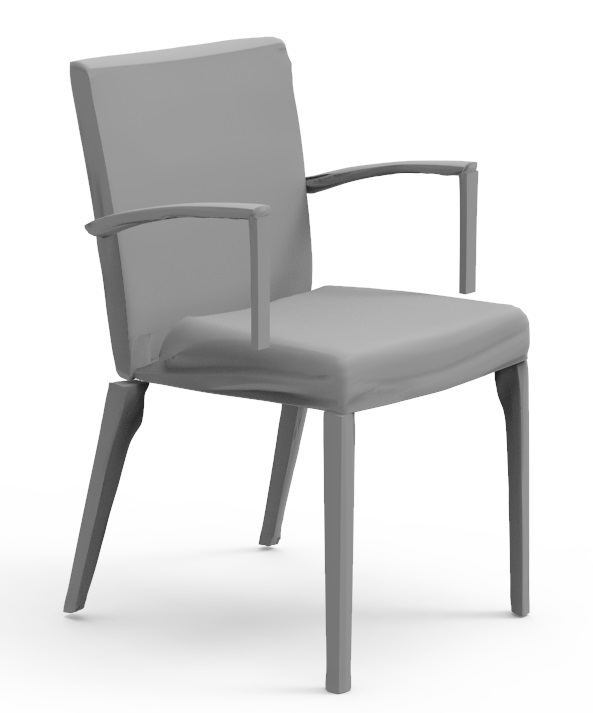}
		\includegraphics[width=0.15\linewidth]{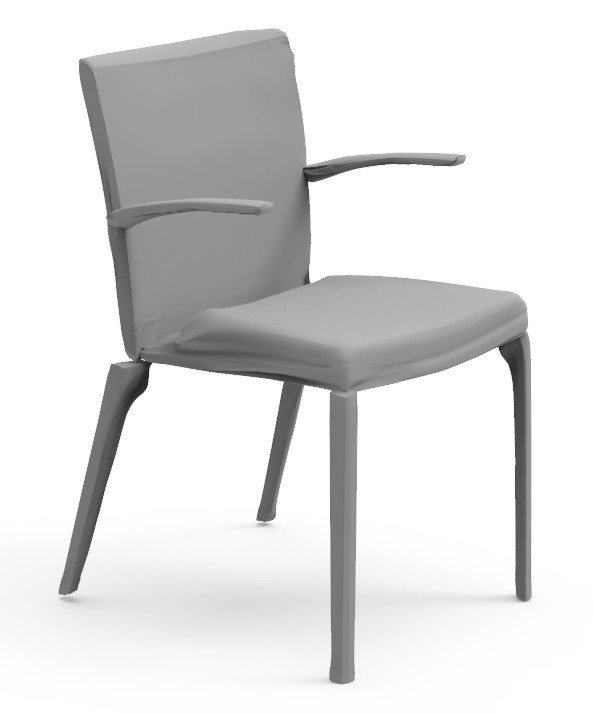}
		\includegraphics[width=0.15\linewidth]{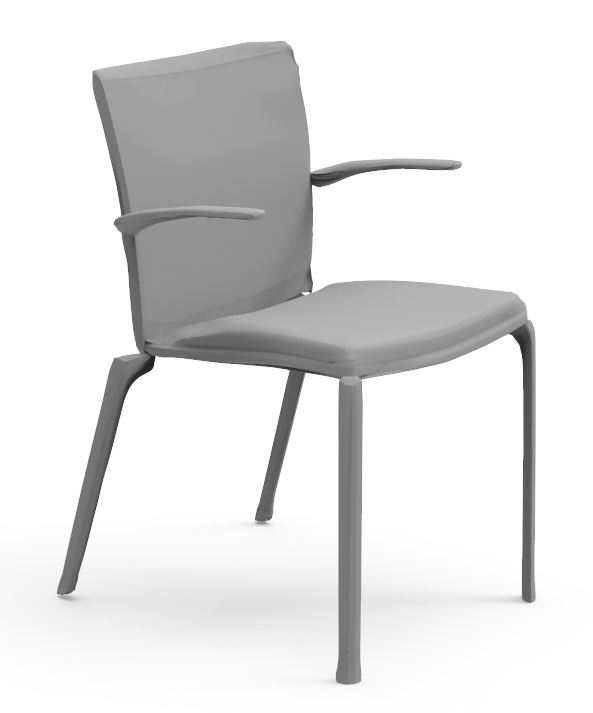}
		\includegraphics[width=0.15\linewidth]{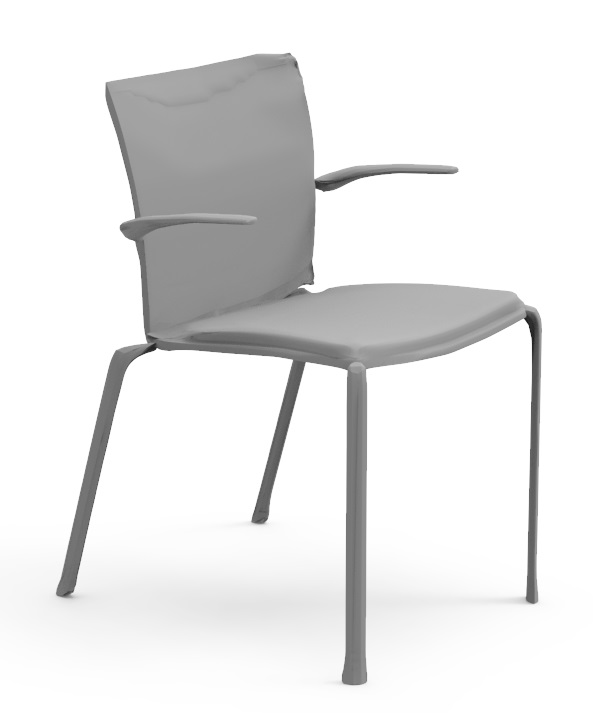}
	}\caption{{Shape interpolation of cups and chairs with different topologies using our method. The first and last columns are the input shapes for interpolation.}}\label{fig:inter-top}
\end{figure}

\textbf{Shape Generation}.
In Figure~\ref{fig:g2l}, we make a qualitative comparison between our technique and the global-to-local method \cite{Wang2018TOG} by randomly generating shapes of airplanes. Their method  
uses an unconditional GAN architecture and thus cannot reconstruct a specific shape. So two randomly generated, visually similar planes are selected for comparisons. Their voxel based method fails to represent smooth, fine details of 3D shapes. \YL{We make further comparison with the global-to-local method~\cite{Wang2018TOG} as well as 3DGAN~\cite{3dgan2016} in Figure~\ref{fig:3dgan}. Again, we select visually similar shapes for comparison, and our method produces high quality shapes with plausible structure and fine details, whereas alternative methods have clear artifacts including fragmented output and rough surfaces.}
We also compare our technique with GRASS~\cite{li_sig17} for random shape generation. As shown in Figure~\ref{fig:grassrandom}, the structures synthesized by GRASS might be problematic, containing parts which are disjoint and/or not well supported. %
\rv{In addition, since it is trained on symmetry hierarchies constructed on top of the shape segmentations,  
once trained, for new inputs GRASS utilizes 
automatically generated symmetry hierarchies, which, however, can be inconsistent. This is one of the main causes for GRASS to \YLN{produce results with} a greater level of structural noise including disconnections and asymmetries.}
In contrast, our results are physically stable and well connected. \YL{Note that our refinement step is an integral part of our pipeline and requires structure relations, so cannot be directly applied to GRASS.}

As a generative model, our technique is able to generate new shapes. Because our architecture consists of two VAEs, i.e., PartVAE and SP-VAE, we can acquire different information from their latent spaces. Specifically, we extract various types of parts and structural information from the latent space of SP-VAE, and combine them with the deformation information from PartVAE, to produce novel shapes. 
Figure~\ref{fig:random_monitor} gives an example, where our method is used to generate computer monitors with various shapes by sampling in the learned latent space. In this example, the training data is obtained from ModelNet~\cite{Wu_2015_CVPR}.

\textbf{Shape Interpolation}. Shape interpolation is a useful technique to generate gradually changing shape sequences between a source shape and a target shape. 
With the help of SP-VAE, we first encode the source and target shapes into latent vectors and then perform linear interpolation in the latent space of VAE. A sequence of shapes between the input shape pairs are finally decoded from the linearly interpolated latent vectors.
In Figure~\ref{fig:vaecomp}, we compare our technique with AtlasNet~\cite{AtlasNet2018} for their performance on shape interpolation.
{It can be easily seen that the results by AtlasNet suffer from patch artifacts and the surfaces of the interpolated shapes are often not very smooth.
The  interpolation in our latent space leads to much more realistic results. For example, the armrests gradually become thinner and then disappear in a more natural manner. This is because we combine the geometry and structure during the training of SDM-NET, which thus learns the implicit joint distribution of the geometry and structure. }

\begin{figure*}
	\centering
	{
		\includegraphics[width=0.16\linewidth]{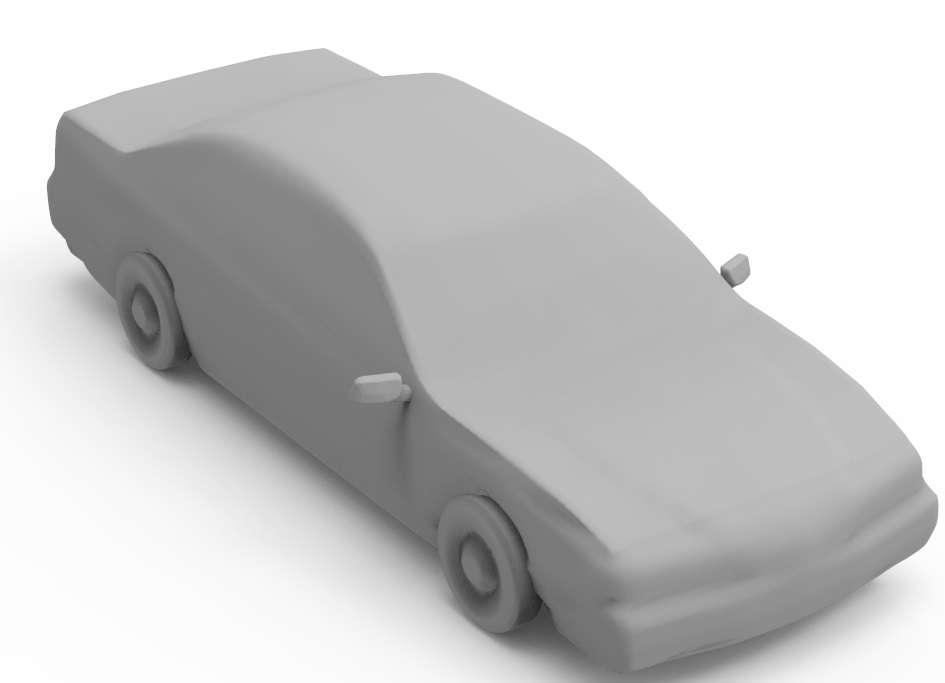}
		\includegraphics[width=0.16\linewidth]{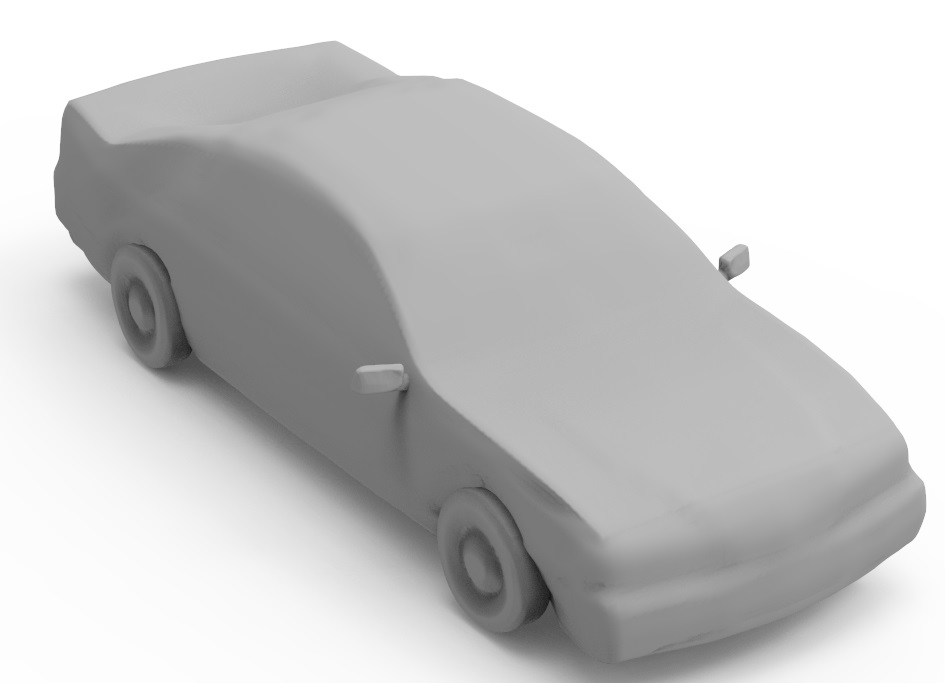}
		\includegraphics[width=0.16\linewidth]{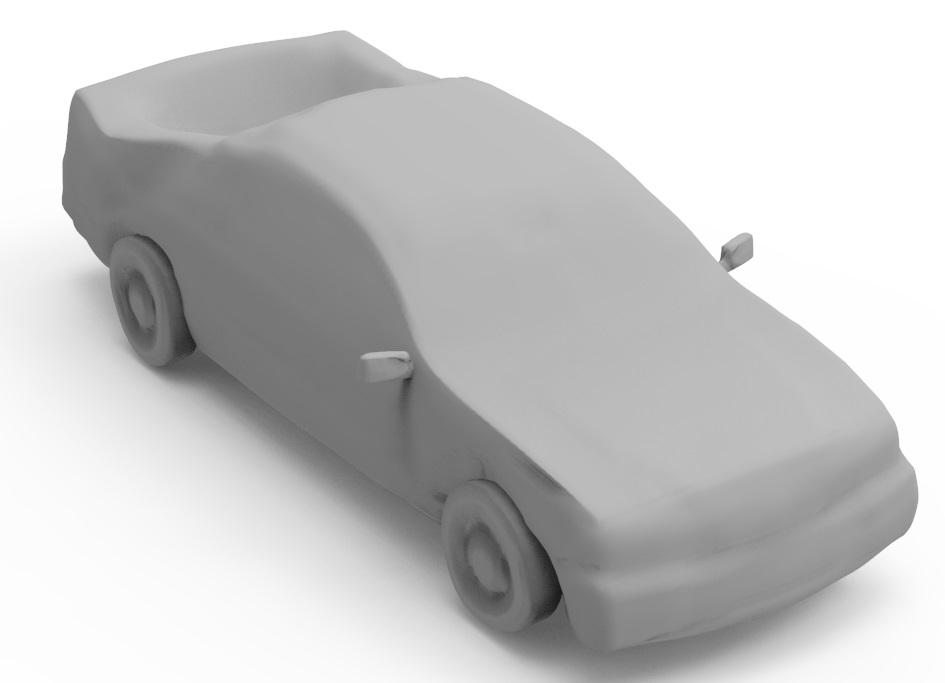}
		\includegraphics[width=0.16\linewidth]{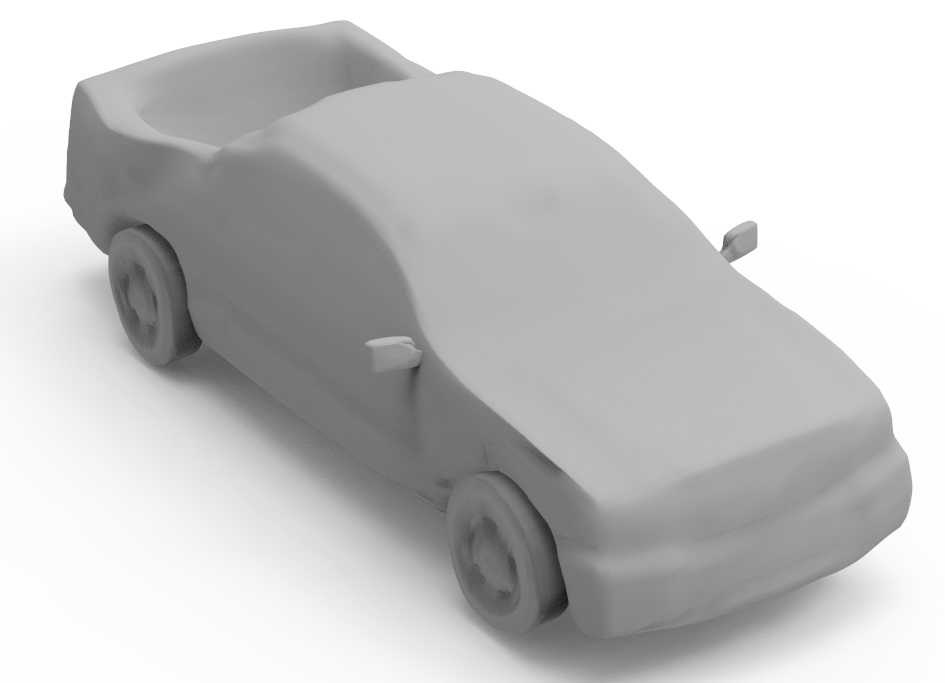}
		\includegraphics[width=0.16\linewidth]{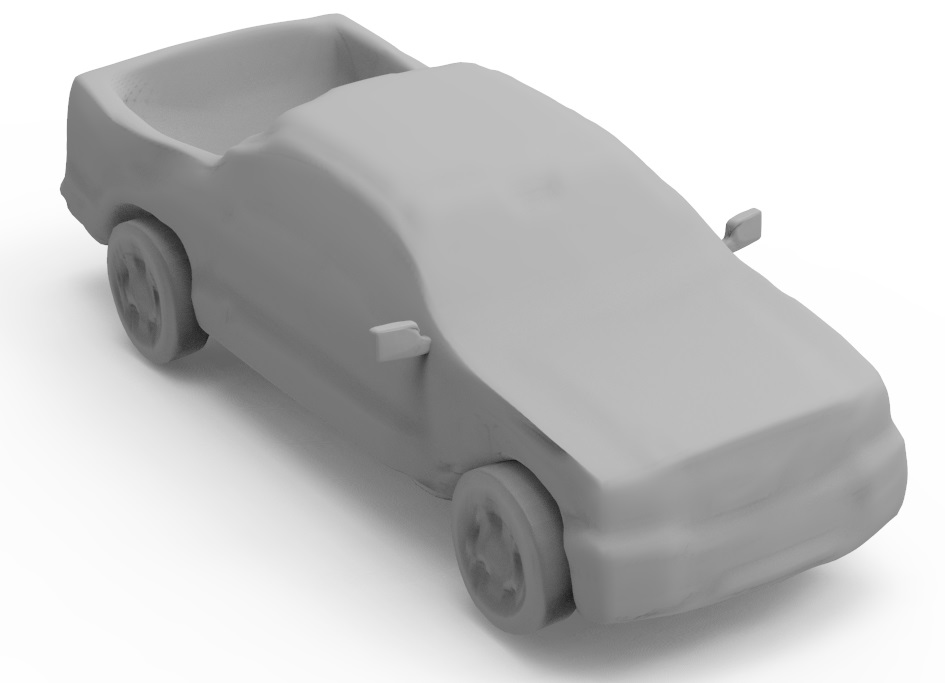}
		\includegraphics[width=0.16\linewidth]{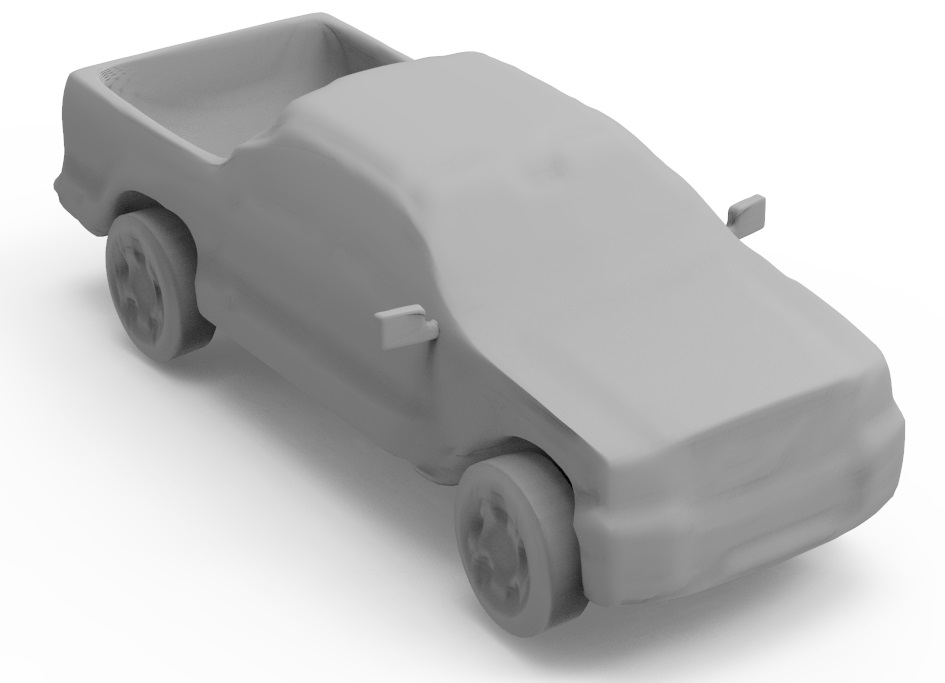}
	}\vspace{2mm}\caption{{Interpolating cars with different geometries using our method. The first and last columns are the shapes to be interpolated. The other columns are the in-between models by linear interpolation in the latent space.}}\label{fig:inter-car}
\end{figure*}

The effectiveness of shape interpolation with SDM-NET is consistently observed with additional experiments on different datasets. 
Figure~\ref{fig:inter-top} shows two examples of natural interpolation between shapes with different topologies, thanks to our flexible structure representation. 
\YL{Figure~\ref{fig:inter-car} shows an additional interpolation result with substantial change of  geometry.}

\begin{table}
  \centering  
  \begin{tabular}{c||cccccccc}
  \hline%
    Dataset     & Car             &  Chair           & Guitar         & Airplane           & Table     \\
    \hline\hline%
    Separate    & \textbf{2.77} & \textbf{3.89} & \textbf{3.58}& \textbf{4.87}  & \textbf{1.85}\\
    \hline%
    End-to-End  & 5.07         & 6.73          & 7.44          & 11.38         & 4.86    \\
  \hline%
    \end{tabular}%
  \caption{{Comparison of reconstruction errors ($\times 10^{-3}$) under the metric of {bidirectional Chamfer distance} with two different training strategies, i.e., separate training vs. end-to-end training.}}
  \label{tab:trainingstrategy}%
\end{table}%

\begin{figure}
	{
		\subfigure[Generated shape]{\includegraphics[width=0.3\linewidth]{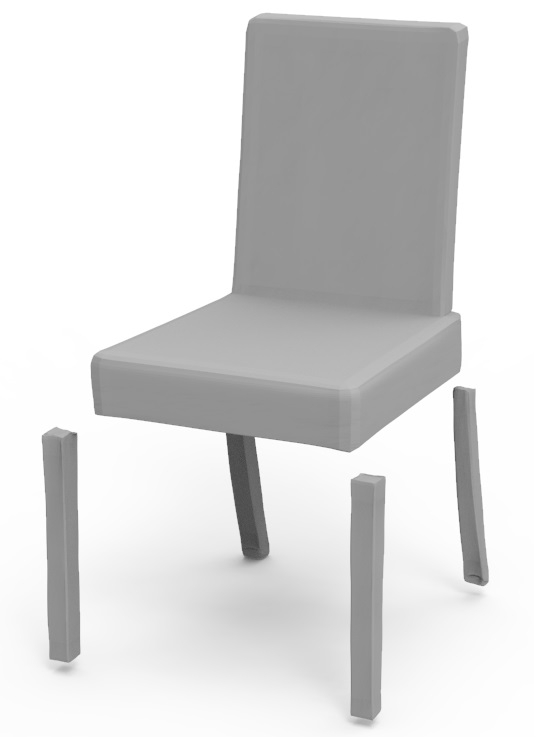}}
		\subfigure[Adjacency constraint]{\includegraphics[width=0.3\linewidth]{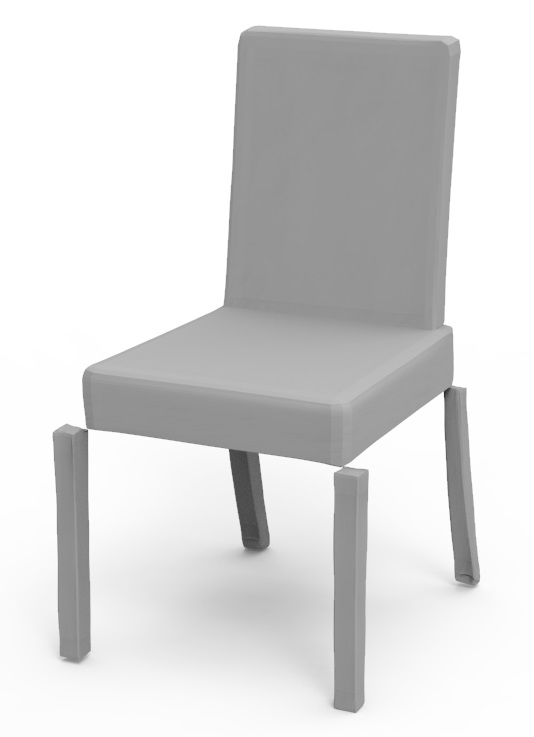}}
		\subfigure[Support constraint]{\includegraphics[width=0.3\linewidth]{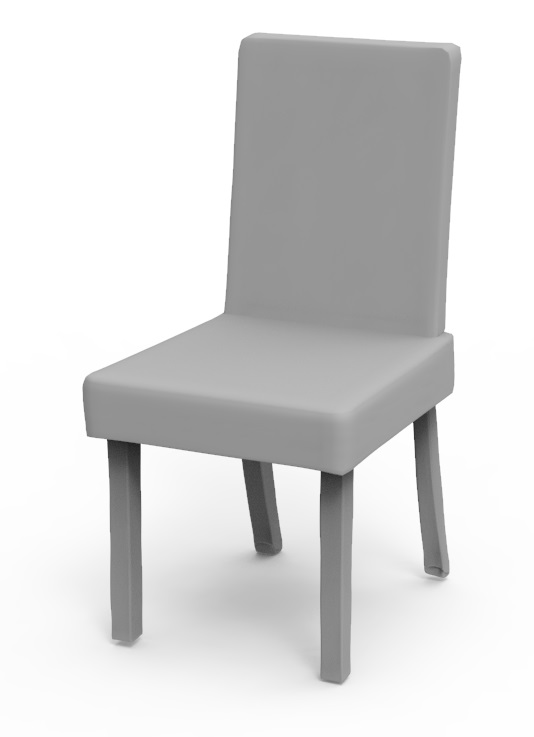}}
	}\caption{{Comparison between adjacency relationship constraint and support relationship constraint. (a) is a randomly generated \YLN{extreme} case by SDM-NET before refinement. (b) and (c) are the results after refinement with only the adjacency constraint and the support constraint, respectively.}}
	\label{fig:adjacencycomp}
\end{figure}

\begin{figure}
	{
	    \subfigure[Input shape]{\includegraphics[width=0.32\linewidth]{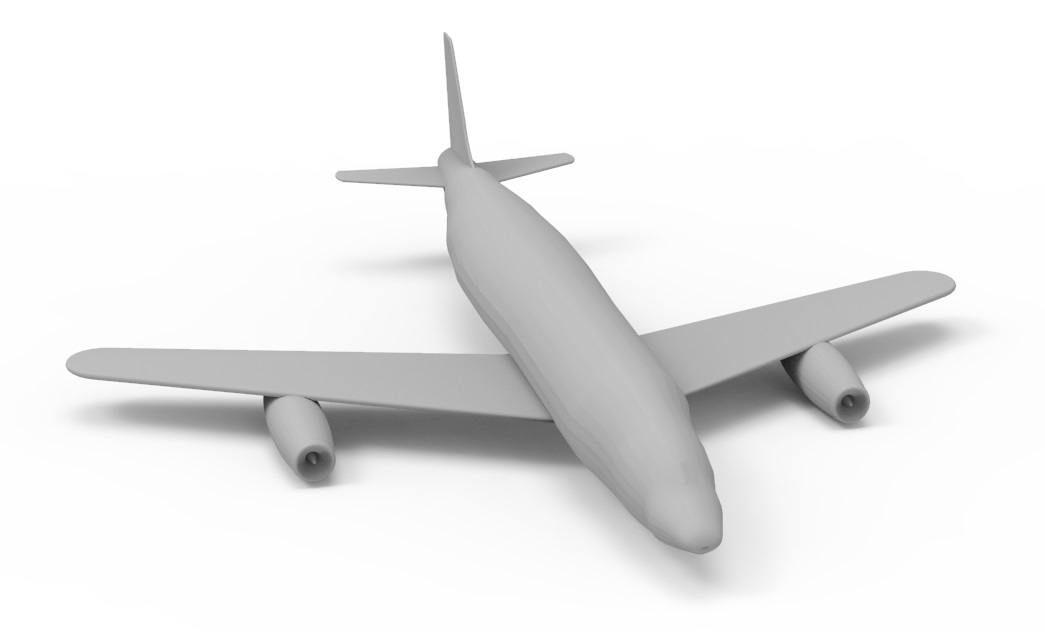}}
	\subfigure[Separate training]	{\includegraphics[width=0.32\linewidth]{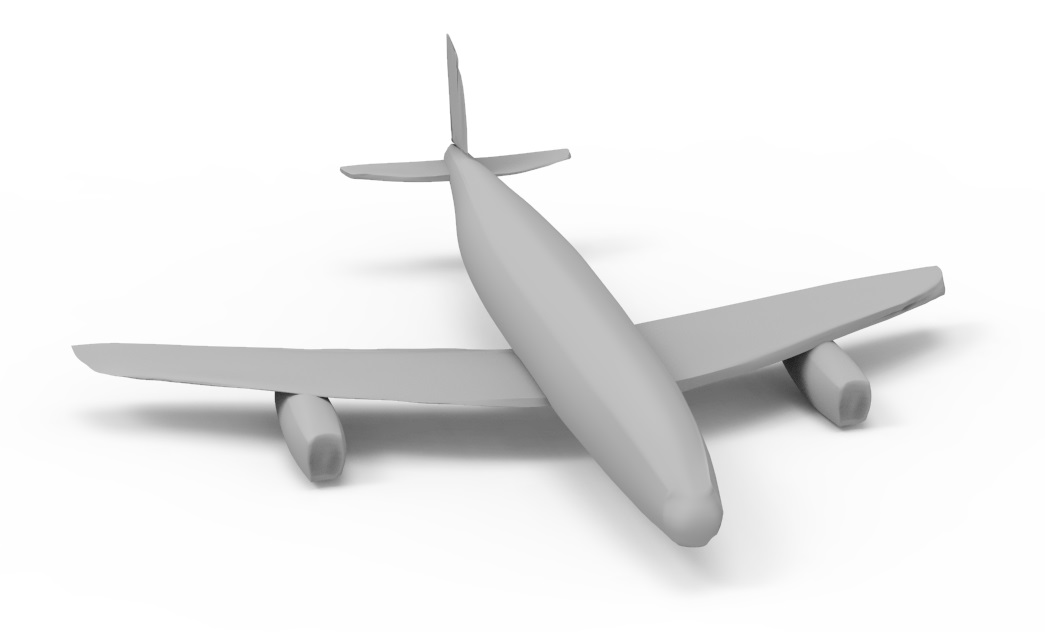}}
	\subfigure[Joint training]	{\includegraphics[width=0.32\linewidth]{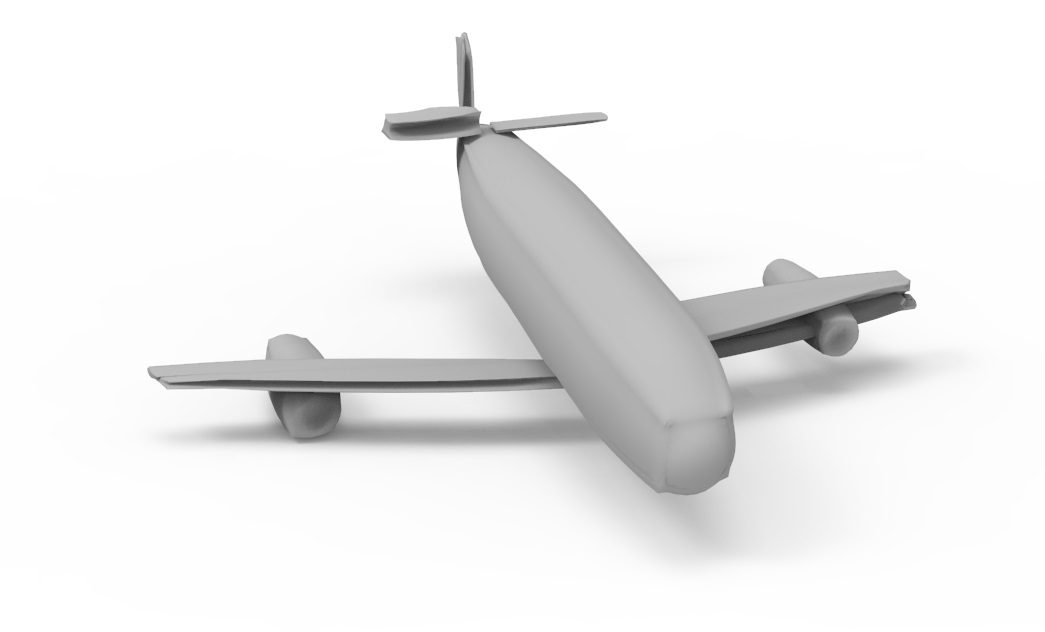}}
	}\caption{{Qualitative comparison between two different training strategies, i.e., separate training vs. end-to-end training. It can be seen that separate training keeps more geometric details.}}
	\label{fig:trainingstrategy}
\end{figure}

\begin{figure}
	\centering
	{
	    \subfigure[Decoupled structure and geometry]{\includegraphics[width=0.23\linewidth]{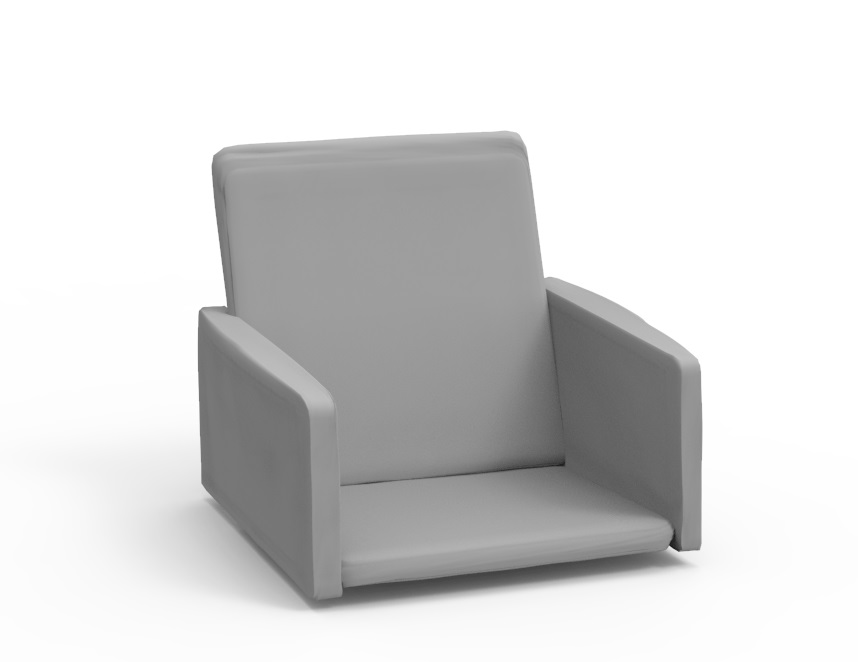}
	    \includegraphics[width=0.23\linewidth]{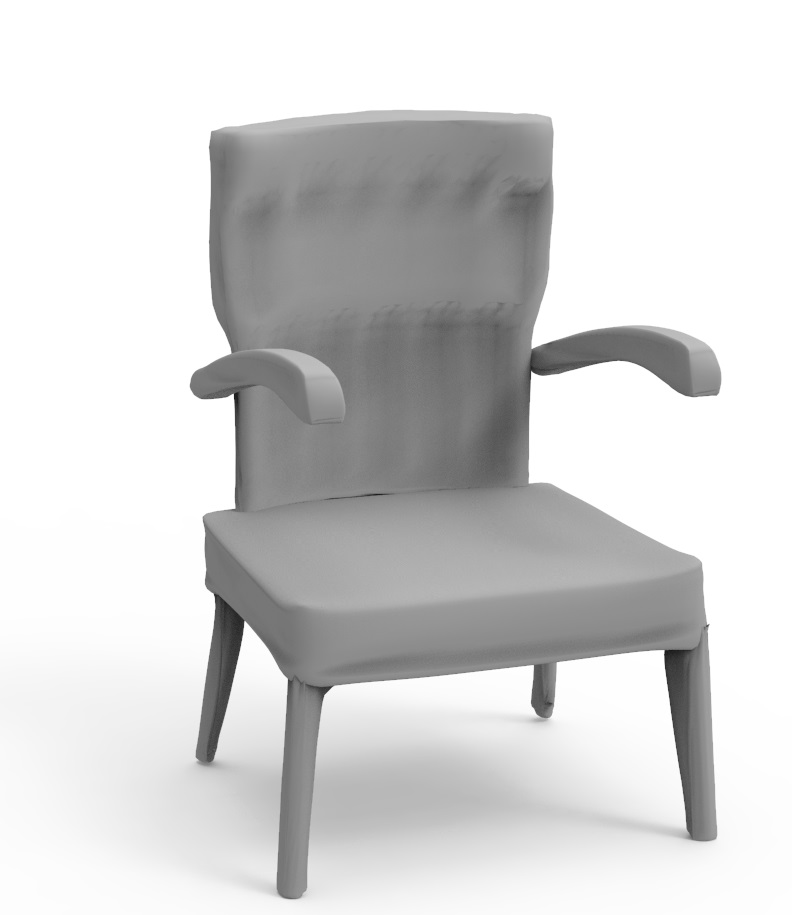}}
	    \subfigure[Our joint encoding]{\includegraphics[width=0.23\linewidth]{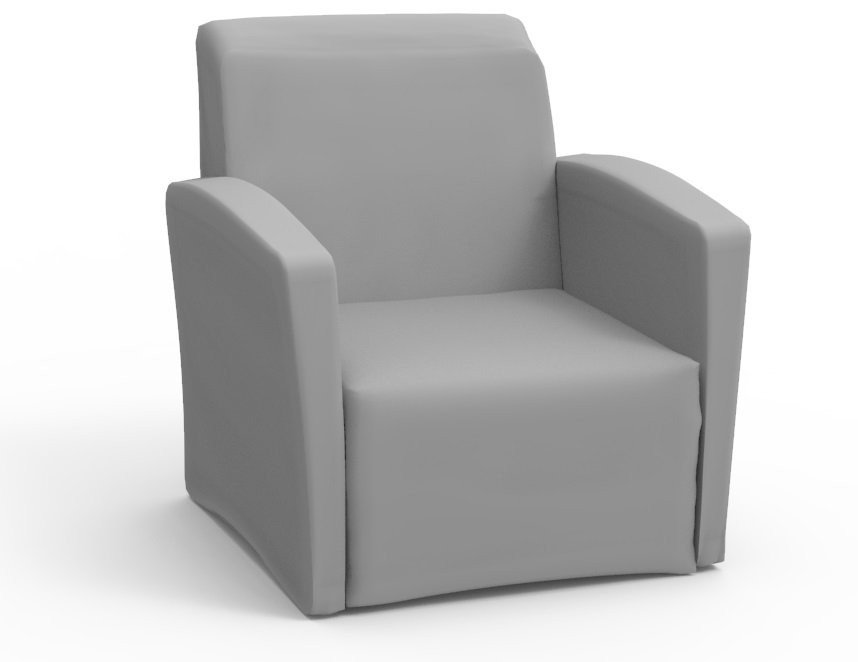}
	    \includegraphics[width=0.23\linewidth]{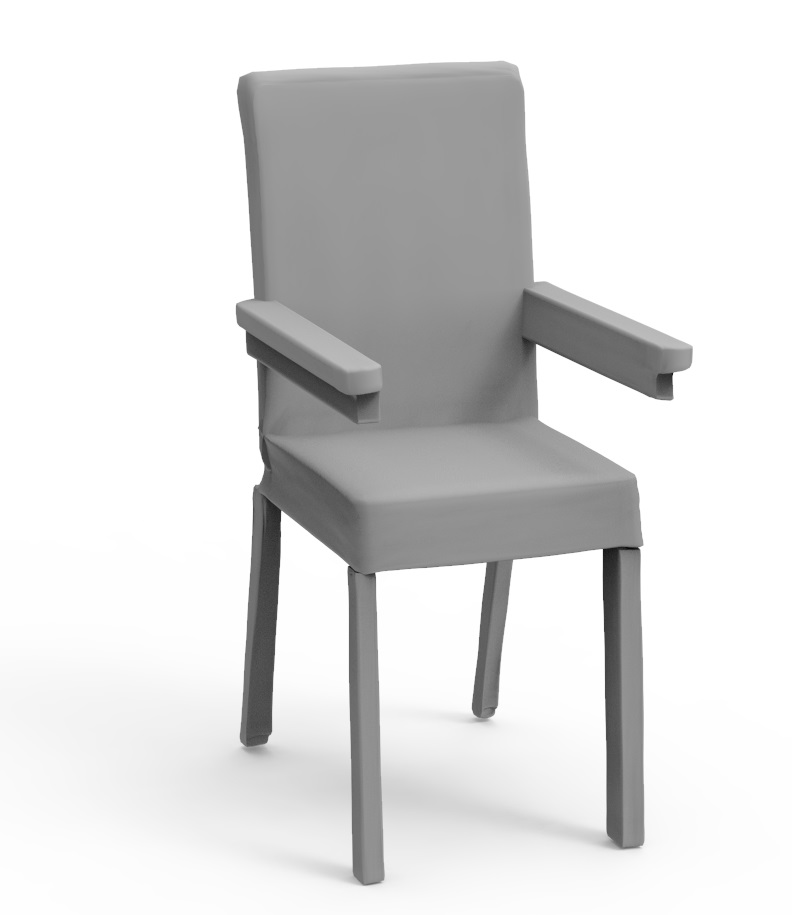}}
	}\caption{\YL{Results of our method (b), compared with decoupled structure and geometry (a).
	The latter is produced by decoupling the geometry information from SP-VAE, and generating the geometry of individual parts independently. }
	} %
	\label{fig:onlystructure}
\end{figure}

\textbf{Ablation Studies}. We perform several ablation studies to demonstrate the necessity of key components of our architecture. 

\paragraph{Support vs. adjacency relationships.} We adopt support relationships in our method, rather than adjacency relationships to get well connected shapes, because support relationships ensure generating physically stable shapes, and provide
 a natural order which is useful to simplify the structure refinement optimization. In contrast, using a bidirectional adjacency, it would be much more complicated to formulate and optimize constraints between two adjacent parts. To evaluate the effectiveness of the support relationships, we replace the support  constraints by simply minimizing the distance between every pair of adjacent parts to approximate the adjacency relationships. The effects \YLN{of} using support and adjacency constraints are shown in Figure~\ref{fig:adjacencycomp}. It can be seen that the support constraints lead to a physically more stable result. 

\paragraph{Separate vs. end-to-end training.}
\YL{We adopt separate training for the two-level VAE, i.e. PartVAEs are trained for individual part types first, before training SP-VAE where the geometries of parts are encoded with the trained PartVAEs. The two-level VAE could also be trained end-to-end, i.e., optimizing both PartVAEs and SP-VAE simultaneously.
We compare the average \YL{bidirectional Chamfer distance} of the reconstruction of each part between end-to-end training and separate training adopted in our solution, as given in Table~\ref{tab:trainingstrategy}.
The visual comparisons are shown in Figure~\ref{fig:trainingstrategy}. Since without the help of the well-trained distribution of the latent space of individual parts, end-to-end training would result in optimization stuck at a poor local minimum, leading to higher reconstruction errors and visually poor results. }

\begin{figure*}
	\centering
	{
		\subfigure[]{\includegraphics[width=0.18\linewidth]{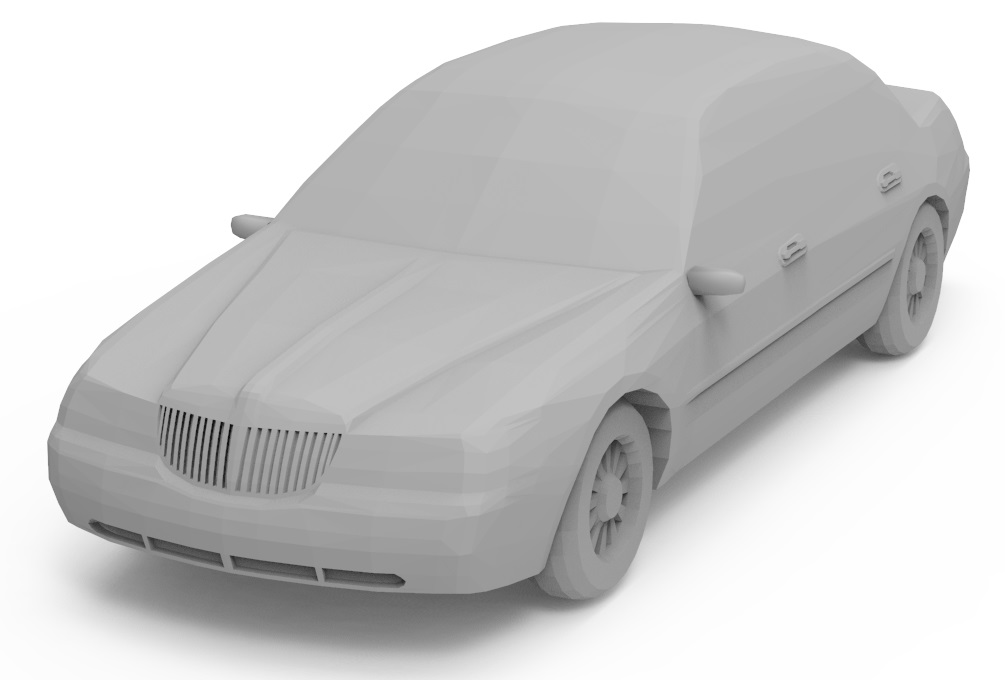}}
		\subfigure[]{\includegraphics[width=0.18\linewidth]{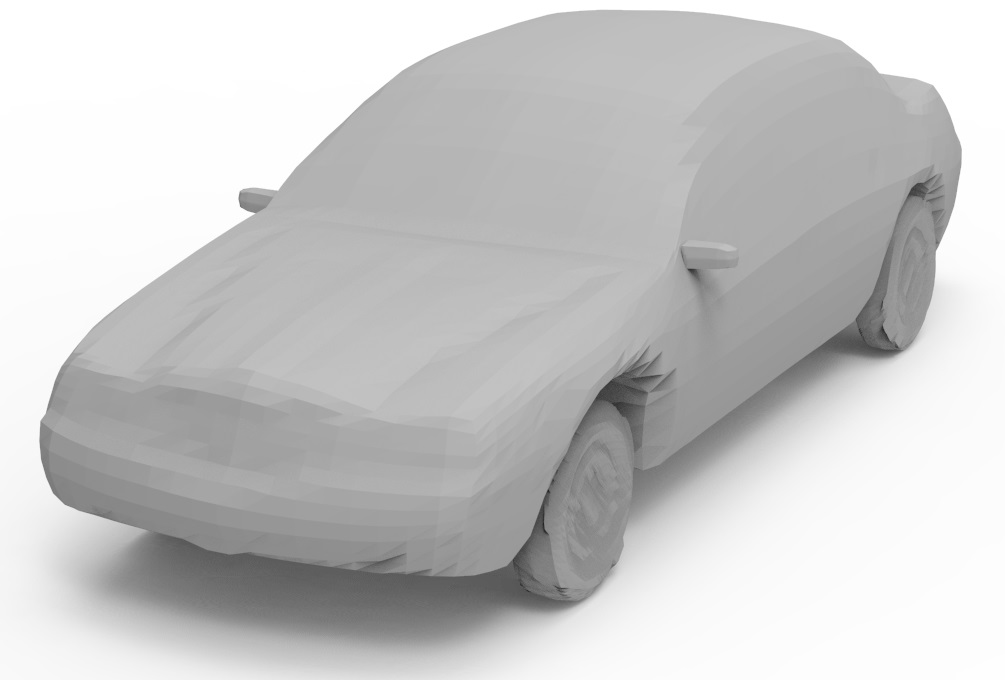}}
		\subfigure[]{\includegraphics[width=0.18\linewidth]{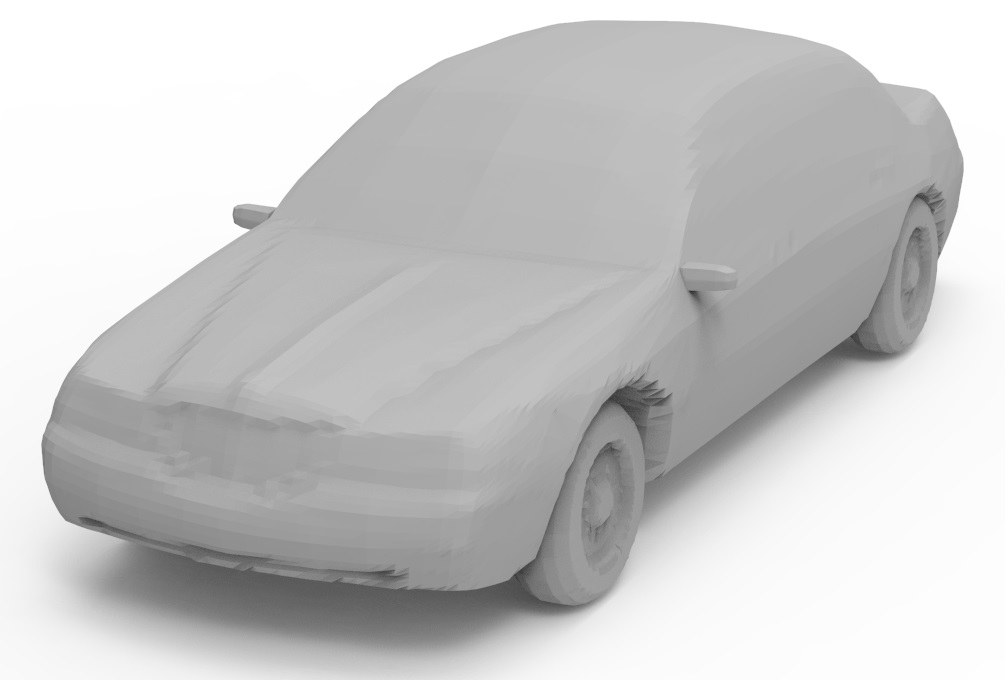}}
		\subfigure[]{\includegraphics[width=0.18\linewidth]{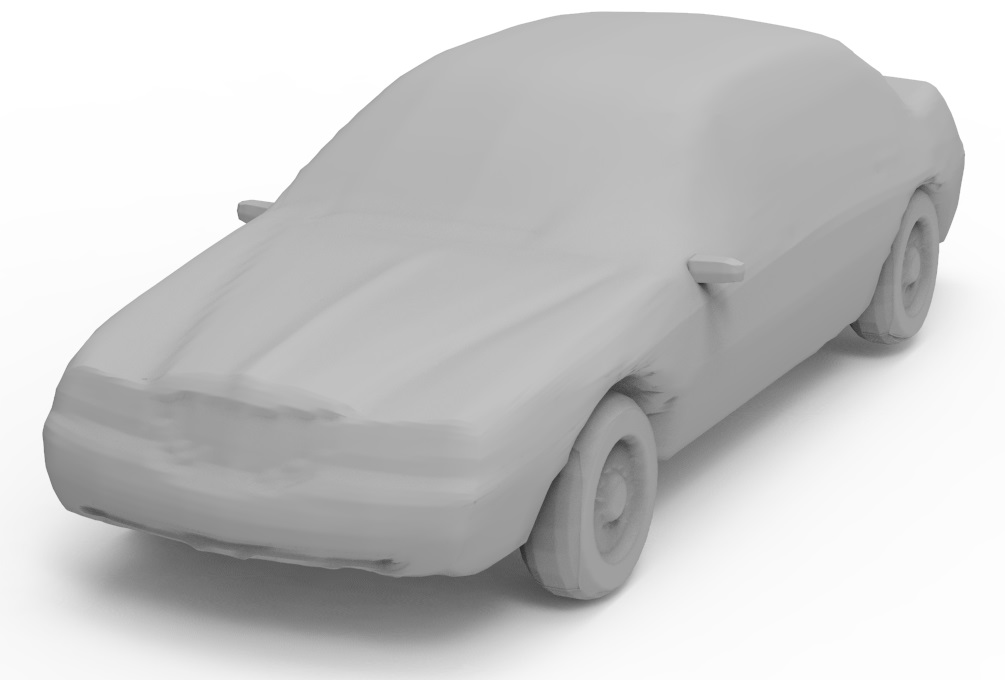}}
		\subfigure[]{\includegraphics[width=0.18\linewidth]{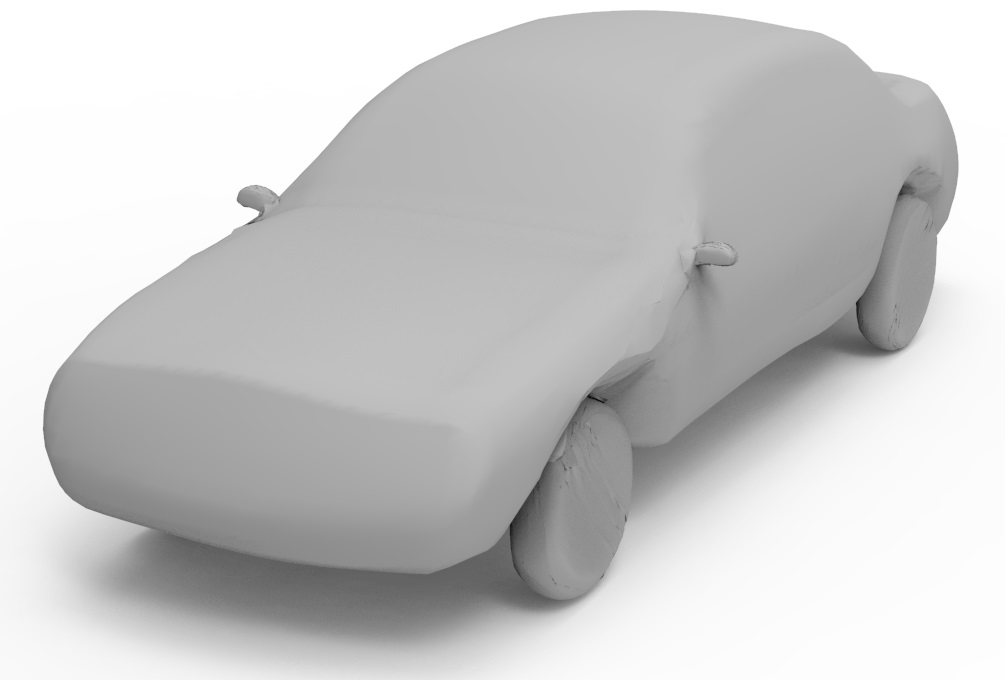}}
	}
	\caption{\YL{Shape reconstruction using our SDM-NET with changing bounding box resolutions and using a single PartVAE for all part categories. (a) input shape, (b) our method with low-resolution bounding boxes ($4.8K$ triangles), (c) our method with default resolution bounding boxes ($19.2K$ triangles), (d) our method with high-resolution bounding boxes ($76.8K$ triangles), (e) result with a single PartVAE for all part categories.}}
	\label{fig:resolution}
\end{figure*}
\begin{figure}
	\centering
	{
		\subfigure[Input Shape]{\includegraphics[width=0.3\linewidth]{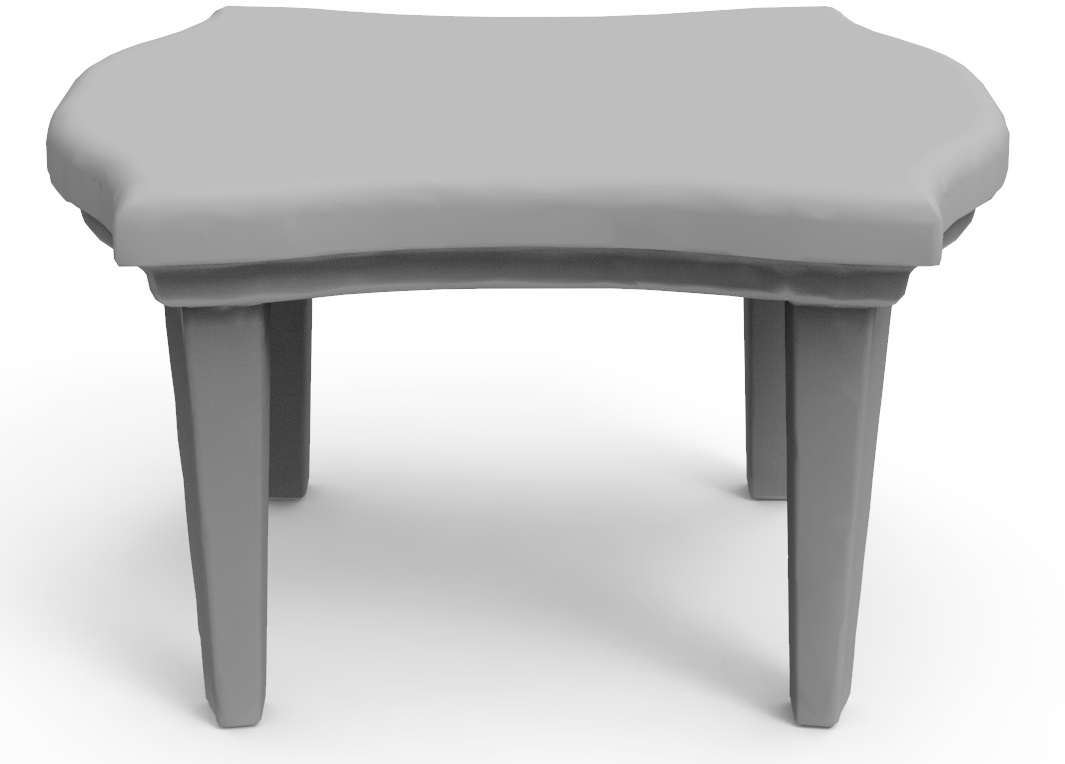}}
		\subfigure[Segmentation]{\includegraphics[width=0.3\linewidth]{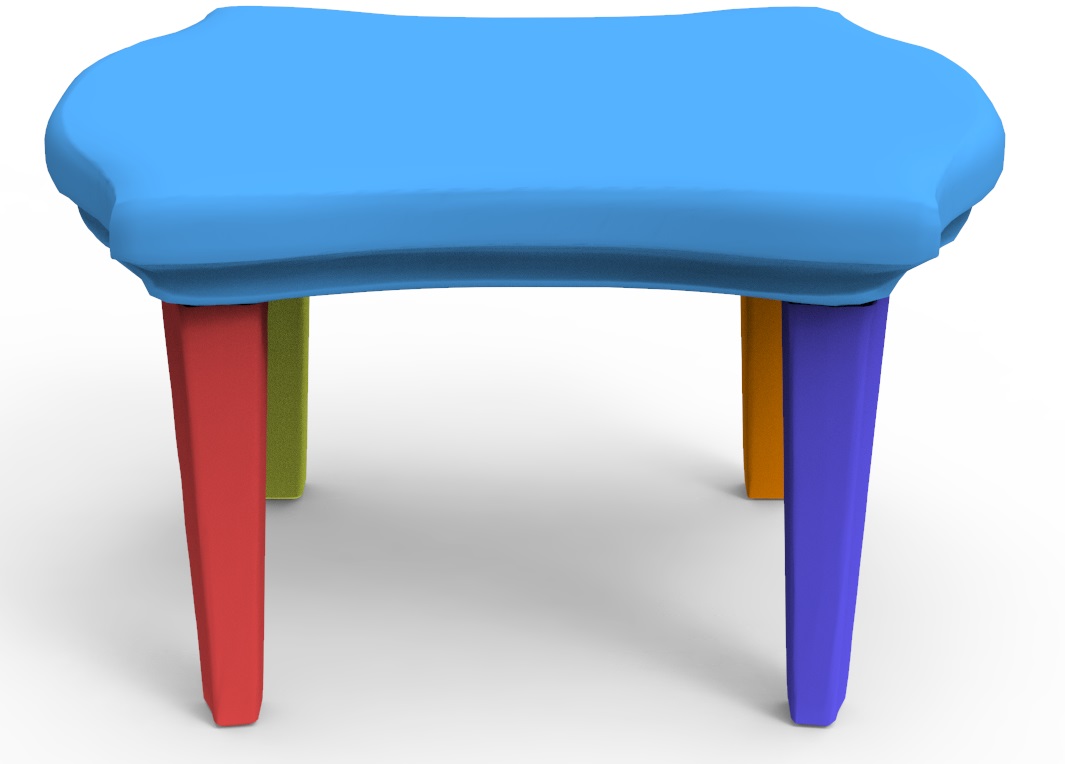}}
		\subfigure[Reconstruction]{\includegraphics[width=0.3\linewidth]{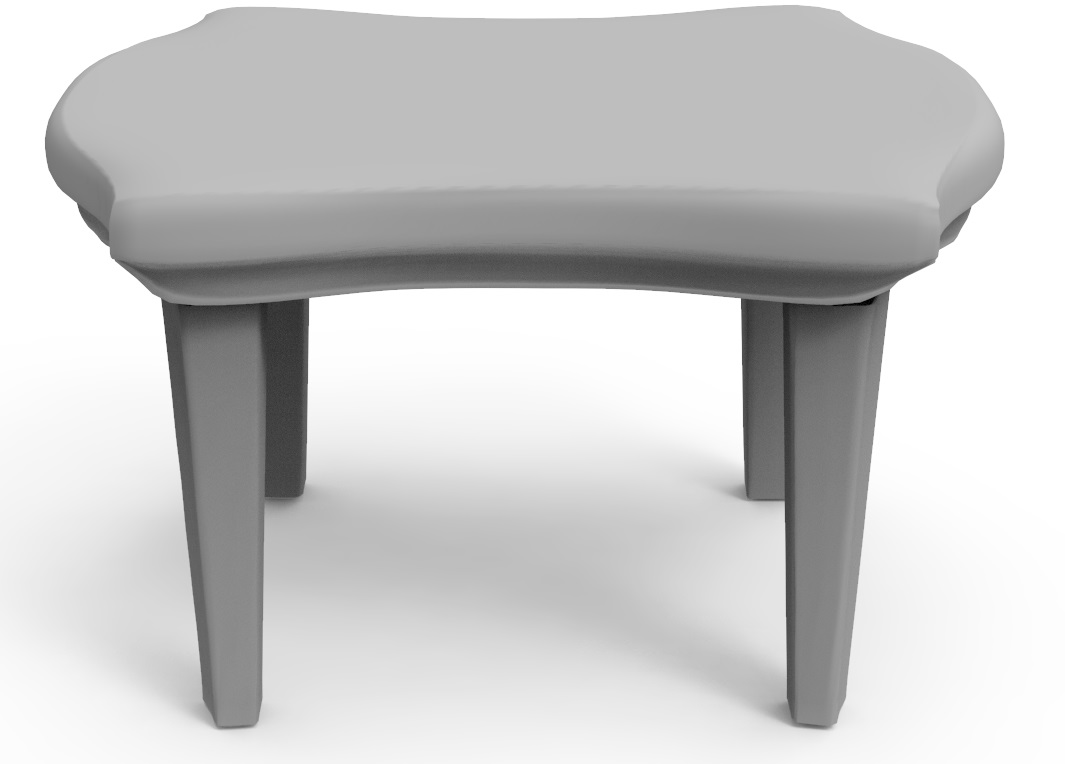}}
	}\caption{{An example demonstrating the generalizability of our method.  (a) input shape without semantic segmentation, (b)  segmentation by PointNet++, (c) the reconstruction result by our method.}}
	\label{fig:Scalability}
\end{figure}

\paragraph{Joint vs. decoupled structure and part geometry encoding.}
In this paper, the geometry details represented as part deformations are encoded into the SP-VAE embedding space jointly (see Section~\ref{sec:SPVAE}). This approach ensures 
that generated shapes have consistent structure and geometry, and the geometries of different parts are also coherent.
We compare our solution with an alternative approach where the structure and geometry encodings are decoupled: SP-VAE only encodes the structure and the geometry of each part is separately encoded using a PartVAE.  %
\YLN{Figure~\ref{fig:onlystructure} shows randomly generated shapes with both approaches.}
The structure \YLN{of the first example} implies the shape is a sofa, but the geometry of the seat part in (a) does not look like a sofa, whereas our method generates part geometry consistent with the structure. For the \YLN{second} example, our method produces parts with coherent geometry (b), whereas using decoupled structure and geometry leads to inconsistent part styles (a).

\paragraph{Resolution of bounding boxes.} \YL{By default, our method uses bounding boxes each with $19.2K$ triangles. We also try using lower and higher resolution bounding boxes. As shown in Figure~\ref{fig:resolution}, using lower resolution (b) cannot capture the details of the shape, and using higher resolution (d) produces very similar result as our default setting (c), but takes longer time. Our default setting (c) provides a good balance between efficiency and quality.}

\paragraph{PartVAE per part type vs. single PartVAE} \YL{In our paper, we train a PartVAE for each part type. We compare this with an alternative approach where a single PartVAE is trained for all part categories. As shown in Figure~\ref{fig:resolution} (e), this approach is not effective in capturing unique geometric features of different parts, leading to poor geometric reconstruction.}

\textbf{Generalizability.}
Figure~\ref{fig:Scalability} shows an example that demonstrates the generalizability of our method, 
\YL{to process new shapes of the same category without input semantic segmentation.
We first train PointNet++~\cite{Qi2017nips} on our labeled dataset, which is then used for semantic segmentation of the new shape. Finally, we obtain the reconstruction result by our SDM-NET. An example is shown in Figure~\ref{fig:Scalability}, which demonstrates that semantic segmentation obtained automatically can be effectively used as input to our method.}

\begin{figure}
	\centering
	{
		\subfigure[Decoded models]
		{
		\includegraphics[width=0.23\linewidth]{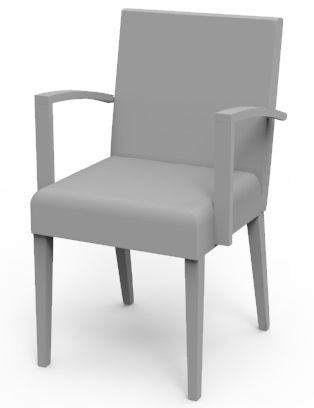}
		\includegraphics[width=0.23\linewidth]{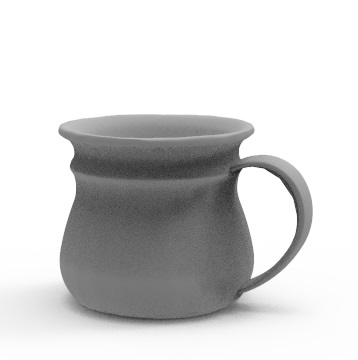}
		}
		\subfigure[Watertight models]
		{
		\includegraphics[width=0.23\linewidth]{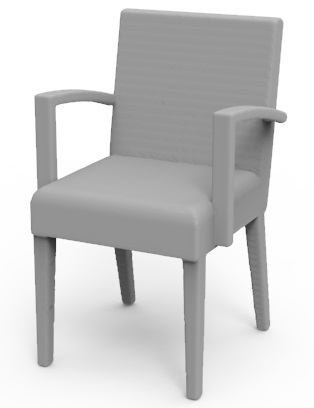}
		\includegraphics[width=0.23\linewidth]{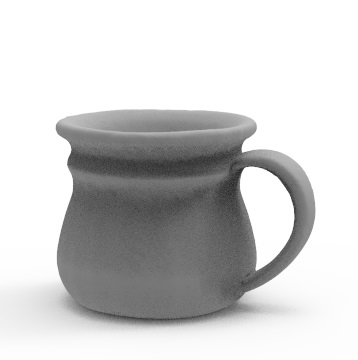}
		}
	}
	\caption{Watertight models derived from our decoded models using \cite{huang2018robust}}
	\label{fig:watertight}
\end{figure}

\textbf{Watertight models.} \YL{The direct output of our method includes watertight meshes for individual parts, but not the shape as a whole. As demonstrated in Figure~\ref{fig:watertight}, by applying a watertight reconstruction technique~\cite{huang2018robust}, watertight meshes can be obtained, which benefit certain downstream applications.}

\begin{figure}
	\centering
	{
		\subfigure[]{\includegraphics[width=0.18\linewidth]{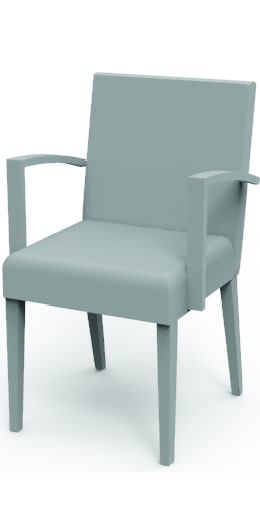}}
		\subfigure[]{\includegraphics[width=0.18\linewidth]{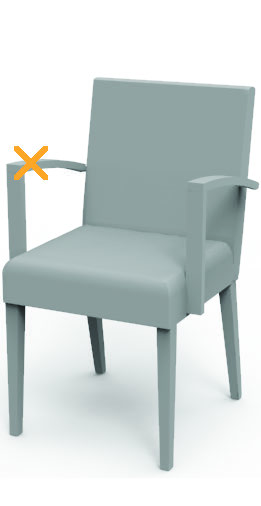}}
		\subfigure[]{\includegraphics[width=0.18\linewidth]{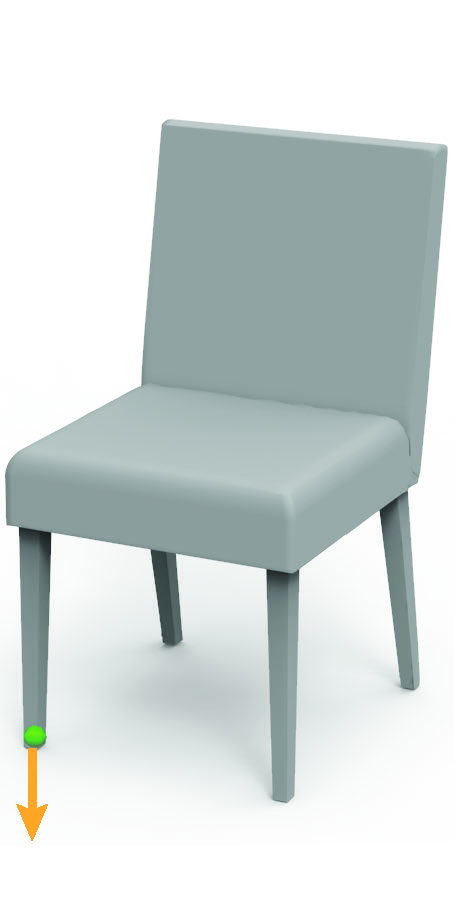}}
		\subfigure[]{\includegraphics[width=0.18\linewidth]{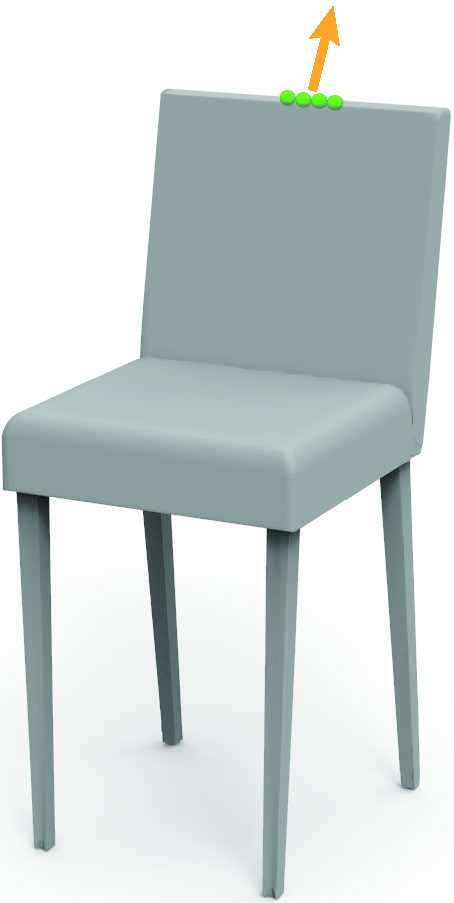}}
		\subfigure[]{\includegraphics[width=0.18\linewidth]{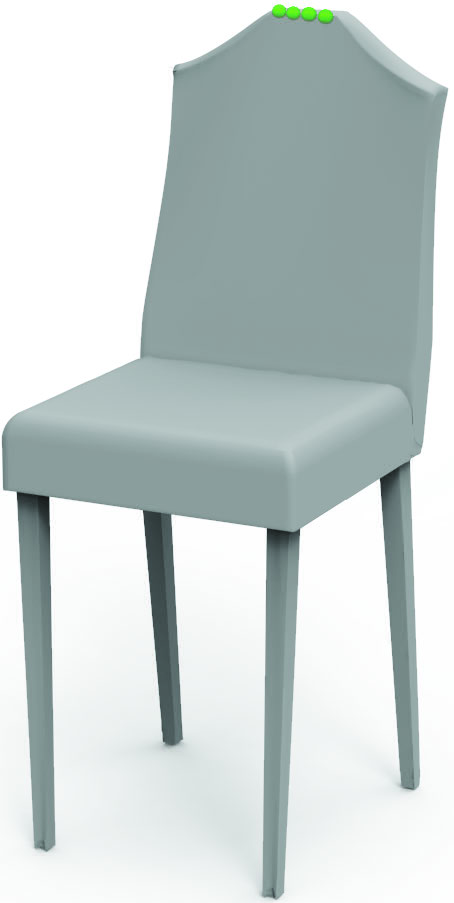}}
	}
	\caption{\YL{Our generated Structured Deformable Mesh is directly editable. We show an example editing sequence. (a) is the decoded shape of our method. After applying the deletion operation (b), we obtain a chair without armrests (c). Dragging a leg of the chair makes all four legs longer due to the equal length constraint (d). Finally, we deform the back of the chair to obtain (e). }}
	\label{fig:edit}
\end{figure}

\textbf{Editability.} \YL{Our generative model produces Structure Deformable Meshes, which are immediately editable in a structure-aware manner. This is difficult for other generative methods (e.g. \cite{3dgan2016,li_sig17}).
An example is given in Figure~\ref{fig:edit}, which shows an editing sequence, including removing parts (when a part is removed, its symmetric part is also removed), making a chair leg longer, which also affects other chair legs due to the equal length constraint, and further deforming the chair back using an off-the-shelf deformation method~\cite{ARAP2007}.} \rv{During shape deformation, the editing constraints are treated as hard constraints and the equal length constraints are used in the refinement step (see Section~\ref{sec:shapegen}).}

\begin{figure}
    \centering
    {
        \includegraphics[width=0.22\linewidth]{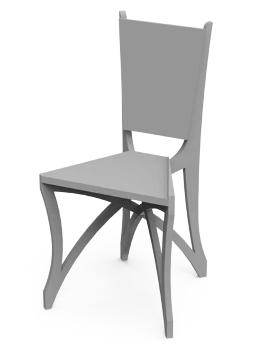}
        \includegraphics[width=0.22\linewidth]{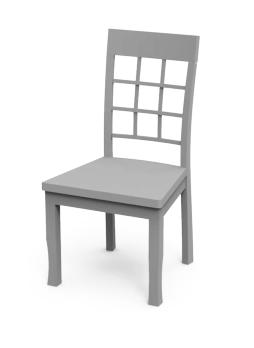}
        \includegraphics[width=0.22\linewidth]{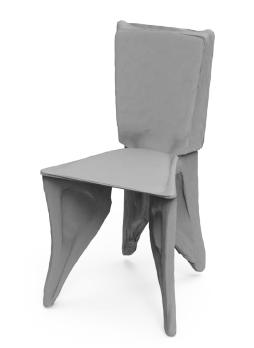}
        \includegraphics[width=0.22\linewidth]{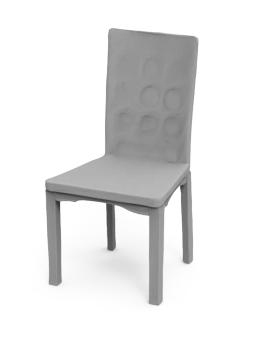}\\
        \subfigure[Input Shapes]{    \includegraphics[width=0.45\linewidth]{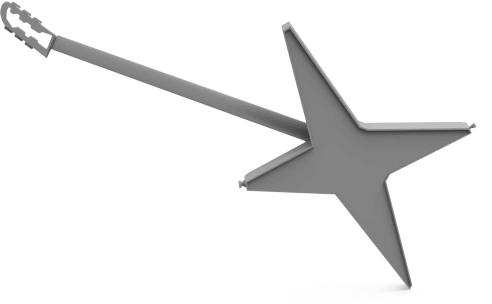}
        }
        \subfigure[Reconstructed Shapes]{
        \includegraphics[width=0.45\linewidth]{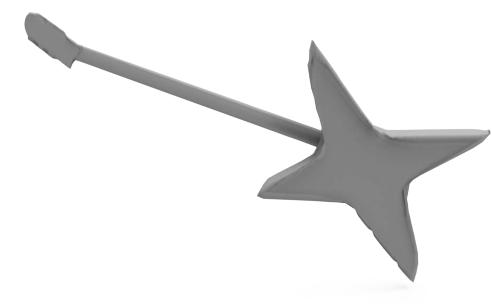}
        }
    }
    \caption{\rv{Failure cases: the headstock of the guitar with a hole, the chair with complex legs {and the chair with grid back} could not be decoded by our SDM-NET since {it requires each part to have} the fixed topology, i.e., the same as a genus-zero box. 
    }}
    \label{fig:failure_case}
\end{figure}

\textbf{Limitations.}
Although our method can handle a large variety of shapes with flexible structures and fine details, it still suffers from several  limitations. While our method can handle shapes with holes formed by multiple parts, if a part itself has holes in it, our deformable box is unable to represent it exactly as the topology of parts cannot be different from genus-zero boxes. In this case, our method will try to preserve the mesh geometry but cannot maintain the hole. \rv{
For certain parts which are unusual (e.g. the legs and back of the chair \YLN{and} the headstock of the \YLN{guitar} in Figure~\ref{fig:failure_case}),
our VAE architecture considers such cases as outliers, and  ``projects'' them back to deformations consistent with the training set. 
}
\YL{Another limitation is that currently SDM-NET is trained using a collection of shapes with the same category. It thus cannot be used for interpolating shapes of different categories.}

\section{Conclusions and Future Work}
\label{sec:conc}

In this paper, we have presented SDM-NET, a novel deep generative model that generates 3D shapes as Structured Deformable Meshes. 
A shape is represented using a set of deformable boxes, and a two-level VAE is built to encode local geometry variations of individual parts, and global structure and geometries of all parts, respectively. Our representation achieves both flexible topology and fine geometric details, outperforming the state-of-the-art methods for both shape generation and shape interpolation.
 
As \YLN{future} work, our method could be generalized to reconstruct shapes from images. {Similar to~\cite{Xin2019}, which uses a deep neural network to learn the segmentation masks of cylinder regions from a given image for reconstructing 3D models composed of cylindrical shapes, a possible approach to extend our method is to learn the segmentation of different parts in images and use such segmentation results as the conditions of the SP-VAE for 3D shape reconstruction.}
In this case, our SDM-NET makes it possible to generate rich shapes with details to better match given images. By exploiting the latent space of our network, our approach could also be generalized for data-driven deformation by incorporating user editing constraints in the optimization framework. \YL{It is also interesting to investigate how to extend our method to encode shapes of different categories using a single network.}
\rv{Our current approach can generate parts with the same resolution as the primitive bounding box mesh. We currently utilize a high-resolution mesh with fixed size, 
and therefore our generated shapes take up \YLN{large} storage space because of richer geometric details and higher resolution of meshes. However, since different kinds of parts have different geometric richness, it would be better to exploit \YLN{(possibly different \hongbo{types of})} primitives with adaptive resolutions for different parts so that we can preserve the same level of details but with significantly less storage space.}

\begin{acks}
{This work was supported by National Natural Science Foundation
of China (No. 61828204 and No. 61872440), Beijing Municipal Natural Science Foundation (No. L182016), Youth Innovation Promotion Association CAS, CCF-Tencent Open Fund, SenseTime Research Fund. %
Hongbo Fu was partially supported by grants from the Research Grants Council of the Hong Kong Special Administrative Region, China (Project No. CityU 11237116, CityU 11300615), and the Centre for Applied Computing and Interactive Media (ACIM) of School of Creative Media, CityU.}

\end{acks}
\bibliographystyle{ACM-Reference-Format}
\bibliography{bibliography}


\begin{thebibliography}{00}


\ifx \showCODEN    \undefined \def \showCODEN     #1{\unskip}     \fi
\ifx \showDOI      \undefined \def \showDOI       #1{#1}\fi
\ifx \showISBNx    \undefined \def \showISBNx     #1{\unskip}     \fi
\ifx \showISBNxiii \undefined \def \showISBNxiii  #1{\unskip}     \fi
\ifx \showISSN     \undefined \def \showISSN      #1{\unskip}     \fi
\ifx \showLCCN     \undefined \def \showLCCN      #1{\unskip}     \fi
\ifx \shownote     \undefined \def \shownote      #1{#1}          \fi
\ifx \showarticletitle \undefined \def \showarticletitle #1{#1}   \fi
\ifx \showURL      \undefined \def \showURL       {\relax}        \fi
\providecommand\bibfield[2]{#2}
\providecommand\bibinfo[2]{#2}
\providecommand\natexlab[1]{#1}
\providecommand\showeprint[2][]{arXiv:#2}

\bibitem[\protect\citeauthoryear{Achlioptas, Diamanti, Mitliagkas, and
  Guibas}{Achlioptas et~al\mbox{.}}{2018}]%
        {achlioptas18a}
\bibfield{author}{\bibinfo{person}{Panos Achlioptas}, \bibinfo{person}{Olga
  Diamanti}, \bibinfo{person}{Ioannis Mitliagkas}, {and}
  \bibinfo{person}{Leonidas Guibas}.} \bibinfo{year}{2018}\natexlab{}.
\newblock \showarticletitle{Learning Representations and Generative Models for
  3{D} Point Clouds}. In \bibinfo{booktitle}{{\em International Conference on
  Machine Learning (ICML)}}, Vol.~\bibinfo{volume}{80}.
  \bibinfo{pages}{40--49}.
\newblock


\bibitem[\protect\citeauthoryear{Anguelov, Srinivasan, Koller, Thrun, Rodgers,
  and Davis}{Anguelov et~al\mbox{.}}{2005}]%
        {anguelov2005scape}
\bibfield{author}{\bibinfo{person}{Dragomir Anguelov}, \bibinfo{person}{Praveen
  Srinivasan}, \bibinfo{person}{Daphne Koller}, \bibinfo{person}{Sebastian
  Thrun}, \bibinfo{person}{Jim Rodgers}, {and} \bibinfo{person}{James Davis}.}
  \bibinfo{year}{2005}\natexlab{}.
\newblock \showarticletitle{{SCAPE}: shape completion and animation of people}.
\newblock \bibinfo{journal}{{\em ACM Trans. Graph.\/}} \bibinfo{volume}{24},
  \bibinfo{number}{3} (\bibinfo{year}{2005}), \bibinfo{pages}{408--416}.
\newblock


\bibitem[\protect\citeauthoryear{Averkiou, Kim, Zheng, and Mitra}{Averkiou
  et~al\mbox{.}}{2014}]%
        {Averkiou2014EG}
\bibfield{author}{\bibinfo{person}{Melinos Averkiou},
  \bibinfo{person}{Vladimir~G Kim}, \bibinfo{person}{Youyi Zheng}, {and}
  \bibinfo{person}{Niloy~J Mitra}.} \bibinfo{year}{2014}\natexlab{}.
\newblock \showarticletitle{Shapesynth: Parameterizing model collections for
  coupled shape exploration and synthesis}. In \bibinfo{booktitle}{{\em
  Computer Graphics Forum}}, Vol.~\bibinfo{volume}{33}. Wiley Online Library,
  \bibinfo{pages}{125--134}.
\newblock


\bibitem[\protect\citeauthoryear{Chang, Funkhouser, Guibas, Hanrahan, Huang,
  Li, Savarese, Savva, Song, Su, Xiao, Yi, and Yu}{Chang et~al\mbox{.}}{2015}]%
        {shapenet}
\bibfield{author}{\bibinfo{person}{Angel~X. Chang}, \bibinfo{person}{Thomas
  Funkhouser}, \bibinfo{person}{Leonidas Guibas}, \bibinfo{person}{Pat
  Hanrahan}, \bibinfo{person}{Qixing Huang}, \bibinfo{person}{Zimo Li},
  \bibinfo{person}{Silvio Savarese}, \bibinfo{person}{Manolis Savva},
  \bibinfo{person}{Shuran Song}, \bibinfo{person}{Hao Su},
  \bibinfo{person}{Jianxiong Xiao}, \bibinfo{person}{Li Yi}, {and}
  \bibinfo{person}{Fisher Yu}.} \bibinfo{year}{2015}\natexlab{}.
\newblock \bibinfo{booktitle}{{\em {{ShapeNet}: An Information-Rich {3D} Model
  Repository}}}.
\newblock \bibinfo{type}{{T}echnical {R}eport} arXiv:1512.03012 [cs.GR].
  \bibinfo{institution}{Stanford University --- Princeton University --- Toyota
  Technological Institute at Chicago}.
\newblock


\bibitem[\protect\citeauthoryear{Chaudhuri, Ritchie, Xu, and Zhang}{Chaudhuri
  et~al\mbox{.}}{2019}]%
        {chaudhuri2019}
\bibfield{author}{\bibinfo{person}{Siddhartha Chaudhuri},
  \bibinfo{person}{Daniel Ritchie}, \bibinfo{person}{Kai Xu}, {and}
  \bibinfo{person}{Hao Zhang}.} \bibinfo{year}{2019}\natexlab{}.
\newblock \showarticletitle{Learning Generative Models of {3D} Structures}. In
  \bibinfo{booktitle}{{\em Eurographics Tutorial}}.
\newblock


\bibitem[\protect\citeauthoryear{Chen and Zhang}{Chen and Zhang}{2019}]%
        {chen2019-IMNET}
\bibfield{author}{\bibinfo{person}{Zhiqin Chen} {and} \bibinfo{person}{Hao
  Zhang}.} \bibinfo{year}{2019}\natexlab{}.
\newblock \showarticletitle{Learning implicit fields for generative shape
  modeling}. In \bibinfo{booktitle}{{\em Proceedings of the IEEE Conference on
  Computer Vision and Pattern Recognition}}. \bibinfo{pages}{5939--5948}.
\newblock


\bibitem[\protect\citeauthoryear{Fan, Su, and Guibas}{Fan
  et~al\mbox{.}}{2017}]%
        {fan2016point}
\bibfield{author}{\bibinfo{person}{Haoqiang Fan}, \bibinfo{person}{Hao Su},
  {and} \bibinfo{person}{Leonidas~J Guibas}.} \bibinfo{year}{2017}\natexlab{}.
\newblock \showarticletitle{A point set generation network for {3D} object
  reconstruction from a single image}. In \bibinfo{booktitle}{{\em Proceedings
  of the IEEE conference on computer vision and pattern recognition}}.
  \bibinfo{pages}{605--613}.
\newblock


\bibitem[\protect\citeauthoryear{Gao, Lai, Yang, Zhang, Xia, and Kobbelt}{Gao
  et~al\mbox{.}}{2019}]%
        {Gao2017}
\bibfield{author}{\bibinfo{person}{Lin Gao}, \bibinfo{person}{Yu{-}Kun Lai},
  \bibinfo{person}{Jie Yang}, \bibinfo{person}{Ling{-}Xiao Zhang},
  \bibinfo{person}{Shihong Xia}, {and} \bibinfo{person}{Leif Kobbelt}.}
  \bibinfo{year}{2019}\natexlab{}.
\newblock \showarticletitle{Sparse Data Driven Mesh Deformation}.
\newblock \bibinfo{journal}{{\em {IEEE} Trans. Vis. Comput. Graph.\/}}
  (\bibinfo{year}{2019}).
\newblock


\bibitem[\protect\citeauthoryear{Gao, Yang, Qiao, Lai, Rosin, Xu, and Xia}{Gao
  et~al\mbox{.}}{2018}]%
        {Gao2018}
\bibfield{author}{\bibinfo{person}{Lin Gao}, \bibinfo{person}{Jie Yang},
  \bibinfo{person}{Yi-Ling Qiao}, \bibinfo{person}{Yu-Kun Lai},
  \bibinfo{person}{Paul~L. Rosin}, \bibinfo{person}{Weiwei Xu}, {and}
  \bibinfo{person}{Shihong Xia}.} \bibinfo{year}{2018}\natexlab{}.
\newblock \showarticletitle{Automatic unpaired shape deformation transfer.}
\newblock \bibinfo{journal}{{\em ACM Trans. Graph.\/}} \bibinfo{volume}{37},
  \bibinfo{number}{6} (\bibinfo{year}{2018}), \bibinfo{pages}{237:1--237:15}.
\newblock


\bibitem[\protect\citeauthoryear{Girdhar, Fouhey, Rodriguez, and Gupta}{Girdhar
  et~al\mbox{.}}{2016}]%
        {Girdhar16b}
\bibfield{author}{\bibinfo{person}{Rohit Girdhar}, \bibinfo{person}{David~F
  Fouhey}, \bibinfo{person}{Mikel Rodriguez}, {and} \bibinfo{person}{Abhinav
  Gupta}.} \bibinfo{year}{2016}\natexlab{}.
\newblock \showarticletitle{Learning a predictable and generative vector
  representation for objects}. In \bibinfo{booktitle}{{\em European Conference
  on Computer Vision}}. Springer, \bibinfo{pages}{484--499}.
\newblock


\bibitem[\protect\citeauthoryear{Goodfellow, Pouget-Abadie, Mirza, Xu,
  Warde-Farley, Ozair, Courville, and Bengio}{Goodfellow et~al\mbox{.}}{2014}]%
        {goodfellow2014generative}
\bibfield{author}{\bibinfo{person}{Ian Goodfellow}, \bibinfo{person}{Jean
  Pouget-Abadie}, \bibinfo{person}{Mehdi Mirza}, \bibinfo{person}{Bing Xu},
  \bibinfo{person}{David Warde-Farley}, \bibinfo{person}{Sherjil Ozair},
  \bibinfo{person}{Aaron Courville}, {and} \bibinfo{person}{Yoshua Bengio}.}
  \bibinfo{year}{2014}\natexlab{}.
\newblock \showarticletitle{Generative adversarial nets}. In
  \bibinfo{booktitle}{{\em Advances in neural information processing systems
  (NIPS)}}. \bibinfo{pages}{2672--2680}.
\newblock


\bibitem[\protect\citeauthoryear{Groueix, Fisher, Kim, Russell, and
  Aubry}{Groueix et~al\mbox{.}}{2018}]%
        {AtlasNet2018}
\bibfield{author}{\bibinfo{person}{Thibault Groueix}, \bibinfo{person}{Matthew
  Fisher}, \bibinfo{person}{Vladimir~G. Kim}, \bibinfo{person}{Bryan Russell},
  {and} \bibinfo{person}{Mathieu Aubry}.} \bibinfo{year}{2018}\natexlab{}.
\newblock \showarticletitle{{AtlasNet: A Papier-M{\^a}ch{\'e} Approach to
  Learning {3D} Surface Generation}}. In \bibinfo{booktitle}{{\em IEEE
  Conference on Computer Vision and Pattern Recognition (CVPR)}}.
  \bibinfo{address}{Salt Lake City, United States}.
\newblock


\bibitem[\protect\citeauthoryear{Hamu, Maron, Kezurer, Avineri, and
  Lipman}{Hamu et~al\mbox{.}}{2018}]%
        {Hamu2018}
\bibfield{author}{\bibinfo{person}{Heli~Ben Hamu}, \bibinfo{person}{Haggai
  Maron}, \bibinfo{person}{Itay Kezurer}, \bibinfo{person}{Gal Avineri}, {and}
  \bibinfo{person}{Yaron Lipman}.} \bibinfo{year}{2018}\natexlab{}.
\newblock \showarticletitle{Multi-chart generative surface modeling.}
\newblock \bibinfo{journal}{{\em ACM Trans. Graph.\/}} \bibinfo{volume}{37},
  \bibinfo{number}{6} (\bibinfo{year}{2018}), \bibinfo{pages}{215:1--215:15}.
\newblock


\bibitem[\protect\citeauthoryear{Hanocka, Hertz, Fish, Giryes, Fleishman, and
  Cohen-Or}{Hanocka et~al\mbox{.}}{2019}]%
        {Hanocka2019}
\bibfield{author}{\bibinfo{person}{Rana Hanocka}, \bibinfo{person}{Amir Hertz},
  \bibinfo{person}{Noa Fish}, \bibinfo{person}{Raja Giryes},
  \bibinfo{person}{Shachar Fleishman}, {and} \bibinfo{person}{Daniel
  Cohen-Or}.} \bibinfo{year}{2019}\natexlab{}.
\newblock \showarticletitle{MeshCNN: A Network with an Edge}.
\newblock \bibinfo{journal}{{\em ACM Trans. Graph.\/}} \bibinfo{volume}{38},
  \bibinfo{number}{4}, Article \bibinfo{articleno}{90} (\bibinfo{year}{2019}),
  \bibinfo{numpages}{90:1--90:12}~pages.
\newblock


\bibitem[\protect\citeauthoryear{Holmstr{\"o}m and Edvall}{Holmstr{\"o}m and
  Edvall}{2004}]%
        {holmstrom2004tomlab}
\bibfield{author}{\bibinfo{person}{Kenneth Holmstr{\"o}m} {and}
  \bibinfo{person}{Marcus~M Edvall}.} \bibinfo{year}{2004}\natexlab{}.
\newblock \showarticletitle{The TOMLAB optimization environment}.
\newblock In \bibinfo{booktitle}{{\em Modeling Languages in Mathematical
  Optimization}}. \bibinfo{publisher}{Springer}, \bibinfo{pages}{369--376}.
\newblock


\bibitem[\protect\citeauthoryear{Huang, Kalogerakis, and Marlin}{Huang
  et~al\mbox{.}}{2015}]%
        {Huang2015CGF}
\bibfield{author}{\bibinfo{person}{Haibin Huang}, \bibinfo{person}{Evangelos
  Kalogerakis}, {and} \bibinfo{person}{Benjamin Marlin}.}
  \bibinfo{year}{2015}\natexlab{}.
\newblock \showarticletitle{Analysis and synthesis of {3D} shape families via
  deep-learned generative models of surfaces}. In \bibinfo{booktitle}{{\em
  Computer Graphics Forum}}, Vol.~\bibinfo{volume}{34}. Wiley Online Library,
  \bibinfo{pages}{25--38}.
\newblock


\bibitem[\protect\citeauthoryear{Huang, Su, and Guibas}{Huang
  et~al\mbox{.}}{2018}]%
        {huang2018robust}
\bibfield{author}{\bibinfo{person}{Jingwei Huang}, \bibinfo{person}{Hao Su},
  {and} \bibinfo{person}{Leonidas Guibas}.} \bibinfo{year}{2018}\natexlab{}.
\newblock \showarticletitle{Robust watertight manifold surface generation
  method for {ShapeNet} models}.
\newblock \bibinfo{journal}{{\em arXiv preprint arXiv:1802.01698\/}}
  (\bibinfo{year}{2018}).
\newblock


\bibitem[\protect\citeauthoryear{Huang, Fu, Wei, and Hu}{Huang
  et~al\mbox{.}}{2016}]%
        {Huang2016TVCG}
\bibfield{author}{\bibinfo{person}{Shi{-}Sheng Huang}, \bibinfo{person}{Hongbo
  Fu}, \bibinfo{person}{Ling{-}Yu Wei}, {and} \bibinfo{person}{Shi{-}Min Hu}.}
  \bibinfo{year}{2016}\natexlab{}.
\newblock \showarticletitle{Support Substructures: Support-Induced Part-Level
  Structural Representation}.
\newblock \bibinfo{journal}{{\em {IEEE} Trans. Vis. Comput. Graph.\/}}
  \bibinfo{volume}{22}, \bibinfo{number}{8} (\bibinfo{year}{2016}),
  \bibinfo{pages}{2024--2036}.
\newblock


\bibitem[\protect\citeauthoryear{Jack, Pontes, Sridharan, Fookes, Shirazi,
  Maire, and Eriksson}{Jack et~al\mbox{.}}{2018}]%
        {Dominic2018}
\bibfield{author}{\bibinfo{person}{Dominic Jack}, \bibinfo{person}{Jhony~K
  Pontes}, \bibinfo{person}{Sridha Sridharan}, \bibinfo{person}{Clinton
  Fookes}, \bibinfo{person}{Sareh Shirazi}, \bibinfo{person}{Frederic Maire},
  {and} \bibinfo{person}{Anders Eriksson}.} \bibinfo{year}{2018}\natexlab{}.
\newblock \showarticletitle{Learning free-form deformations for 3d object
  reconstruction}. In \bibinfo{booktitle}{{\em Asian Conference on Computer
  Vision}}. Springer, \bibinfo{pages}{317--333}.
\newblock


\bibitem[\protect\citeauthoryear{Kim, Li, Mitra, Chaudhuri, DiVerdi, and
  Funkhouser}{Kim et~al\mbox{.}}{2013}]%
        {Kim2013TOG}
\bibfield{author}{\bibinfo{person}{Vladimir~G. Kim}, \bibinfo{person}{Wilmot
  Li}, \bibinfo{person}{Niloy~J. Mitra}, \bibinfo{person}{Siddhartha
  Chaudhuri}, \bibinfo{person}{Stephen DiVerdi}, {and} \bibinfo{person}{Thomas
  Funkhouser}.} \bibinfo{year}{2013}\natexlab{}.
\newblock \showarticletitle{Learning Part-based Templates from Large
  Collections of {3D} Shapes}.
\newblock \bibinfo{journal}{{\em ACM Trans. Graph.\/}} \bibinfo{volume}{32},
  \bibinfo{number}{4}, Article \bibinfo{articleno}{70} (\bibinfo{date}{July}
  \bibinfo{year}{2013}), \bibinfo{numpages}{12}~pages.
\newblock


\bibitem[\protect\citeauthoryear{Kingma and Ba}{Kingma and Ba}{2015}]%
        {adamsolver}
\bibfield{author}{\bibinfo{person}{Diederik~P. Kingma} {and}
  \bibinfo{person}{Jimmy Ba}.} \bibinfo{year}{2015}\natexlab{}.
\newblock \showarticletitle{Adam: A Method for Stochastic Optimization}. In
  \bibinfo{booktitle}{{\em International Conference on Learning Representations
  (ICLR)}}.
\newblock


\bibitem[\protect\citeauthoryear{Kingma and Welling}{Kingma and
  Welling}{2013}]%
        {kingma2013auto}
\bibfield{author}{\bibinfo{person}{Diederik~P Kingma} {and}
  \bibinfo{person}{Max Welling}.} \bibinfo{year}{2013}\natexlab{}.
\newblock \showarticletitle{Auto-encoding variational bayes}.
\newblock \bibinfo{journal}{{\em arXiv preprint arXiv:1312.6114\/}}
  (\bibinfo{year}{2013}).
\newblock


\bibitem[\protect\citeauthoryear{Li, Xu, Chaudhuri, Yumer, Zhang, and
  Guibas}{Li et~al\mbox{.}}{2017}]%
        {li_sig17}
\bibfield{author}{\bibinfo{person}{Jun Li}, \bibinfo{person}{Kai Xu},
  \bibinfo{person}{Siddhartha Chaudhuri}, \bibinfo{person}{Ersin Yumer},
  \bibinfo{person}{Hao Zhang}, {and} \bibinfo{person}{Leonidas~J. Guibas}.}
  \bibinfo{year}{2017}\natexlab{}.
\newblock \showarticletitle{{GRASS}: Generative Recursive Autoencoders for
  Shape Structures}.
\newblock \bibinfo{journal}{{\em ACM Trans. Graph.\/}} \bibinfo{volume}{36},
  \bibinfo{number}{4} (\bibinfo{year}{2017}), \bibinfo{pages}{52:1--52:14}.
\newblock


\bibitem[\protect\citeauthoryear{Maron, Galun, Aigerman, Trope, Dym, Yumer,
  Kim, and Lipman}{Maron et~al\mbox{.}}{2017}]%
        {Maron2017TOG}
\bibfield{author}{\bibinfo{person}{Haggai Maron}, \bibinfo{person}{Meirav
  Galun}, \bibinfo{person}{Noam Aigerman}, \bibinfo{person}{Miri Trope},
  \bibinfo{person}{Nadav Dym}, \bibinfo{person}{Ersin Yumer},
  \bibinfo{person}{Vladimir~G Kim}, {and} \bibinfo{person}{Yaron Lipman}.}
  \bibinfo{year}{2017}\natexlab{}.
\newblock \showarticletitle{Convolutional neural networks on surfaces via
  seamless toric covers.}
\newblock \bibinfo{journal}{{\em ACM Trans. Graph.\/}} \bibinfo{volume}{36},
  \bibinfo{number}{4} (\bibinfo{year}{2017}), \bibinfo{pages}{71:1--71:10}.
\newblock


\bibitem[\protect\citeauthoryear{Meng, Gao, Lai, and Manocha}{Meng
  et~al\mbox{.}}{2019}]%
        {meng2018vv}
\bibfield{author}{\bibinfo{person}{Hsien-Yu Meng}, \bibinfo{person}{Lin Gao},
  \bibinfo{person}{Yu-Kun Lai}, {and} \bibinfo{person}{Dinesh Manocha}.}
  \bibinfo{year}{2019}\natexlab{}.
\newblock \showarticletitle{VV-Net: Voxel VAE Net with Group Convolutions for
  Point Cloud Segmentation}. In \bibinfo{booktitle}{{\em 2019 IEEE
  International Conference on Computer Vision (ICCV)}}.
\newblock


\bibitem[\protect\citeauthoryear{Mescheder, Oechsle, Niemeyer, Nowozin, and
  Geiger}{Mescheder et~al\mbox{.}}{2019}]%
        {mescheder2019-Occupancy}
\bibfield{author}{\bibinfo{person}{Lars Mescheder}, \bibinfo{person}{Michael
  Oechsle}, \bibinfo{person}{Michael Niemeyer}, \bibinfo{person}{Sebastian
  Nowozin}, {and} \bibinfo{person}{Andreas Geiger}.}
  \bibinfo{year}{2019}\natexlab{}.
\newblock \showarticletitle{Occupancy networks: Learning 3d reconstruction in
  function space}. In \bibinfo{booktitle}{{\em Proceedings of the IEEE
  Conference on Computer Vision and Pattern Recognition}}.
  \bibinfo{pages}{4460--4470}.
\newblock


\bibitem[\protect\citeauthoryear{Mitra, Wand, Zhang, Cohen-Or, Kim, and
  Huang}{Mitra et~al\mbox{.}}{2013}]%
        {Mitra2013}
\bibfield{author}{\bibinfo{person}{Niloy Mitra}, \bibinfo{person}{Michael
  Wand}, \bibinfo{person}{Hao~(Richard) Zhang}, \bibinfo{person}{Daniel
  Cohen-Or}, \bibinfo{person}{Vladimir Kim}, {and} \bibinfo{person}{Qi-Xing
  Huang}.} \bibinfo{year}{2013}\natexlab{}.
\newblock \showarticletitle{Structure-aware Shape Processing}. In
  \bibinfo{booktitle}{{\em SIGGRAPH Asia 2013 Courses}}.
  \bibinfo{publisher}{ACM}, \bibinfo{address}{New York, NY, USA}, Article
  \bibinfo{articleno}{1}, \bibinfo{numpages}{20}~pages.
\newblock


\bibitem[\protect\citeauthoryear{Mo, Guerrero, Yi, Su, Wonka, Mitra, and
  Guibas}{Mo et~al\mbox{.}}{2019a}]%
        {mo2019structurenet}
\bibfield{author}{\bibinfo{person}{Kaichun Mo}, \bibinfo{person}{Paul
  Guerrero}, \bibinfo{person}{Li Yi}, \bibinfo{person}{Hao Su},
  \bibinfo{person}{Peter Wonka}, \bibinfo{person}{Niloy Mitra}, {and}
  \bibinfo{person}{Leonidas Guibas}.} \bibinfo{year}{2019}\natexlab{a}.
\newblock \showarticletitle{{StructureNet}: Hierarchical Graph Networks for
  {3D} Shape Generation}.
\newblock \bibinfo{journal}{{\em ACM Trans. Graph.\/}} \bibinfo{volume}{38},
  \bibinfo{number}{6} (\bibinfo{year}{2019}).
\newblock


\bibitem[\protect\citeauthoryear{Mo, Zhu, Chang, Yi, Tripathi, Guibas, and
  Su}{Mo et~al\mbox{.}}{2019b}]%
        {mo2018partnet}
\bibfield{author}{\bibinfo{person}{Kaichun Mo}, \bibinfo{person}{Shilin Zhu},
  \bibinfo{person}{Angel~X Chang}, \bibinfo{person}{Li Yi},
  \bibinfo{person}{Subarna Tripathi}, \bibinfo{person}{Leonidas~J Guibas},
  {and} \bibinfo{person}{Hao Su}.} \bibinfo{year}{2019}\natexlab{b}.
\newblock \showarticletitle{PartNet: A Large-scale Benchmark for Fine-grained
  and Hierarchical Part-level 3D Object Understanding}. In
  \bibinfo{booktitle}{{\em Proceedings of the IEEE Conference on Computer
  Vision and Pattern Recognition}}. \bibinfo{pages}{909--918}.
\newblock


\bibitem[\protect\citeauthoryear{Nash and Williams}{Nash and Williams}{2017}]%
        {nash2017shape}
\bibfield{author}{\bibinfo{person}{Charlie Nash} {and}
  \bibinfo{person}{Christopher~KI Williams}.} \bibinfo{year}{2017}\natexlab{}.
\newblock \showarticletitle{The shape variational autoencoder: A deep
  generative model of part-segmented {3D} objects}. In \bibinfo{booktitle}{{\em
  Computer Graphics Forum}}, Vol.~\bibinfo{volume}{36}. Wiley Online Library,
  \bibinfo{pages}{1--12}.
\newblock


\bibitem[\protect\citeauthoryear{Ovsjanikov, Li, Guibas, and Mitra}{Ovsjanikov
  et~al\mbox{.}}{2011}]%
        {Ovsjanikov2011}
\bibfield{author}{\bibinfo{person}{Maks Ovsjanikov}, \bibinfo{person}{Wilmot
  Li}, \bibinfo{person}{Leonidas Guibas}, {and} \bibinfo{person}{Niloy~J.
  Mitra}.} \bibinfo{year}{2011}\natexlab{}.
\newblock \showarticletitle{Exploration of Continuous Variability in
  Collections of {3D} Shapes}.
\newblock \bibinfo{journal}{{\em ACM Trans. Graph.\/}} \bibinfo{volume}{30},
  \bibinfo{number}{4}, Article \bibinfo{articleno}{33} (\bibinfo{date}{July}
  \bibinfo{year}{2011}), \bibinfo{numpages}{10}~pages.
\newblock


\bibitem[\protect\citeauthoryear{Park, Florence, Straub, Newcombe, and
  Lovegrove}{Park et~al\mbox{.}}{2019}]%
        {park2019-DeepSDF}
\bibfield{author}{\bibinfo{person}{Jeong~Joon Park}, \bibinfo{person}{Peter
  Florence}, \bibinfo{person}{Julian Straub}, \bibinfo{person}{Richard
  Newcombe}, {and} \bibinfo{person}{Steven Lovegrove}.}
  \bibinfo{year}{2019}\natexlab{}.
\newblock \showarticletitle{DeepSDF: Learning Continuous Signed Distance
  Functions for Shape Representation}. In \bibinfo{booktitle}{{\em Proceedings
  of the IEEE Conference on Computer Vision and Pattern Recognition}}.
  \bibinfo{pages}{165--174}.
\newblock


\bibitem[\protect\citeauthoryear{Podolak, Shilane, Golovinskiy, Rusinkiewicz,
  and Funkhouser}{Podolak et~al\mbox{.}}{2006}]%
        {podolak2006planar}
\bibfield{author}{\bibinfo{person}{Joshua Podolak}, \bibinfo{person}{Philip
  Shilane}, \bibinfo{person}{Aleksey Golovinskiy}, \bibinfo{person}{Szymon
  Rusinkiewicz}, {and} \bibinfo{person}{Thomas Funkhouser}.}
  \bibinfo{year}{2006}\natexlab{}.
\newblock \showarticletitle{A planar-reflective symmetry transform for {3D}
  shapes}.
\newblock \bibinfo{journal}{{\em ACM Trans. Graph.\/}} \bibinfo{volume}{25},
  \bibinfo{number}{3} (\bibinfo{year}{2006}), \bibinfo{pages}{549--559}.
\newblock


\bibitem[\protect\citeauthoryear{Pons-Moll, Romero, Mahmood, and
  Black}{Pons-Moll et~al\mbox{.}}{2015}]%
        {pons2015dyna}
\bibfield{author}{\bibinfo{person}{Gerard Pons-Moll}, \bibinfo{person}{Javier
  Romero}, \bibinfo{person}{Naureen Mahmood}, {and} \bibinfo{person}{Michael~J.
  Black}.} \bibinfo{year}{2015}\natexlab{}.
\newblock \showarticletitle{Dyna: A model of dynamic human shape in motion}.
\newblock \bibinfo{journal}{{\em ACM Trans. Graph.\/}} \bibinfo{volume}{34},
  \bibinfo{number}{4} (\bibinfo{year}{2015}), \bibinfo{pages}{120:1--120:14}.
\newblock


\bibitem[\protect\citeauthoryear{Poulenard and Ovsjanikov}{Poulenard and
  Ovsjanikov}{2018}]%
        {Poulenard2018}
\bibfield{author}{\bibinfo{person}{Adrien Poulenard} {and}
  \bibinfo{person}{Maks Ovsjanikov}.} \bibinfo{year}{2018}\natexlab{}.
\newblock \showarticletitle{Multi-directional Geodesic Neural Networks via
  Equivariant Convolution}.
\newblock \bibinfo{journal}{{\em ACM Trans. Graph.\/}} \bibinfo{volume}{37},
  \bibinfo{number}{6} (\bibinfo{year}{2018}), \bibinfo{pages}{236:1--236:14}.
\newblock


\bibitem[\protect\citeauthoryear{Qi, Su, Mo, and Guibas}{Qi
  et~al\mbox{.}}{2017}]%
        {Qi2017cvpr}
\bibfield{author}{\bibinfo{person}{Charles~Ruizhongtai Qi},
  \bibinfo{person}{Hao Su}, \bibinfo{person}{Kaichun Mo}, {and}
  \bibinfo{person}{Leonidas~J. Guibas}.} \bibinfo{year}{2017}\natexlab{}.
\newblock \showarticletitle{{PointNet}: Deep Learning on Point Sets for {3D}
  Classification and Segmentation.}. In \bibinfo{booktitle}{{\em IEEE
  Conference on Computer Vision and Pattern Recognition (CVPR)}}.
  \bibinfo{pages}{77--85}.
\newblock


\bibitem[\protect\citeauthoryear{Qi, Su, Nie{\ss}ner, Dai, Yan, and Guibas}{Qi
  et~al\mbox{.}}{2016}]%
        {Qi2016}
\bibfield{author}{\bibinfo{person}{Charles~Ruizhongtai Qi},
  \bibinfo{person}{Hao Su}, \bibinfo{person}{Matthias Nie{\ss}ner},
  \bibinfo{person}{Angela Dai}, \bibinfo{person}{Mengyuan Yan}, {and}
  \bibinfo{person}{Leonidas~J. Guibas}.} \bibinfo{year}{2016}\natexlab{}.
\newblock \showarticletitle{Volumetric and Multi-view {CNNs} for Object
  Classification on {3D} Data}. In \bibinfo{booktitle}{{\em IEEE Conference on
  Computer Vision and Pattern Recognition (CVPR)}}.
  \bibinfo{pages}{5648--5656}.
\newblock


\bibitem[\protect\citeauthoryear{Qi, Yi, Su, and Guibas}{Qi
  et~al\mbox{.}}{2017}]%
        {Qi2017nips}
\bibfield{author}{\bibinfo{person}{Charles~Ruizhongtai Qi}, \bibinfo{person}{Li
  Yi}, \bibinfo{person}{Hao Su}, {and} \bibinfo{person}{Leonidas~J. Guibas}.}
  \bibinfo{year}{2017}\natexlab{}.
\newblock \showarticletitle{{PointNet++}: Deep Hierarchical Feature Learning on
  Point Sets in a Metric Space.}. In \bibinfo{booktitle}{{\em Advances in
  Neural Information Processing Systems (NIPS)}}. \bibinfo{pages}{5105--5114}.
\newblock


\bibitem[\protect\citeauthoryear{Sinha, Bai, and Ramani}{Sinha
  et~al\mbox{.}}{2016}]%
        {Sinha2016DeepL3}
\bibfield{author}{\bibinfo{person}{Ayan Sinha}, \bibinfo{person}{Jing Bai},
  {and} \bibinfo{person}{Karthik Ramani}.} \bibinfo{year}{2016}\natexlab{}.
\newblock \showarticletitle{Deep Learning {3D} Shape Surfaces Using Geometry
  Images}. In \bibinfo{booktitle}{{\em European Conference on Computer Vision
  (ECCV)}}. Springer, \bibinfo{pages}{223--240}.
\newblock


\bibitem[\protect\citeauthoryear{Soltani, Huang, Wu, Kulkarni, and
  Tenenbaum}{Soltani et~al\mbox{.}}{2017}]%
        {3DVAE}
\bibfield{author}{\bibinfo{person}{Amir~Arsalan Soltani},
  \bibinfo{person}{Haibin Huang}, \bibinfo{person}{Jiajun Wu},
  \bibinfo{person}{Tejas~D Kulkarni}, {and} \bibinfo{person}{Joshua~B
  Tenenbaum}.} \bibinfo{year}{2017}\natexlab{}.
\newblock \showarticletitle{Synthesizing {3D} Shapes via Modeling Multi-View
  Depth Maps and Silhouettes with Deep Generative Networks}. In
  \bibinfo{booktitle}{{\em IEEE Conference on Computer Vision and Pattern
  Recognition (CVPR)}}. \bibinfo{pages}{1511--1519}.
\newblock


\bibitem[\protect\citeauthoryear{Sorkine and Alexa}{Sorkine and Alexa}{2007}]%
        {ARAP2007}
\bibfield{author}{\bibinfo{person}{Olga Sorkine} {and} \bibinfo{person}{Marc
  Alexa}.} \bibinfo{year}{2007}\natexlab{}.
\newblock \showarticletitle{As-Rigid-As-Possible Surface Modeling}. In
  \bibinfo{booktitle}{{\em Proceedings of EUROGRAPHICS/ACM SIGGRAPH Symposium
  on Geometry Processing}}. \bibinfo{pages}{109--116}.
\newblock


\bibitem[\protect\citeauthoryear{Su, Maji, Kalogerakis, and Learned-Miller}{Su
  et~al\mbox{.}}{2015}]%
        {Su2015ICCV}
\bibfield{author}{\bibinfo{person}{Hang Su}, \bibinfo{person}{Subhransu Maji},
  \bibinfo{person}{Evangelos Kalogerakis}, {and} \bibinfo{person}{Erik
  Learned-Miller}.} \bibinfo{year}{2015}\natexlab{}.
\newblock \showarticletitle{Multi-view convolutional neural networks for {3D}
  shape recognition}. In \bibinfo{booktitle}{{\em Proceedings of the IEEE
  international conference on computer vision}}. \bibinfo{pages}{945--953}.
\newblock


\bibitem[\protect\citeauthoryear{Tan, Gao, Lai, and Xia}{Tan
  et~al\mbox{.}}{2018}]%
        {meshvae2017}
\bibfield{author}{\bibinfo{person}{Qingyang Tan}, \bibinfo{person}{Lin Gao},
  \bibinfo{person}{Yu{-}Kun Lai}, {and} \bibinfo{person}{Shihong Xia}.}
  \bibinfo{year}{2018}\natexlab{}.
\newblock \showarticletitle{Variational Autoencoders for Deforming {3D} Mesh
  Models}. In \bibinfo{booktitle}{{\em IEEE Conference on Computer Vision and
  Pattern Recognition (CVPR)}}. \bibinfo{pages}{5841--5850}.
\newblock


\bibitem[\protect\citeauthoryear{Tatarchenko, Dosovitskiy, and
  Brox}{Tatarchenko et~al\mbox{.}}{2017}]%
        {Octree2017iccv}
\bibfield{author}{\bibinfo{person}{Maxim Tatarchenko}, \bibinfo{person}{Alexey
  Dosovitskiy}, {and} \bibinfo{person}{Thomas Brox}.}
  \bibinfo{year}{2017}\natexlab{}.
\newblock \showarticletitle{Octree Generating Networks: Efficient Convolutional
  Architectures for High-resolution {3D} Outputs}. In \bibinfo{booktitle}{{\em
  IEEE International Conference on Computer Vision (ICCV)}}.
  \bibinfo{pages}{2088--2096}.
\newblock


\bibitem[\protect\citeauthoryear{Wang, Schor, Hu, Huang, Cohen-Or, and
  Huang}{Wang et~al\mbox{.}}{2018a}]%
        {Wang2018TOG}
\bibfield{author}{\bibinfo{person}{Hao Wang}, \bibinfo{person}{Nadav Schor},
  \bibinfo{person}{Ruizhen Hu}, \bibinfo{person}{Haibin Huang},
  \bibinfo{person}{Daniel Cohen-Or}, {and} \bibinfo{person}{Hui Huang}.}
  \bibinfo{year}{2018}\natexlab{a}.
\newblock \showarticletitle{Global-to-local Generative Model for {3D} Shapes}.
\newblock \bibinfo{journal}{{\em ACM Trans. Graph.\/}} \bibinfo{volume}{37},
  \bibinfo{number}{6}, Article \bibinfo{articleno}{214} (\bibinfo{date}{Dec.}
  \bibinfo{year}{2018}), \bibinfo{numpages}{10}~pages.
\newblock
\showISSN{0730-0301}


\bibitem[\protect\citeauthoryear{Wang, Zhang, Li, Fu, Liu, and Jiang}{Wang
  et~al\mbox{.}}{2018c}]%
        {wang2018pixel2mesh}
\bibfield{author}{\bibinfo{person}{Nanyang Wang}, \bibinfo{person}{Yinda
  Zhang}, \bibinfo{person}{Zhuwen Li}, \bibinfo{person}{Yanwei Fu},
  \bibinfo{person}{Wei Liu}, {and} \bibinfo{person}{Yu-Gang Jiang}.}
  \bibinfo{year}{2018}\natexlab{c}.
\newblock \showarticletitle{Pixel2mesh: Generating 3d mesh models from single
  rgb images}. In \bibinfo{booktitle}{{\em European Conference on Computer
  Vision (ECCV)}}. \bibinfo{pages}{52--67}.
\newblock


\bibitem[\protect\citeauthoryear{Wang, Liu, Guo, Sun, and Tong}{Wang
  et~al\mbox{.}}{2017}]%
        {Wang2017}
\bibfield{author}{\bibinfo{person}{Peng-Shuai Wang}, \bibinfo{person}{Yang
  Liu}, \bibinfo{person}{Yu-Xiao Guo}, \bibinfo{person}{Chun-Yu Sun}, {and}
  \bibinfo{person}{Xin Tong}.} \bibinfo{year}{2017}\natexlab{}.
\newblock \showarticletitle{{O-CNN}: Octree-based Convolutional Neural Networks
  for {3D} Shape Analysis}.
\newblock \bibinfo{journal}{{\em ACM Trans. Graph.\/}} \bibinfo{volume}{36},
  \bibinfo{number}{4} (\bibinfo{year}{2017}), \bibinfo{pages}{72:1--72:11}.
\newblock


\bibitem[\protect\citeauthoryear{Wang, Sun, Liu, and Tong}{Wang
  et~al\mbox{.}}{2018b}]%
        {Wang2018ocnn}
\bibfield{author}{\bibinfo{person}{Peng-Shuai Wang}, \bibinfo{person}{Chun-Yu
  Sun}, \bibinfo{person}{Yang Liu}, {and} \bibinfo{person}{Xin Tong}.}
  \bibinfo{year}{2018}\natexlab{b}.
\newblock \showarticletitle{Adaptive {O-CNN}: A Patch-based Deep Representation
  of {3D} Shapes}.
\newblock \bibinfo{journal}{{\em ACM Trans. Graph.\/}} \bibinfo{volume}{37},
  \bibinfo{number}{6}, Article \bibinfo{articleno}{217} (\bibinfo{date}{Dec.}
  \bibinfo{year}{2018}), \bibinfo{numpages}{11}~pages.
\newblock
\showISSN{0730-0301}


\bibitem[\protect\citeauthoryear{Wu, Zhang, Xue, Freeman, and Tenenbaum}{Wu
  et~al\mbox{.}}{2016}]%
        {3dgan2016}
\bibfield{author}{\bibinfo{person}{Jiajun Wu}, \bibinfo{person}{Chengkai
  Zhang}, \bibinfo{person}{Tianfan Xue}, \bibinfo{person}{Bill Freeman}, {and}
  \bibinfo{person}{Josh Tenenbaum}.} \bibinfo{year}{2016}\natexlab{}.
\newblock \showarticletitle{Learning a Probabilistic Latent Space of Object
  Shapes via {3D} Generative-Adversarial Modeling.}. In
  \bibinfo{booktitle}{{\em Advances in Neural Information Processing Systems
  (NIPS)}}. \bibinfo{pages}{82--90}.
\newblock


\bibitem[\protect\citeauthoryear{Wu, Song, Khosla, Yu, Zhang, Tang, and
  Xiao}{Wu et~al\mbox{.}}{2015}]%
        {Wu_2015_CVPR}
\bibfield{author}{\bibinfo{person}{Zhirong Wu}, \bibinfo{person}{Shuran Song},
  \bibinfo{person}{Aditya Khosla}, \bibinfo{person}{Fisher Yu},
  \bibinfo{person}{Linguang Zhang}, \bibinfo{person}{Xiaoou Tang}, {and}
  \bibinfo{person}{Jianxiong Xiao}.} \bibinfo{year}{2015}\natexlab{}.
\newblock \showarticletitle{3D ShapeNets: A deep representation for volumetric
  shapes}. In \bibinfo{booktitle}{{\em IEEE Conference on Computer Vision and
  Pattern Recognition (CVPR)}}. \bibinfo{pages}{1912--1920}.
\newblock


\bibitem[\protect\citeauthoryear{Wu, Wang, Lin, Lischinski, Cohen-Or, and
  Huang}{Wu et~al\mbox{.}}{2019}]%
        {pageSAGnet19}
\bibfield{author}{\bibinfo{person}{Zhijie Wu}, \bibinfo{person}{Xiang Wang},
  \bibinfo{person}{Di Lin}, \bibinfo{person}{Dani Lischinski},
  \bibinfo{person}{Daniel Cohen-Or}, {and} \bibinfo{person}{Hui Huang}.}
  \bibinfo{year}{2019}\natexlab{}.
\newblock \showarticletitle{SAGNet: Structure-aware Generative Network for
  3D-Shape Modeling}.
\newblock \bibinfo{journal}{{\em ACM Trans. Graph.\/}} \bibinfo{volume}{38},
  \bibinfo{number}{4} (\bibinfo{year}{2019}), \bibinfo{pages}{91:1--91:14}.
\newblock


\bibitem[\protect\citeauthoryear{Xin, Li, Luo, Shao, Yu, Zhou, and Zheng}{Xin
  et~al\mbox{.}}{2018}]%
        {Xin2019}
\bibfield{author}{\bibinfo{person}{Chen Xin}, \bibinfo{person}{Yuwei Li},
  \bibinfo{person}{Xi Luo}, \bibinfo{person}{Tianjia Shao},
  \bibinfo{person}{Jingyi Yu}, \bibinfo{person}{Kun Zhou}, {and}
  \bibinfo{person}{Youyi Zheng}.} \bibinfo{year}{2018}\natexlab{}.
\newblock \showarticletitle{{AutoSweep}: Recovering {3D} Editable Objects from
  a Single Photograph}.
\newblock \bibinfo{journal}{{\em IEEE Transactions on Visualization and
  Computer Graphics\/}} (\bibinfo{year}{2018}).
\newblock
\showISSN{1077-2626}


\bibitem[\protect\citeauthoryear{Yang, Xu, Chen, and Fu}{Yang
  et~al\mbox{.}}{2017}]%
        {yang2017view}
\bibfield{author}{\bibinfo{person}{Sheng Yang}, \bibinfo{person}{Jie Xu},
  \bibinfo{person}{Kang Chen}, {and} \bibinfo{person}{Hongbo Fu}.}
  \bibinfo{year}{2017}\natexlab{}.
\newblock \showarticletitle{View suggestion for interactive segmentation of
  indoor scenes}.
\newblock \bibinfo{journal}{{\em Computational Visual Media\/}}
  \bibinfo{volume}{3}, \bibinfo{number}{2} (\bibinfo{year}{2017}),
  \bibinfo{pages}{131--146}.
\newblock


\bibitem[\protect\citeauthoryear{Yi, Kim, Ceylan, Shen, Yan, Su, Lu, Huang,
  Sheffer, Guibas, et~al\mbox{.}}{Yi et~al\mbox{.}}{2016}]%
        {Yi16}
\bibfield{author}{\bibinfo{person}{Li Yi}, \bibinfo{person}{Vladimir~G Kim},
  \bibinfo{person}{Duygu Ceylan}, \bibinfo{person}{I Shen},
  \bibinfo{person}{Mengyan Yan}, \bibinfo{person}{Hao Su},
  \bibinfo{person}{Cewu Lu}, \bibinfo{person}{Qixing Huang},
  \bibinfo{person}{Alla Sheffer}, \bibinfo{person}{Leonidas Guibas},
  {et~al\mbox{.}}} \bibinfo{year}{2016}\natexlab{}.
\newblock \showarticletitle{A scalable active framework for region annotation
  in 3d shape collections}.
\newblock \bibinfo{journal}{{\em ACM Trans. Graph.\/}} \bibinfo{volume}{35},
  \bibinfo{number}{6} (\bibinfo{year}{2016}), \bibinfo{pages}{210:1--210:12}.
\newblock


\bibitem[\protect\citeauthoryear{Yin, Huang, Cohen-Or, and Zhang}{Yin
  et~al\mbox{.}}{2018}]%
        {yin_sig18}
\bibfield{author}{\bibinfo{person}{Kangxue Yin}, \bibinfo{person}{Hui Huang},
  \bibinfo{person}{Daniel Cohen-Or}, {and} \bibinfo{person}{Hao Zhang}.}
  \bibinfo{year}{2018}\natexlab{}.
\newblock \showarticletitle{{P2P-NET}: Bidirectional Point Displacement Net for
  Shape Transform}.
\newblock \bibinfo{journal}{{\em ACM Trans. Graph.\/}} \bibinfo{volume}{37},
  \bibinfo{number}{4} (\bibinfo{year}{2018}), \bibinfo{pages}{152:1--152:13}.
\newblock


\bibitem[\protect\citeauthoryear{Zhou, Synder, Guo, and Shum}{Zhou
  et~al\mbox{.}}{2004}]%
        {Zhou2004}
\bibfield{author}{\bibinfo{person}{Kun Zhou}, \bibinfo{person}{John Synder},
  \bibinfo{person}{Baining Guo}, {and} \bibinfo{person}{Heung-Yeung Shum}.}
  \bibinfo{year}{2004}\natexlab{}.
\newblock \showarticletitle{Iso-charts: stretch-driven mesh parameterization
  using spectral analysis}. In \bibinfo{booktitle}{{\em Proceedings of the 2004
  Eurographics/ACM SIGGRAPH symposium on Geometry processing}}. ACM,
  \bibinfo{pages}{45--54}.
\newblock


\bibitem[\protect\citeauthoryear{Zollh\"{o}fer, Nie{\ss}ner, Izadi, Rehmann,
  Zach, Fisher, Wu, Fitzgibbon, Loop, Theobalt, and Stamminger}{Zollh\"{o}fer
  et~al\mbox{.}}{2014}]%
        {Zollhofer2014}
\bibfield{author}{\bibinfo{person}{Michael Zollh\"{o}fer},
  \bibinfo{person}{Matthias Nie{\ss}ner}, \bibinfo{person}{Shahram Izadi},
  \bibinfo{person}{Christoph Rehmann}, \bibinfo{person}{Christopher Zach},
  \bibinfo{person}{Matthew Fisher}, \bibinfo{person}{Chenglei Wu},
  \bibinfo{person}{Andrew Fitzgibbon}, \bibinfo{person}{Charles Loop},
  \bibinfo{person}{Christian Theobalt}, {and} \bibinfo{person}{Marc
  Stamminger}.} \bibinfo{year}{2014}\natexlab{}.
\newblock \showarticletitle{Real-time Non-rigid Reconstruction Using an RGB-D
  Camera}.
\newblock \bibinfo{journal}{{\em ACM Trans. Graph.\/}} \bibinfo{volume}{33},
  \bibinfo{number}{4}, Article \bibinfo{articleno}{156} (\bibinfo{date}{July}
  \bibinfo{year}{2014}), \bibinfo{numpages}{12}~pages.
\newblock
\showISSN{0730-0301}


\end{thebibliography}

\section*{Appendix: support relationship formulation.}

Let $\tilde{b}_i$ and $\tilde{b}_j$ be the bounding boxes of parts $i$ and $j$ projected onto \YL{the plane orthogonal to the supporting direction}. They should satisfy either $\tilde{b}_i \subseteq \tilde{b}_j$ or $\tilde{b}_j \subseteq \tilde{b}_i$. This constraint can be formulated as an integer programming problem and solved efficiently during the optimization as follows:

Let $t_1$ and $t_2$ be the two directions in the tangential plane. Denote by $\delta^{i,j}_1$ and $\delta^{i,j}_2$ two auxiliary binary variables,  $\delta^{i,j}_1, \delta^{i,j}_2 \in \{0, 1\}$,
this is equivalent to
\begin{align}
\mathbf{p}^{'}_j[t_1] - \mathbf{q}^{'}_j[t_1] \leq  \mathbf{p}^{'}_i[t_1] - \mathbf{q}^{'}_i[t_1] + M \delta^{i,j}_1, \nonumber \\
\mathbf{p}^{'}_i[t_1] + \mathbf{q}^{'}_i[t_1] \leq \mathbf{p}^{'}_j[t_1] + \mathbf{q}^{'}_j[t_1] + M \delta^{i,j}_1, \nonumber \\
\mathbf{p}^{'}_j[t_2] - \mathbf{q}^{'}_j[t_2] \leq  \mathbf{p}^{'}_i[t_2] - \mathbf{q}^{'}_i[t_2] + M \delta^{i,j}_1, \nonumber \\
\mathbf{p}^{'}_i[t_2] + \mathbf{q}^{'}_i[t_2] \leq \mathbf{p}^{'}_j[t_2] + \mathbf{q}^{'}_j[t_2] + M \delta^{i,j}_1,  \label{eqn:d1} \\
\mathbf{p}^{'}_i[t_1] - \mathbf{q}^{'}_i[t_1] \leq  \mathbf{p}^{'}_j[t_1] - \mathbf{q}^{'}_j[t_1] + M \delta^{i,j}_2, \nonumber \\
\mathbf{p}^{'}_j[t_1] + \mathbf{q}^{'}_j[t_1] \leq \mathbf{p}^{'}_i[t_1] + \mathbf{q}^{'}_i[t_1] + M \delta^{i,j}_2, \nonumber \\
\mathbf{p}^{'}_i[t_2] - \mathbf{q}^{'}_i[t_2] \leq  \mathbf{p}^{'}_j[t_2] - \mathbf{q}^{'}_j[t_2] + M \delta^{i,j}_2, \nonumber \\
\mathbf{p}^{'}_j[t_2] + \mathbf{q}^{'}_j[t_2] \leq \mathbf{p}^{'}_i[t_2] + \mathbf{q}^{'}_i[t_2] + M \delta^{i,j}_2,  \label{eqn:d2} \\
\delta^{i,j}_1 + \delta^{i,j}_2 \leq 1, \label{eqn:dd}
\end{align}
where $M$ is a large positive number (larger than any possible coordinate in the shape), Eq.~(\ref{eqn:dd}) is true if at most one of $\delta^{i,j}_1$ or $\delta^{i,j}_2$ can be 1, i.e., at least one of them is 0. Without loss of generality, assuming $\delta^{i,j}_1 = 0$, then the set of equations in (\ref{eqn:d1}) without the term involving $M$ is true, meaning $\tilde{b}_i \subseteq \tilde{b}_j$. Similarly, when $\delta^{i,j}_2 = 0$, it satisfies that $\tilde{b}_j \subseteq \tilde{b}_i$.

\end{document}